\documentclass[12pt]{spieman}  
\usepackage{amsmath,amsfonts,amssymb}
\usepackage{graphicx}
\usepackage{setspace}
\usepackage{tocloft}
\usepackage{lineno}
\usepackage{titletoc}
\usepackage{float}
\usepackage{tabularx}
\usepackage{array}
\usepackage{ragged2e}

\title{Optical Pin Beams: Research Progresses and Emerging Applications}

\author[1,2,3,†,*]{Ze Zhang}
\author[2,4,†]{Hongwei Jiang}
\author[1,2,5,†]{Hongyue Xiao}
\author[2]{Meiling Guan}
\author[4]{Lu Gao}
\author[6,7,8]{Nikolaos K. Efremidis}
\author[1]{Hairong Xiao}
\author[8,*]{and Zhigang Chen}
\affil[1]{College of Photonics and Optical Engineering, Aerospace Information Technology University, Jinan, 250299, China}
\affil[2]{Aerospace Information Research Institute, Chinese Academy of Sciences, Beijing, 100094, China}
\affil[3]{Shandong Provincial Key Laboratory of Intelligent Photonic Transmission and Sensing, Aerospace Information Technology University, Jinan, 250101, China}
\affil[4]{China University of Geosciences (Beijing), Beijing, 100083, China}
\affil[5]{Shandong University, Jinan, 250100, China}
\affil[6]{Department of Mathematics and Applied Mathematics, University of Crete, Heraklion, Crete, 70013, Greece}
\affil[7]{Institute of Applied and Computational Mathematics, Foundation for Research and Technology-Hellas, Heraklion, Crete, 70013, Greece}
\affil[8]{The MOE Key Laboratory of Weak-Light Nonlinear Photonics, TEDA Applied Physics Institute and School of Physics, Nankai University, Tianjin, 300457, China}

\cftpagenumbersoff{figure}
\cftpagenumbersoff{table}
\begin{document}
\maketitle

\begin{abstract}
Optical pin beams (OPBs) represent a novel class of structured light fields engineered for resilient, long-distance propagation. Their exceptional stability and strong resistance to atmospheric turbulence make them a compelling alternative to conventional Gaussian and other structured beams for free-space optical systems. This review provides a comprehensive overview of the physical principles, generation strategies, experimental realizations, and emerging applications of OPBs. By precise spatial modulation of the optical wave vectors, OPBs achieve highly collimated, self-reconstructing propagation with distinctive pin-like features that confer remarkable robustness and self-healing capability. We further discuss several OPB derivatives—including vortex, inverted, and vortex-inverted OPBs—which expand the functional landscape by enabling flexible control over amplitude, phase, polarization, and orbital angular momentum. Experimentally, OPBs have demonstrated outstanding performance across diverse platforms, ranging from free-space and underwater optical communications to optical trapping and super-resolution imaging. With their unique combination of propagation stability, light-field tunability, and environmental adaptability, OPBs hold strong promise for next-generation optical communication, precision sensing, and advanced imaging technologies. This review summarizes recent research progresses in OPBs and highlights key opportunities and prospects for advancing their scientific discoveries and practical applications.
\end{abstract}


{\noindent \footnotesize\textbf{†}These authors contributed equally.}

{\noindent \footnotesize\textbf{*}Corresponding Author:  Ze Zhang, E-mail: zhangze@aircas.ac.cn, Zhigang Chen, E-mail: zgchen@nankai.edu.cn}
\vspace{1cm}
\begin{spacing}{1.0}   

\titlecontents{section}
              [2cm]
              {}%
              {\contentslabel{2.5em}}%
              {}%
              {\titlerule*[0.5pc]{$.$}\contentspage\hspace*{2cm}}%
\titlecontents{subsection}
              [3cm]
              {\small}%
              {\contentslabel{2.5em}}%
              {}%
              {\titlerule*[0.5pc]{$.$}\contentspage\hspace*{2cm}}%
\titlecontents{subsubsection}
              [4cm]
              {\small}%
              {\contentslabel{2.5em}}%
              {}%
              {\titlerule*[0.5pc]{$.$}\contentspage\hspace*{2cm}}%

\noindent \textbf{Table of Contents}\\
\end{spacing}
\renewcommand{\contentsname}{}
\begin{spacing}{1.0}
\tableofcontents
\end{spacing}

\begin{spacing}{1.5}
\section{Introduction and Key Advancements}
\label{sect:intro}  
Structured light-defined by the deliberate tailoring of a beam’s amplitude, phase, polarization, coherence, and topology-has profoundly expanded the frontiers of modern optics and photonics \cite{1,2,3,4,5,6,7,8,9,10,11,12,13,235,236}. Beyond conventional Gaussian beam (GS), diverse families of structured light, including Bessel\cite{14,15,16,17,18,19,20,21}, Airy\cite{2,22,23,24,25,26,27,28,29,30,31,237}, and vortex beams\cite{32,33,34,35,36,37}, have enabled non-diffracting propagation, self-healing, and orbital angular momentum (OAM) transport. These advances have driven transformative developments in optical communications, particle manipulation, and imaging\cite{34,35,36,37,38,39,40,41,42,43,44,45,46,47,48}. Nevertheless, despite their remarkable properties, most structured beams remain vulnerable to intensity spreading, phase distortion, or instability when propagating over long distances or through complex and turbulent environments\cite{49,50,51,52,53,54}. Overcoming these limitations calls for new beam architectures that combine strong directionality, high robustness against perturbations, and versatile light-field control.

The Optical Pin Beam (OPB) has recently emerged as a promising solution to these long-standing limitations (Fig. 1) \cite{55,56}. Conceptually distinct from Bessel or Airy beams, the OPBs are constructed through tailored modulation of spatial wave vectors, effectively suppressing transverse components to confine optical energy along a narrow, “pin-like” axis \cite{55,67,68}. This design endows OPBs with highly collimated, diffraction-resilient, and self-reconstructing propagation characteristics. The underlying mechanism can be understood from both Fourier and modal perspectives: by redistributing and mutually cancelling transverse wave vectors, the OPB realizes a quasi-one-dimensional propagation mode that maintains spatial coherence and robustness even under perturbations\cite{69}. The theoretical foundation of OPBs builds on the principles of beam superposition, angular spectrum engineering, and nonparaxial field evolution, offering a versatile platform for light-field manipulation\cite{7}.

\begin{figure}[H]
\begin{center}
\begin{tabular}{c}
\includegraphics[scale=0.75]{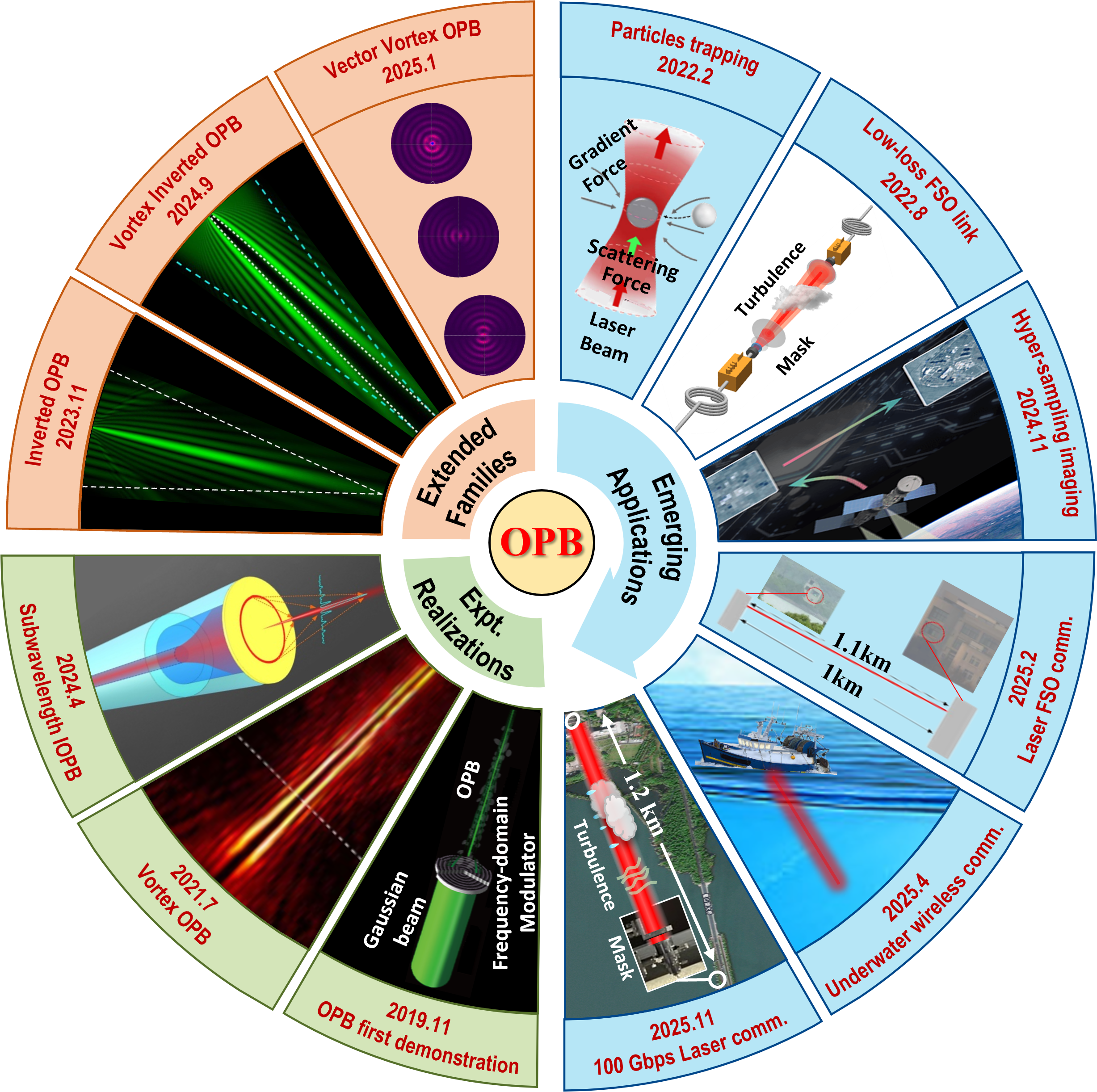}
\end{tabular}
\end{center}
\caption
{ \label{fig:example1}
Timeline of key milestones in the development of Optical Pin Beams (OPBs). OPBs were introduced as a new class of structured light engineered to enable quasi-one-dimensional, pin-like propagation distinct from conventional Bessel and Airy beams. Initial experiments demonstrated highly collimated, diffraction-resilient OPB propagation with exceptional robustness against atmospheric turbulence and scattering (2019-2021). Over the past few years, the OPB framework has been expanded to include vortex, inverted, and vortex-inverted OPBs, integrating phase singularities, intensity inversion, and polarization control while preserving intrinsic propagation stability (2023-2025). Since their first demonstrations, OPBs have been applied to free-space and underwater optical communications, optical trapping, particle manipulation, and super-resolution imaging, benefiting from strong axial confinement, self-reconstruction, and reduced environmental sensitivity (2022-2025)\cite{55,56,57,58,59,60,61,62,63,64,65,66}. Current research efforts are advancing OPBs toward high-power operation, integration with adaptive optics, and propagation in nonlinear and bio–soft-matter systems, targeting turbulence-immune communication, precision sensing and detection, and versatile multifunctional photonic platforms. }
\end{figure}

The first experimental realization of OPBs demonstrated stable beam propagation with exceptional robustness against scattering and turbulence, confirming theoretical predictions\cite{55,67}. Subsequent developments rapidly expanded the concept into a rich family of OPB derivatives, including vortex OPBs, inverted OPBs, and vortex-inverted OPBs\cite{56,58,59}. These variants introduce new degrees of freedom in optical field control by embedding phase singularities, polarization textures, or intensity inversions, thereby integrating the functionalities of structured light with the inherent resilience of OPBs. Such diversity has greatly broadened the scope of OPB-based research, allowing the exploration of novel light–matter interactions and complex propagation dynamics.

Driven by these advances, emerging applications of OPBs have begun to take shape across multiple domains (Fig. 1). In free-space optical communication (FSO), OPBs exhibit exceptional turbulence resilience and strong spatial-coherence retention, enabling more reliable long-distance data transmission\cite{62,65,66,70}. In underwater optical communications, their reduced scattering sensitivity supports high-quality signal transport in turbid environment\cite{65}. More recently, OPBs have been employed as a receiver in a high-performance 100 Gbps free-space optical communication system, highlighting their potential to redefine the boundaries of wireless data transfer and enable next-generation long-distance communication beyond conventional fiber-optic links\cite{66}.Beyond communication, OPBs have been employed in optical trapping, particle manipulation, and super-resolution imaging, benefiting from their strong axial confinement and propagation stability\cite{61,63}. These capabilities point toward broad implications for precision measurement, laser micro-machining, and even quantum light-field control in next-generation photonic systems\cite{71}.

In this review, we present a comprehensive overview of the research progress and application prospects of OPBs. We begin by introducing their physical mechanisms and theoretical foundations, followed by a discussion of experimental generation techniques and characteristic properties. We then summarize the development of OPB variants and review their expanding applications across diverse optical platforms. Finally, we outline current challenges and future directions, highlighting how the OPB paradigm may serve as a foundation for next- generation optical communication, imaging, and photonic information technologies. Given the rapid evolution of this field, this review aims to capture the essential developments while acknowledging that some recent advances may inevitably lie beyond its scope.

\section{Background and Underlying Mechanism of OPBs}

The stable propagation of laser beams and the suppression of disturbances remain long-standing challenges across nearly all laser-based applications, sustaining their place as key frontiers in optical research\cite{60,72,73,74,75,76,77,78,79,80,81,82,83,84}. On one hand, in long-distance laser applications, such as free-space optical communication, ground-based remote sensing, and laser mapping\cite{62,85,86,87,88,89,90,91,92,93,94,95,96,97,98,99,100,101,102,103,104}, stable beam propagation is essential for achieving efficient energy and signal transmission. On the other hand, in short-distance precision measurements, minimizing external disturbances is crucial for maintaining measurement accuracy in systems such as laser interferometers and three-dimensional imaging instruments\cite{105,106,107,108,109,110,111,112,113,114,115,116,117,118,119,120}.

Since the invention of the laser, numerous strategies have been proposed to address these issues, giving rise to research directions such as optical solitons\cite{121,122,123,124,125,126,127}, adaptive optics\cite{85,128,129,130,131,132,133,134,135}, and non-diffracting beams \cite{18,23,24,25,26,74,75,76,136,137,138,139,140,141,142,143,144,145}. However, each of these approaches has inherent limitations. Optical solitons depend on medium nonlinearity and therefore cannot be realized in free space. Adaptive optics, while effective, often involve complex hardware architectures, high implementation costs, and limited suppression of high-frequency turbulence. Non-diffracting beams, though capable of maintaining intensity profiles over extended distances, typically suffer from low modulation efficiency and practical challenges in long-distance propagation. Consequently, achieving stable and disturbance-resistant laser transmission remains a pressing challenge.

The OPBs have recently emerged as a novel beam-construction paradigm that directly addresses these limitations. By precisely engineering the surface profile of the modulation device\cite{146}, the OPBs effectively eliminate transverse wave vectors\cite{147}, thereby realizing highly stable propagation and strong disturbance suppression\cite{55}. In what follows, we first discuss the underlying mechanism and theory behind this new class of optical beams.

During free-space transmission, laser beams are highly vulnerable to various environmental and optical disturbances, including aberrations, rain and fog, and atmospheric turbulence, all of which induce energy attenuation, scattering, and wavefront distortion\cite{148,149,150}. These effects collectively degrade the beam quality, leading to distortion, wandering, and intensity fluctuations, which in turn severely limit the efficiency of energy and information transfer. Because these perturbations are spatially random and temporally varying, they have long been recognized as one of the primary challenges constraining the advancement of laser-based technologies\cite{151,152,153}. Specifically, attenuation and scattering diminish the optical power density at the receiver, thereby reducing the achievable data rate in laser communication systems\cite{154,155,156,157,158}. Wavefront distortion undermines the stability and accuracy of optical systems such as laser communication, remote sensing, mapping, and precision metrology. Meanwhile, beam drift introduces severe challenges to long-distance tracking, alignment, and targeting\cite{159,160}.

\begin{figure}[H]
\begin{center}
\begin{tabular}{c}
\includegraphics[scale=1.25]{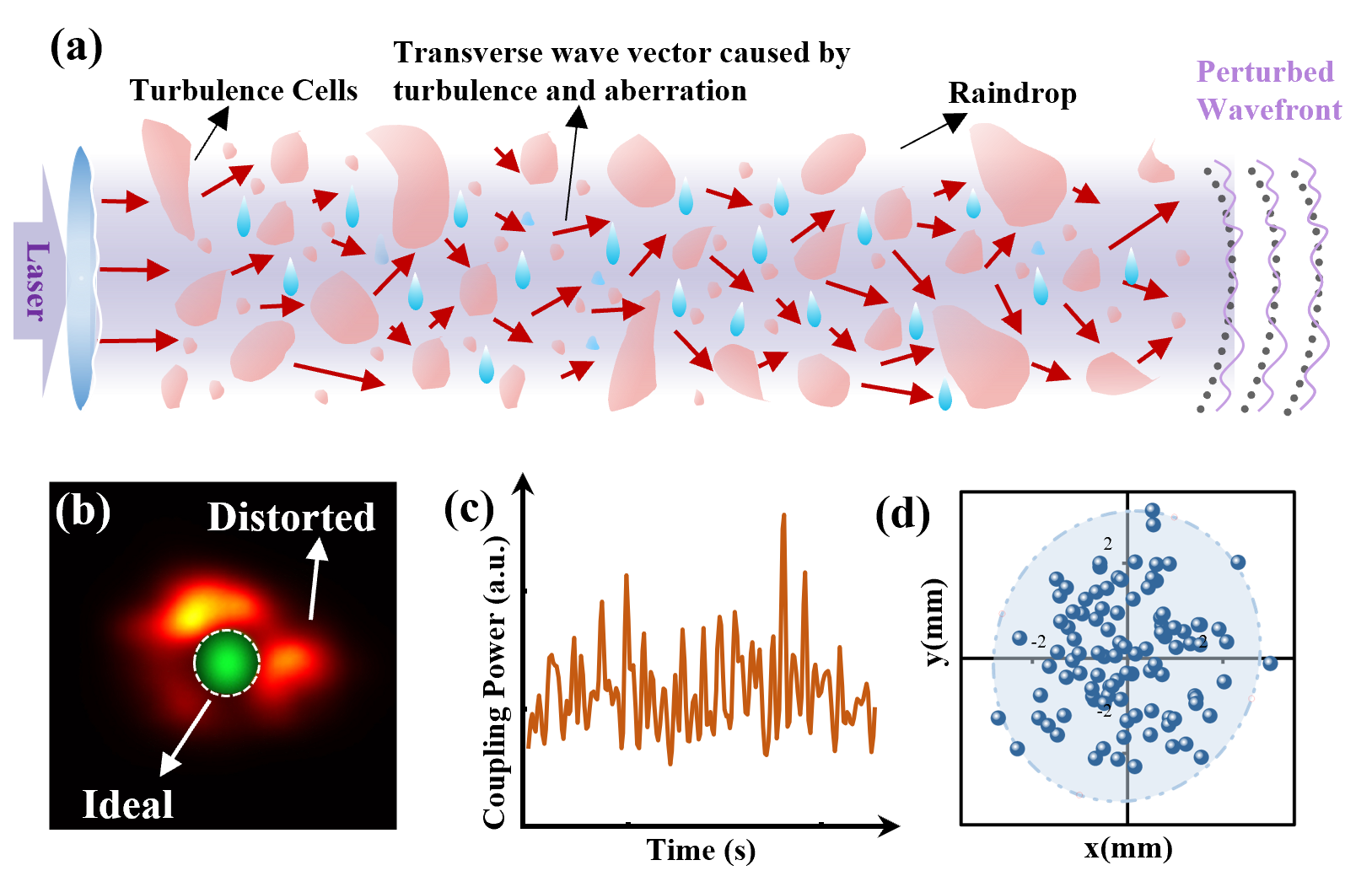}
\end{tabular}
\end{center}
\caption
{ \label{fig:example2}
Mechanisms of atmospheric turbulence affecting laser beam transmission. (a) Schematic illustration of turbulence-induced disturbances during propagation. (b) Distortion of beam spot morphology. (c) Fluctuation of coupling power caused by temporal instability. (d) Centroid drift of the beam spot resulting from spatial randomness.}
\end{figure}

Fundamentally, the propagation behavior of a laser beam is governed by the distribution of its wave vectors, which comprise components transverse to and along the propagation direction (with scattering and reflection occasionally contributing backward components)\cite{161,162,163}. In general, the axial wave vector supports efficient forward transmission of optical energy and information, whereas the transverse wave vector represents detrimental components that deteriorate beam stability and transmission quality\cite{162,164}. In an ideal scenario, a laser beam would generate no additional transverse wave vectors—aside from those due to diffraction—thereby achieving optimal propagation performance. In practice, however, environmental disturbances such as aberrations, aerosols, and turbulence continuously introduce random transverse wave vectors, resulting in beam broadening, deformation, and instability, as illustrated schematically in Fig. 2(a)\cite{165,166,167}.

The underlying mechanisms of these effects are shown in Figs. 2(b-d). In the ideal case, as indicated by the dashed circle in Fig. 2(a), the total beam energy is expected to remain stably confined within the receiver aperture (the colors shown are schematic and not representative of the actual optical field). In reality, however, the beam spot undergoes random spatial dispersion and temporal fluctuations, which reduce both the received energy and its temporal stability, as illustrated in Fig. 2(c). Moreover, the centroid of the beam experiences continuous drift under turbulent conditions, leading to significant pointing errors and reduced tracking precision, as depicted in Fig. 2(d).

To reduce the deterioration of the laser beam transmission quality, it is essentially necessary to eliminate the unwanted transverse wave vectors. First, in the Cartesian coordinate system, assume that there are two one-dimensional optical fields carrying transverse and longitudinal wave vectors respectively and propagating forward\cite{55}:
\begin{equation}
\label{eq:fov1}
\vec{E}(x) = \vec{E}_0 \cdot e^{i(\vec{k}_x x + \vec{k}_z z)}
\end{equation}
\begin{equation}
\label{eq:fov2}
\vec{E}'(x) = \vec{E}_0' \cdot e^{i(\vec{k}_x' x + \vec{k}_z z)},
\end{equation}

where $\vec{E}(x)$ and $\vec{E}'(x) $represent the amplitudes of the optical fields, $\vec{k}_x x $ and  $\vec{k}_x' x $ represent the transverse wave vectors, and $\vec{k}_zz$ represents the forward longitudinal wave vector.
The two light beams meet in space and interfere with each other. The superposed optical field can be expressed as\cite{55}:
\begin{equation}
\label{eq:fov3}
\vec{E}_{\text{sum}}(x) = \vec{E}_0 e^{i(\vec{k}_x x + \vec{k}_z z)} + \vec{E}_0' e^{i(\vec{k}_x' x + \vec{k}_z z)}.
\end{equation}
It can be seen from the above formula that when $\vec{E}(x)=\vec{E}'(x)$,the third term can be eliminated. In the second term, when the directions of $\vec{k}_x x $ and  $\vec{k}_x' x $ are opposite, they will partially cancel each other out. Especially when  $\vec{k}_x x =- \vec{k}_x' x$, the second term can be completely cancelled out, that is, a sinusoidal diffraction-free pattern appears, and there is no transverse dynamics. At this time, the steady-state optical field can be described as\cite{55}:
\begin{equation}
\label{eq:fov4}
\vec{E}_{\text{steady}}(x) = 2\vec{E}_0 e^{i\vec{k}_z z}.
\end{equation}
The simulation of formula (4) is shown in Fig. 3(a). It can be seen that during the superposition process of the beams, an area without transverse wave vectors appears, which we call the steady-state area.
When there are an infinite number of one-dimensional optical fields in the two-dimensional optical field plane moving their transverse energy towards the central axis (with transverse wave vectors towards the central axis), the effect can be regarded as the energy of a circular ring beam moving towards the inner area. At this time, the optical field can be expressed in the cylindrical coordinate system as\cite{168}:
\begin{equation}
\label{eq:fov5}
\psi(r,z) = E_0 \exp\left(i \vec{k}_r \cdot \vec{r} + i k_z z\right) \, .
\end{equation}
Here, $E_0$ represents the amplitude of the optical field, $\vec{k}_r$ represents the transverse wave vector, and $ \vec{k}_z$ represents the longitudinal wave vector. When there are $n$ such optical fields propagating in space, for the sake of simplicity in calculation, the amplitudes of these optical fields are all assumed to be $E_0$, and the superposed optical field can be expressed as\cite{168}:
\begin{equation}
\label{eq:fov6}
\psi (r,z)=\frac{1}{n}\sum\limits_{j=1}^{n}{{{E}_{0}}{{e}^{i{{{\vec{k}}}_{{{r}_{j}}}}\vec{r}+i{{k}_{z}}z}}}\approx {{J}_{0}}\left( {{k}_{r}}r \right){{e}^{i{{k}_{z}}z}}.
\end{equation}
In the above formula, the sum is taken over n plane waves in form of $\exp (i{{\vec{k}}_{r}}\cdot \vec{r})$ with $\left| {{{\vec{k}}}_{r}} \right|={{k}_{r}}$ that are equally spaced on the circle in the limit . Based on the above results, the transmission of five circular ring beams is simulated, and the results are shown in Fig. 3(b). Here, to be consistent with the actual modulation process, the outer boundary of the inner circular ring between two adjacent circular rings is exactly equal to the inner boundary of the outer circular ring. In addition, for comparison, we also simulate the transmission behavior of the GS at the same time, and set the initial GS to have the same beam diameter as the outer boundary of the largest circular ring. The transmission envelope of the GS is shown as the white dotted line in the figure.

The simulation results indicate that before the wave vector cancellation behavior occurs, the transmission envelopes of the superimposed beam and the GS are nearly identical. However, after the wave vector cancellation behavior occurs, the divergence angle of the superimposed beam is significantly reduced, and its energy is basically within the white envelope of the GS. At the far end, the spot diameter of the superimposed beam even reaches about one-third of the width of the Gaussian envelope. Referring to the evaluation method of the Rayleigh length, as shown in the Fig. 3, the Rayleigh length of the superimposed beam can reach nearly three times that of the GS.

\begin{figure}[H]
\begin{center}
\begin{tabular}{c}
\includegraphics[height=11cm]{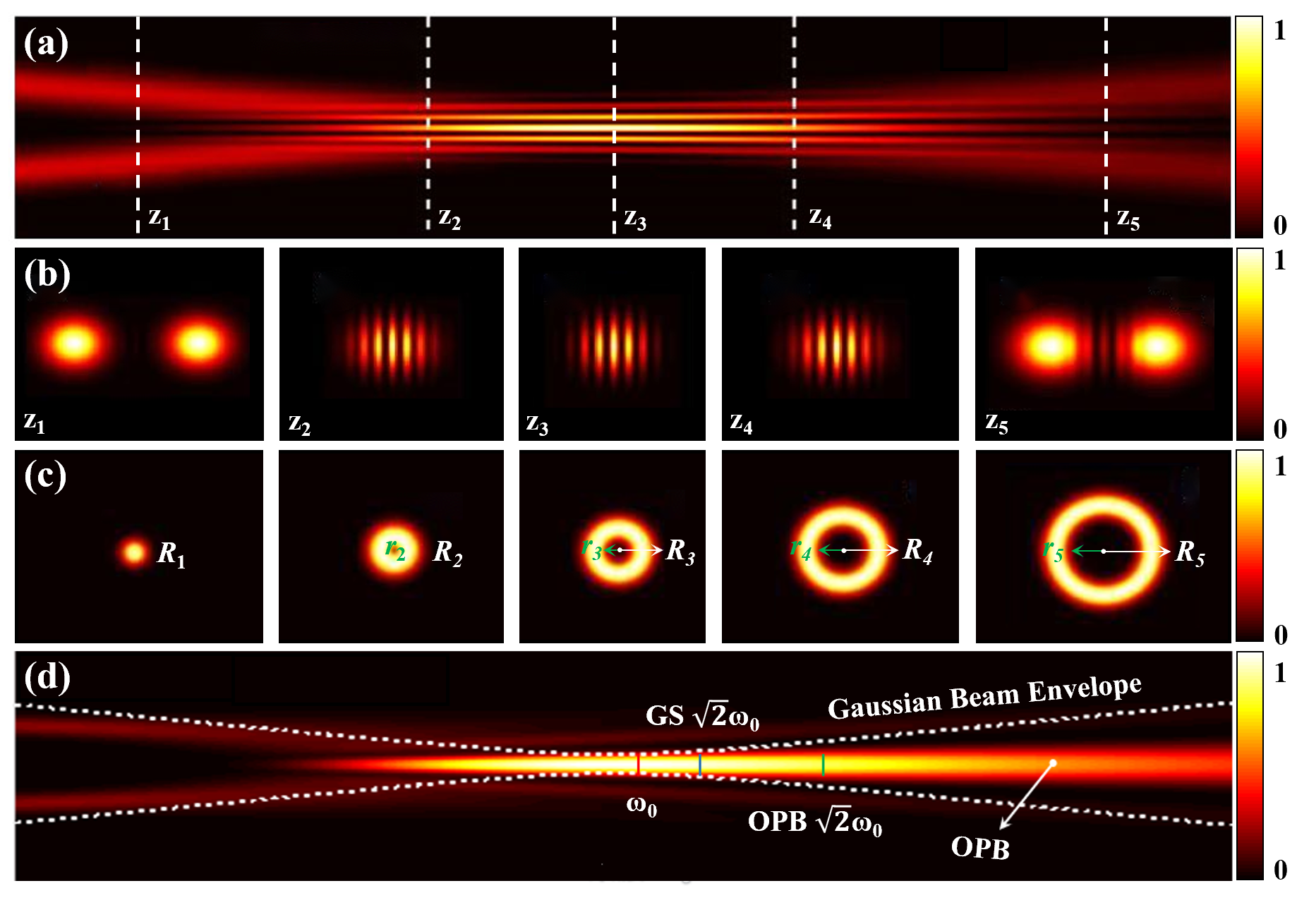}
\end{tabular}
\end{center}
\caption
{ \label{fig:example3}
 Numerical simulation of the mechanism for constructing a stable optical field through the elimination of transverse wave vectors (a, b), and simulation results of generating a stable optical field using multiple ring beams, compared with the transmission behavior of a GS (c, d). In (c), the inner and outer diameters of adjacent rings are identical. In (d), the white dashed line denotes the transmission envelope of a GS with the same initial spot size; the red short lines mark the beam waist positions of the two beams. The region between the red and blue short lines corresponds to the Rayleigh length of the GS, while the region between the red and green short lines represents the Rayleigh length of the stable optical field\cite{168}. }
\end{figure}

Considering that the Airy beam has the characteristic of transverse acceleration, we first employed the method of using an infinite number of one-dimensional Airy beams accelerating towards the center to construct a laser beam that can propagate in a stable fashion, as shown in Fig. 4(a). In order to increase the filling factor of the beam arrangement, we performed truncation on the one-dimensional Airy beams at the initial spot position. Based on the self-healing property, the truncated Airy beams can still restore to the wavefront shape of the Airy beam at the distant end\cite{169}. Use these truncated and ring-arranged Airy beams, OPBs can be constructed in the far end, as shown in Fig. 4(b). It can be seen that as the number of Airy beams N increases, the pin beam quality improves, leading to more uniform propagation of the pin-like feature.

\begin{figure}[H]
\begin{center}
\begin{tabular}{c}
\includegraphics[scale=1.2]{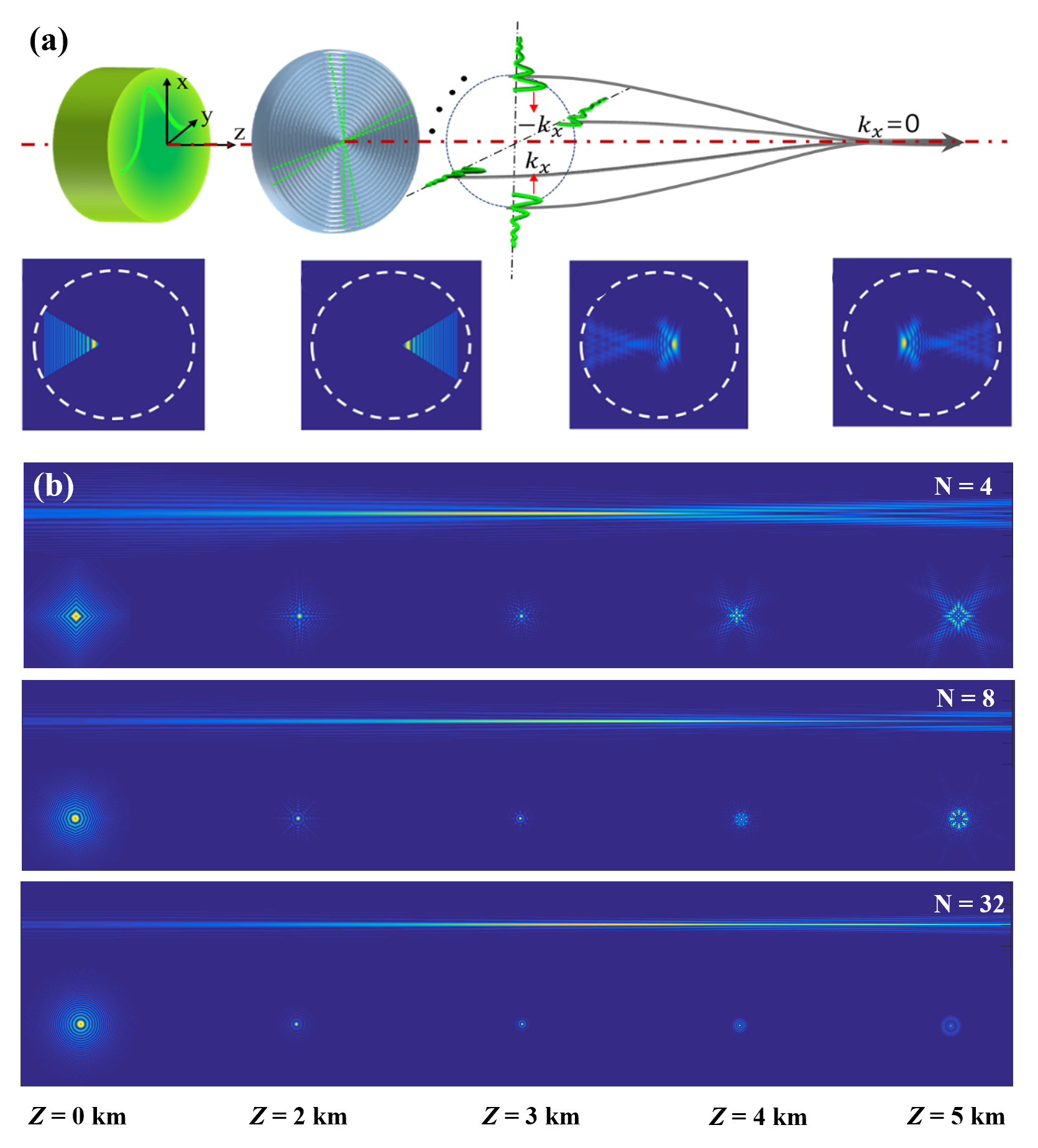}
\end{tabular}
\end{center}
\caption
{ \label{fig:example4}
 (a) Schematic diagram of constructing an OPB using truncated Airy beam ring array and illustration of elimination of transverse wavevectors during subsequent propagation\cite{55}. (b) Numerical simulation results showing the OPB construction with N=4, 8, and 32 Airy-beam ring arrays. }
\end{figure}

\begin{figure}[H]
\begin{center}
\begin{tabular}{c}
\includegraphics[scale=1.1]{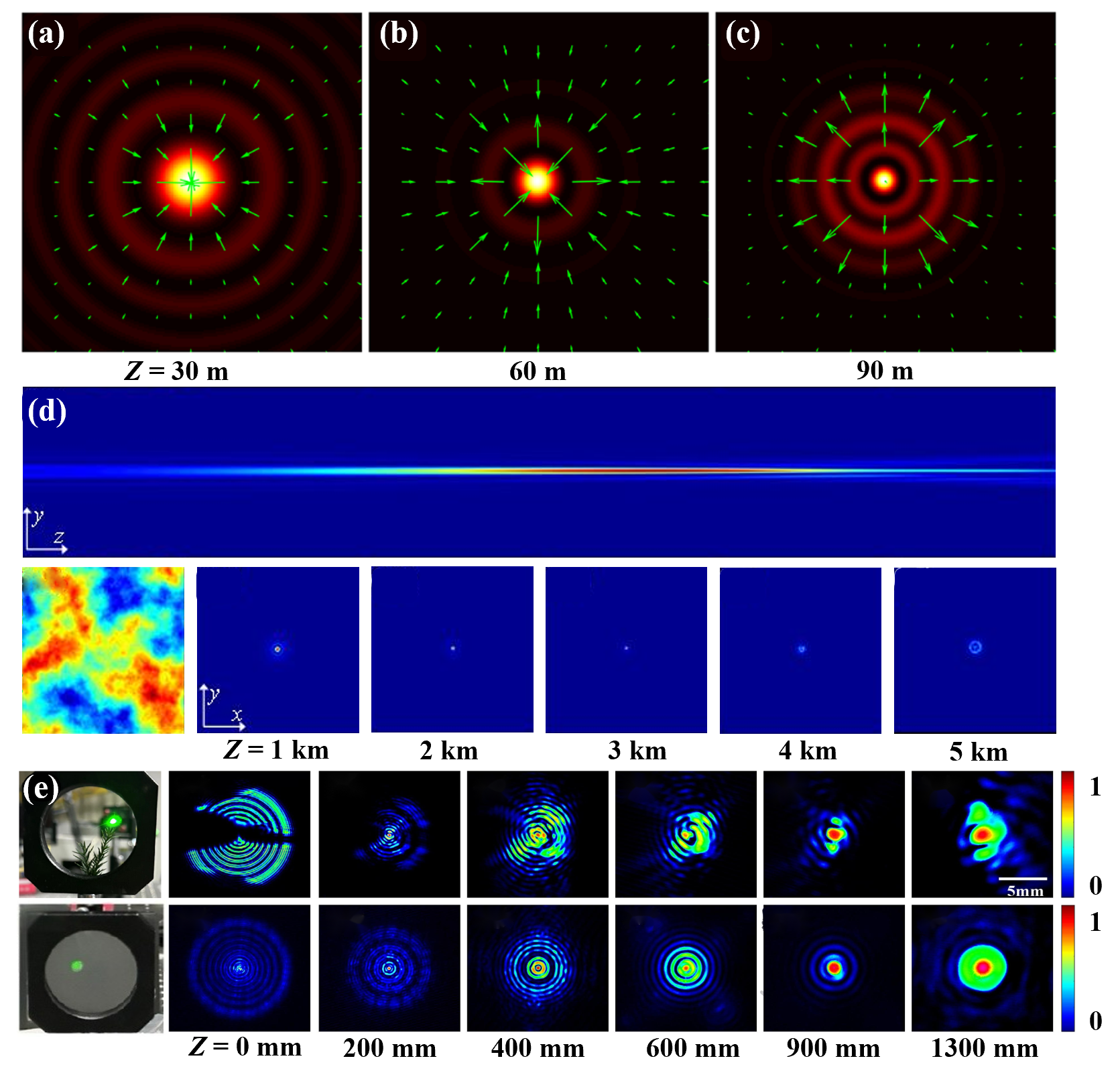}
\end{tabular}
\end{center}
\caption
{ \label{fig:example5}
 (a)–(c) Simulation results showing the Poynting vector distribution of the OPB at different propagation distances\cite{55}. (d) Simulation of OPB propagation in turbulence using the phase screen method\cite{55}. (e) Experimental demonstration of the self-healing characteristic of the beam spot under obstacle occlusion conditions\cite{170}. }
\end{figure}

As shown in Fig. 4, when circularly arranged Airy beams propagate forward and overlap, their transverse wave vectors cancel out, thereby compressing the laser beam divergence angle. Simulations reveal that with only four Airy beams, some critical Poynting vectors cannot be fully eliminated, resulting in suboptimal propagation behavior. As the number of Airy beams gradually increases to 32, the propagation characteristics of the optical field essentially reach their optimal state. Further increasing the beam count to 1024 or more shows no significant improvement. Additionally, analysis of the Poynting vectors in the superimposed optical field (as illustrated in Figs. 5(a-c)) reveals a dynamic evolutionary characteristic during propagation: energy continuously transfers between different regions of the beam profile, sometimes inward and sometimes outward, maintaining the beam morphology. This behavior fundamentally differs from GS, where energy only diffuses outward due to diffraction. It is right this unique characteristic that enables the superimposed optical field to achieve stable propagation and disturbance suppression capabilities in scenarios involving atmospheric turbulence, rain/fog, and random aberrations. Phase-screen simulations of OPB propagation in turbulence, as shown in Fig. 5(d). The result shows that OPB does maintain stable wave field during propagation. Notably, this stability persists not only against turbulence but also demonstrates self-healing properties: even when partially obstructed, the beam can restore its original profile after propagating a certain distance, as demonstrated in Fig. 5(e).

We can describe the dynamics of the system by considering the radially symmetric case which is equivalent to assuming that the number of the Airy-type fragments goes to infinity. The dynamics of the optical beam in the paraxial regime is given by the following Fresnel integral\cite{55,56,58,59}:

\begin{equation}
\label{eq:fov7}
\psi (r,z)=\frac{k{{e}^{\frac{ik{{r}^{2}}}{2z}}}}{iz}\int_{0}^{\infty }{{{\psi }_{0}}(\rho ){{J}_{0}}\left( \frac{kr\rho }{z} \right){{e}^{\frac{ik{{r}^{2}}}{2z}}}}d\rho \,
\end{equation}

where ${\psi }_{0}(\rho )$ is the optical beam at the input $z=0$ or the phase mask plane, $k=\frac{2\pi }{\lambda }$ is the wavenumber, and ${{J}_{0}}$ is the zeroth order Bessel function. We assume an initial condition with the Airy-type phase (proportional to $\rho^{3/2}$ )\cite{55,56,58,59}:
\begin{equation}
\label{eq:fov8}
\psi_0(\rho) = A(\rho) {{e}^{-i\frac{4}{3}k{{\beta }^{\frac{1}{2}}}{{\rho }^{\frac{3}{2}}}}} \, ,
\end{equation}

where $\beta $ is a parameter with inverse length dimensions that determines the local frequency of the phase oscillations of the wave. Following a stationary phase approach to the above Hankel transform we conclude that the dynamics of the optical wave is given by\cite{55,56,58,59}:
\begin{equation}
\label{eq:fov9}
\psi (r,z)=8{{\left( \pi k{{\beta }^{2}}{{z}^{3}} \right)}^{\frac{1}{2}}}A\left( 4\beta {{z}^{2}} \right){{J}_{0}}(4k\beta \rho z){{e}^{i\Phi }},
\end{equation}
where $\Phi = \frac{k r^{2}}{2z} - \frac{8}{3} k \beta^{2} z^{3} - \frac{\pi}{4}$. We see that the resulting wave takes the form of a Bessel-like beam with a width that is inversely proportional to the propagation distance\cite{55,56,58,59}:

\begin{equation}
\label{eq:fov10}
W(z)=\frac{1}{4k\beta z}.
\end{equation}
For this reason, such a beam propagates in a highly confined and stable manner—much like a pin—hence it is referred to as an OPB or simply a pin-like beam. In this review, we will use the term OPB for clarity and consistency.

\section{Experimental Realization of OPBs}
\label{sect:sections3}
Like other structured beams such as Airy beams, OPBs can be experimentally generated using a liquid-crystal spatial light modulator (SLM). In this approach, the desired OPB phase pattern (Fig. 6(a)) is encoded onto the SLM, which—together with a 4f system—converts an incident GS into an OPB. However, inherent limitations of liquid-crystal SLMs, including polarization selectivity, low fill factor, and insufficient phase depth, result in limited modulation efficiency and low power-handling capability, restricting their use in many practical applications. This motivates the design of static phase-delay elements based on transparent materials such as fused quartz, or reflective materials such as aluminum, to realize high-efficiency, high-power OPB modulator\cite{55}.

In 2019, we designed and fabricated a fused-quartz phase modulation device with both high modulation efficiency and high-power tolerance. Experimental characterization showed a modulation efficiency exceeding 90 \% and a power-handling capability on the order of tens of kilowatts. To ensure fabrication accuracy, a phase-folding strategy was adopted, and six overlay lithography steps were performed on the fused-quartz substrate. The resulting quartz-based OPB modulation device is shown in Fig. 6(b, c).

\begin{figure}
\begin{center}
\begin{tabular}{c}
\includegraphics[scale=0.6]{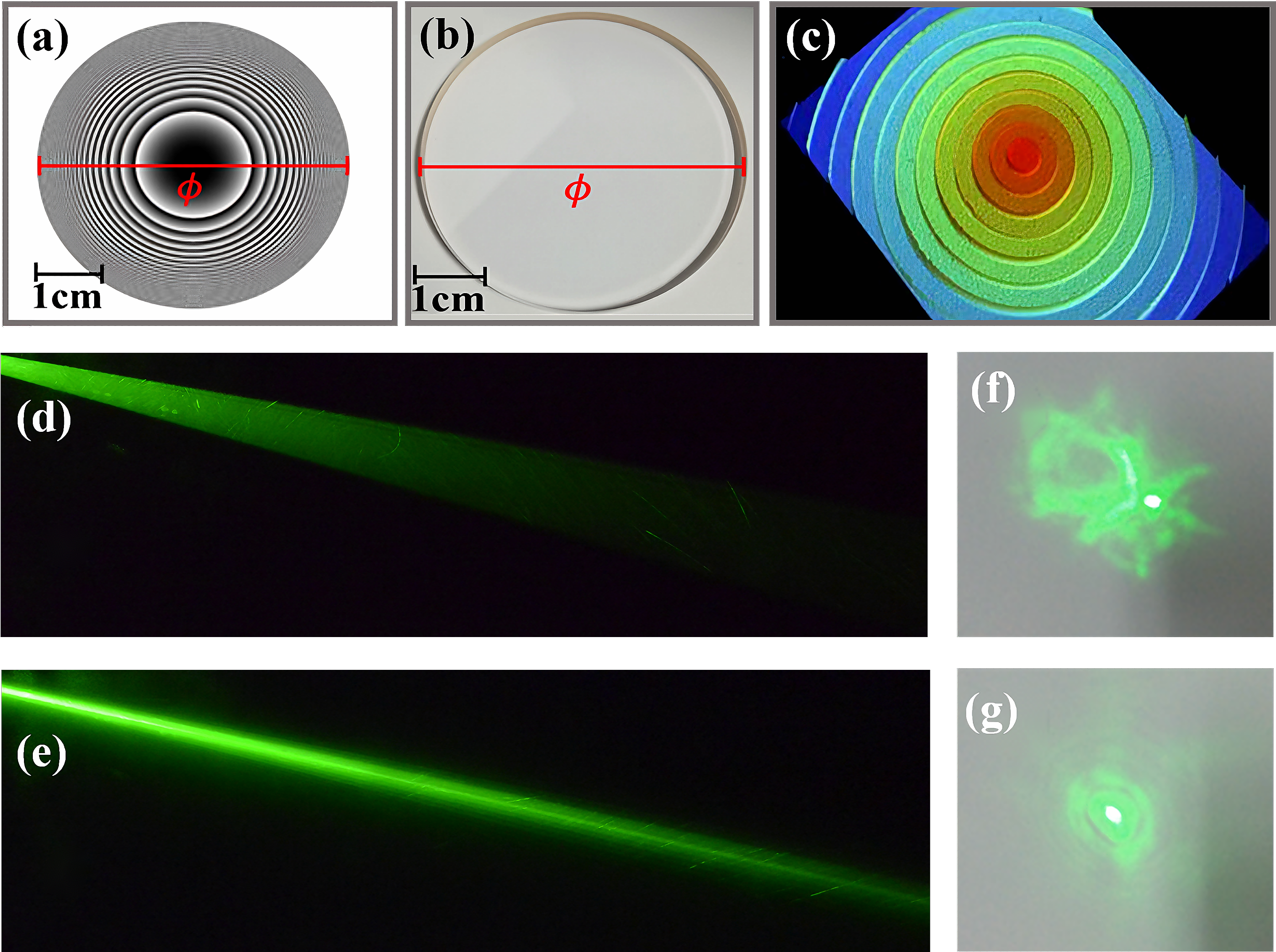}
\end{tabular}
\end{center}
\caption
{ \label{fig:example6}
  Phase distribution (a), physical object (b), and microstructure (c) of the quartz-based OPB modulation diaphragm. The transmission effects of the GS and OPB are compared under turbulent conditions over a 100 m propagation distance. Panels (b) and (e) show side views of the GS and OPB transmission, respectively, while (f) and (g) illustrate the spot distortion effects for the GS and OPB under the same turbulent conditions\cite{55}. }
\end{figure}
To evaluate transmission stability, we constructed a nearly 100-m free-space optical link in the laboratory. The central air-conditioning system was set to 40 °C with maximum airflow to simulate strong turbulence conditions. The measured propagation performance and stability of the OPB under this environment are shown in Figs. 6(d–g). As shown in Fig. 6, under identical transmission conditions, the OPB exhibits a markedly reduced divergence angle, significantly enhanced central-lobe energy concentration, and far more stable beam morphology under turbulence compared to a conventional GS.

\begin{figure}
\begin{center}
\begin{tabular}{c}
\includegraphics[scale=1.1]{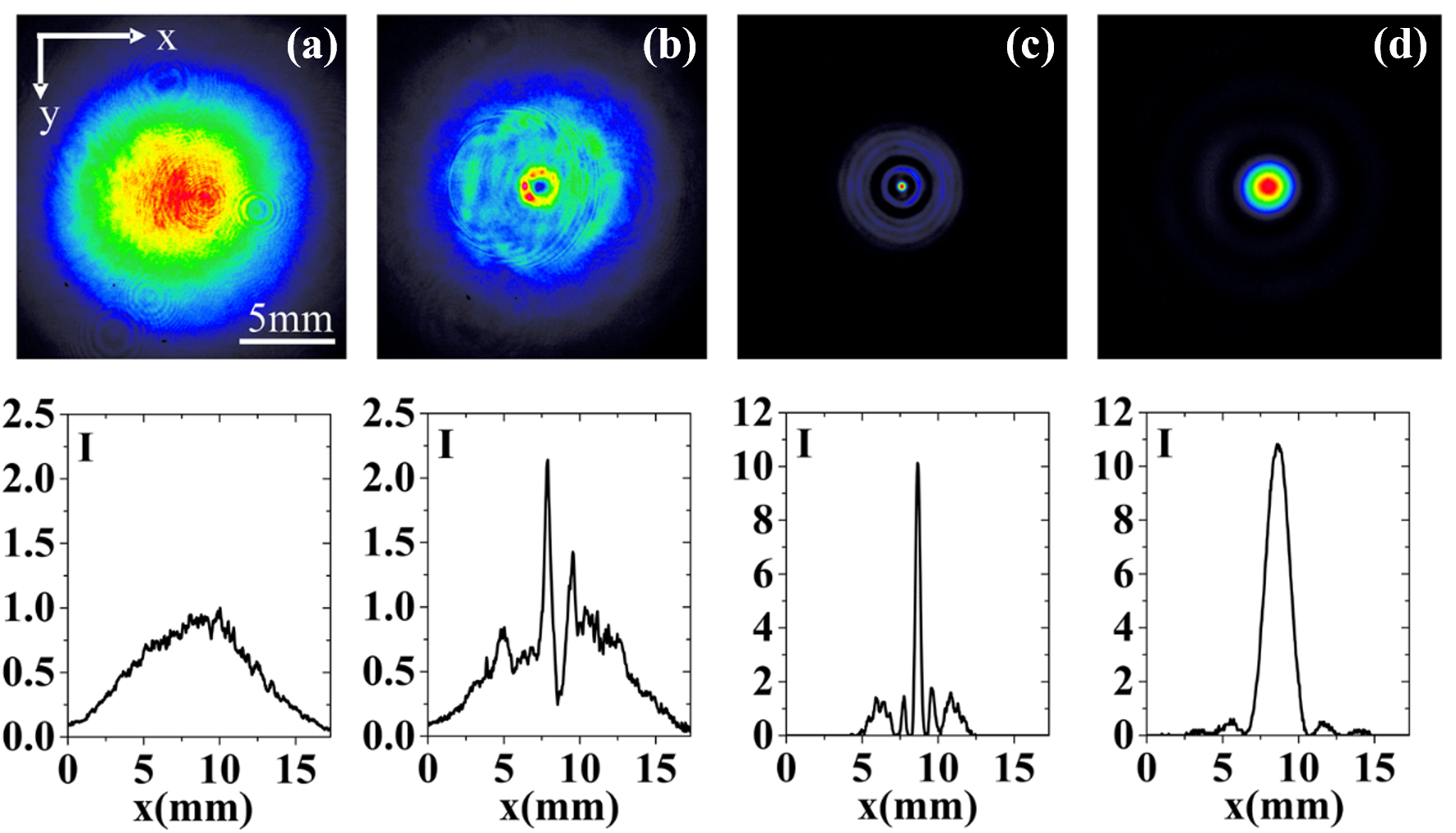}
\end{tabular}
\end{center}
\caption
{ \label{fig:example7}
  Snapshots of the beam profiles of the optical pin beam (OPB) at various propagation distances. The top row shows the transverse intensity patterns, while the bottom row presents the corresponding horizontal beam profiles. The beam profiles are taken at the following distances from the mask: (a) 0 m, (b) 2.4 m, (c) 25 m, and (d) 60 m. Note that the maximum value on the $y$ -axis differs for the intensity profiles in the bottom row, as the peak intensity in the first two panels is significantly lower than in the last two panels. For the last two panels, a stronger attenuation was applied to avoid camera saturation. These figures qualitatively demonstrate that the peak intensity of the OPB increases, while the overall beam width decreases with increasing propagation distance\cite{55}. }
\end{figure}

To quantitatively examine the beam evolution and weak-divergence characteristics of the OPB, we recorded the beam profiles at propagation distances of 0 m, 2.4 m, 25 m, and 60 m in the absence of turbulence, as shown in Figs. 7(a-d). Fig. 7(a) shows that the initial GS has a spot size of approximately 10 mm, whereas after 60 m of propagation in Fig. 7(d), the OPB’s central lobe remains only about ~2.5 mm in diameter. As seen from Figs. 7(c) and 7(d), after nearly 40 m of transmission the central lobe of the OPB expands only from ~1 mm to ~2.5 mm, indicating an extremely small divergence angle. A comparison of the intensity distributions further shows that energy from the side lobes is efficiently redistributed into the central main lobe. It is worth noting that in Fig. 7(d), due to the limited dynamic range of the CCD, the side lobes appear very weak; however, they are indeed present. These side lobes play an essential role in maintaining the stable propagation of the main lobe, contributing to the characteristic robustness of the OPB.

We subsequently designed a method to generate the OPB by using an aspherical modulation device designed with a circular ring wave-front splitting approach, and showed superior propagation stability and reduced divergence of the OPB over the conventional GS\cite{146}. However, this design method still cannot determine the optimal shape of the modulator. In order to achieve the best design effect, we use the genetic algorithm to further optimize the wavefront, as shown in Fig. 8\cite{171}. During the optimization process, we take the minimum divergence angle of the OPB as the constraint objective. By continuously iterating the number of annular wavefronts within a certain aperture, as well as the radius and width of each ring, the final BPP can reach 67\% of that before optimization. It is worth mentioning that the genetic algorithm can also be guided by other objectives to design a OPB with specific performance, such as the ability to suppress turbulence in specific scenarios. This will be a research direction with great practical prospects in the future.

\begin{figure}
\begin{center}
\begin{tabular}{c}
\includegraphics[scale=1.2]{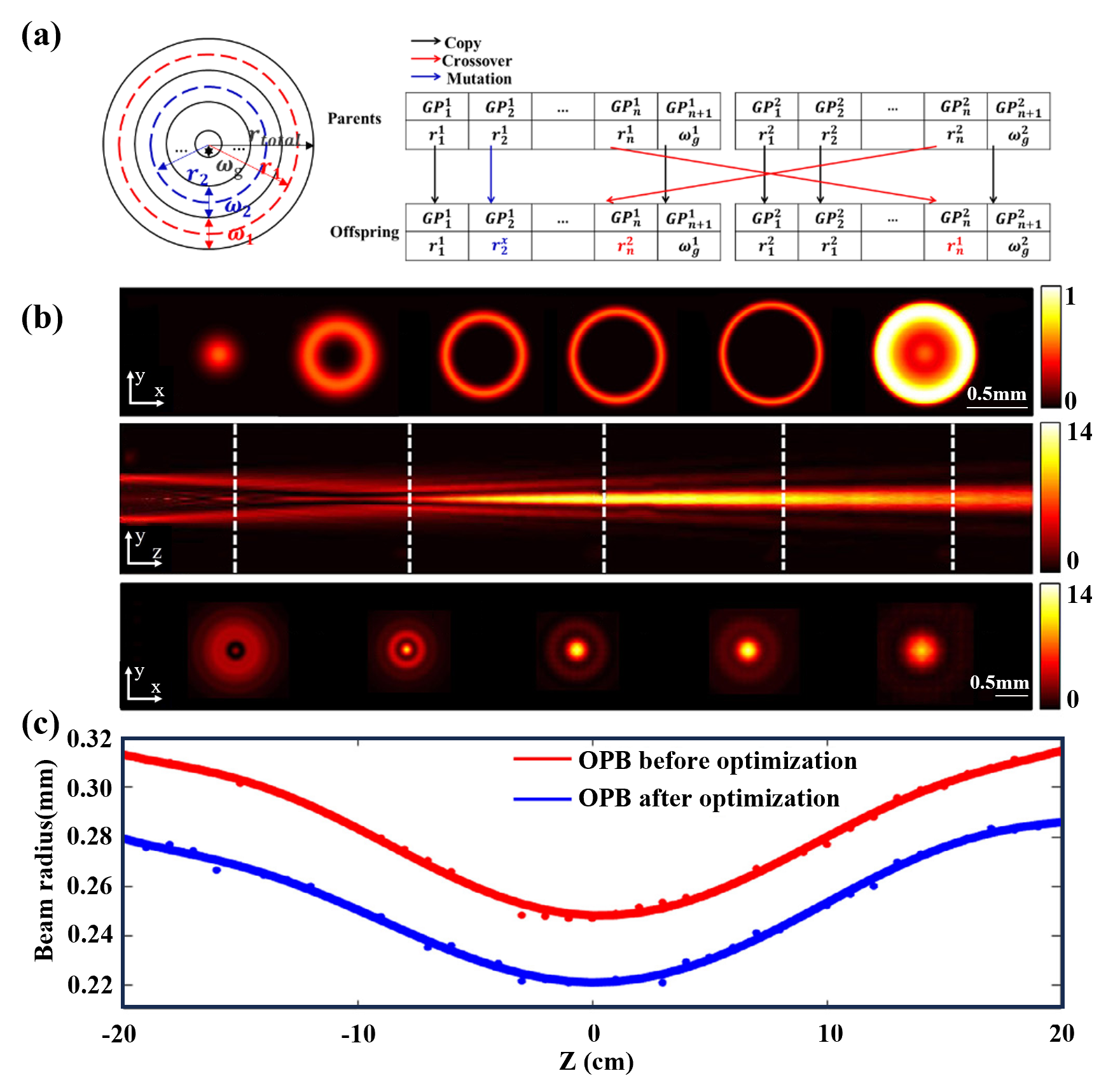}
\end{tabular}
\end{center}
\caption
{ \label{fig:example8}
  (a) Schematic diagram of the evolutionary process for generating optimized OPB through selection, crossover, and mutation operations. (b) Illustration of the initial light-field composition of the optimized OPB, its propagation behavior, and the corresponding transverse light-field distributions at different positions. (c) Measured beam radius of the optimized OPB as a function of propagation distance, demonstrating its enhanced propagation stability\cite{171}. }
\end{figure}

Similar to the distortion of light spots caused by air turbulence and the scattering fragmentation caused by rain and fog, in water, there will also be scattering caused by liquid flow turbulence and water impurities. Moreover, the resulting distortion and scattering fragmentation are significantly greater than those in the air\cite{172,173,174}. Researchers including Y. Yang and X. Kang have investigated the transmission performance of the OPB in the water environment\cite{146,175,176}. The results show that in a water environment with a scale of several meters, the laser spot will undergo significant distortion and fragmentation. However, compared with the GS, the transmission performance of the OPB is still significantly better, mainly manifested in the significant improvement of the peak light intensity, stable transmission distance, focal depth, and the maintenance of the wavefront morphology. Notably, increasing both the number of beam rings and the focal length of the innermost beam is critical for enhancing focal depth, a relationship summarized in Fig. 9.

\begin{figure}
\begin{center}
\begin{tabular}{c}
\includegraphics[scale=0.5]{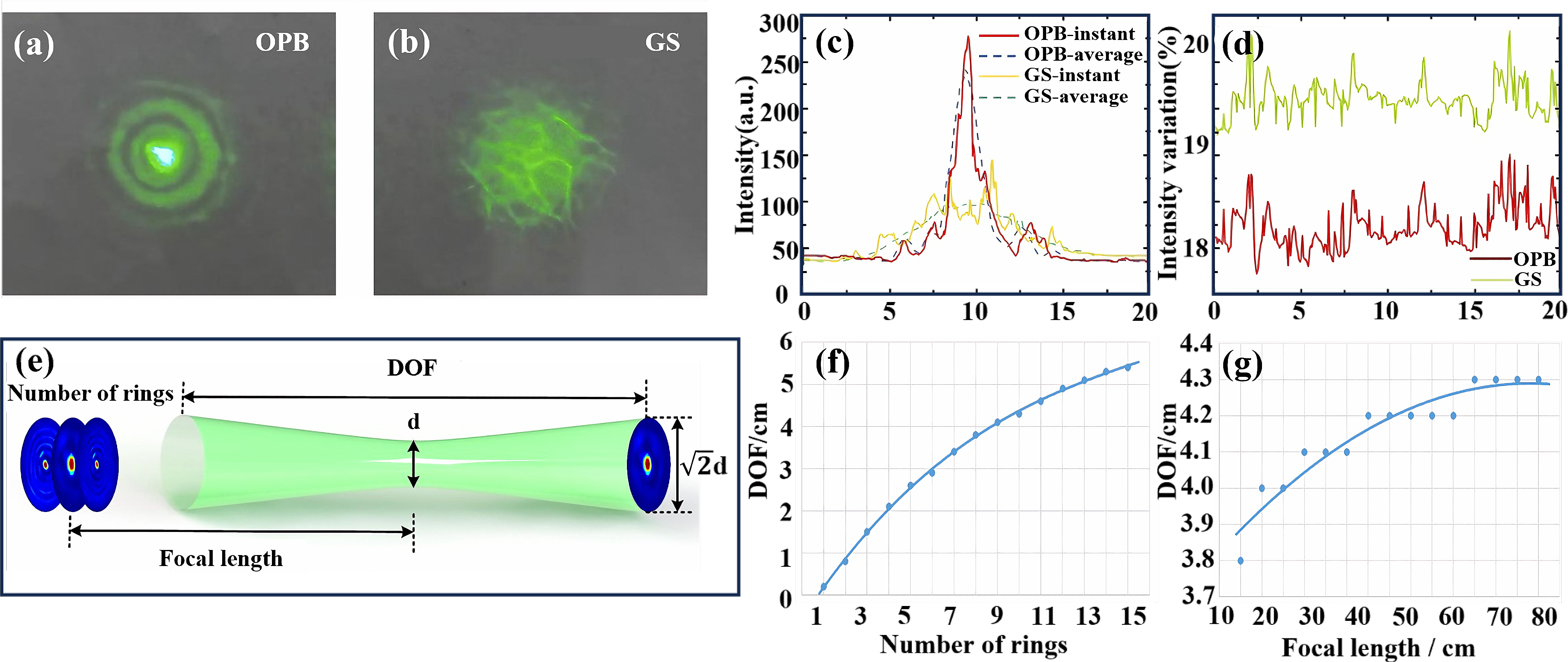}
\end{tabular}
\end{center}
\caption
{ \label{fig:example9}
  Comparison of the OPB and GS in underwater turbulence. (a, b) Beam spot distributions of the OPB and GS after propagating through turbulent water\cite{146}. (c) Comparison of the transverse intensity profiles of the two beams\cite{146}. (d) Relative intensity fluctuation percentage as a function of time for the OPB and GS, showing the superior stability of the OPB\cite{146}. (e) Definition of the depth of focus (DOF) used for performance evaluation\cite{175}. (f, g) Variation trends of the DOF with respect to the number of rings and the focal length of the innermost ring beam, respectively, demonstrating the tunability of propagation characteristics through structural parameter modulation \cite{175}. }
\end{figure}

The above research results are all carried out based on numerical and laboratory simulations. In order to effectively prove the stable transmission and disturbance suppression capabilities of the OPB, it is necessary to verify it in actual turbulent flow and rain-fog environments. In 2019, Zhang Ze and others verified the robust propagation capability of the OPB in actual turbulent flow at the kilometer scale\cite{55}. As shown in Fig. 10(a), the results indicate that both the temporal and spatial stabilities of the OPB have been greatly improved. In addition, in 2025, Xiao Hongyue and others verified the spot recovery capability of the OPB in a kilometer-scale rain-fog environment\cite{170}. The experimental results show that, in the rain-fog environment, compared with the GS, the peak intensity of the beam spot is significantly increased. The average received power on the terminal surface is increased by about 2.11 times, and the stability of the light intensity scintillation is improved by about 46\%, as shown in Fig. 10(c)\cite{55,170}.

\begin{figure}
\begin{center}
\begin{tabular}{c}
\includegraphics[scale=0.52]{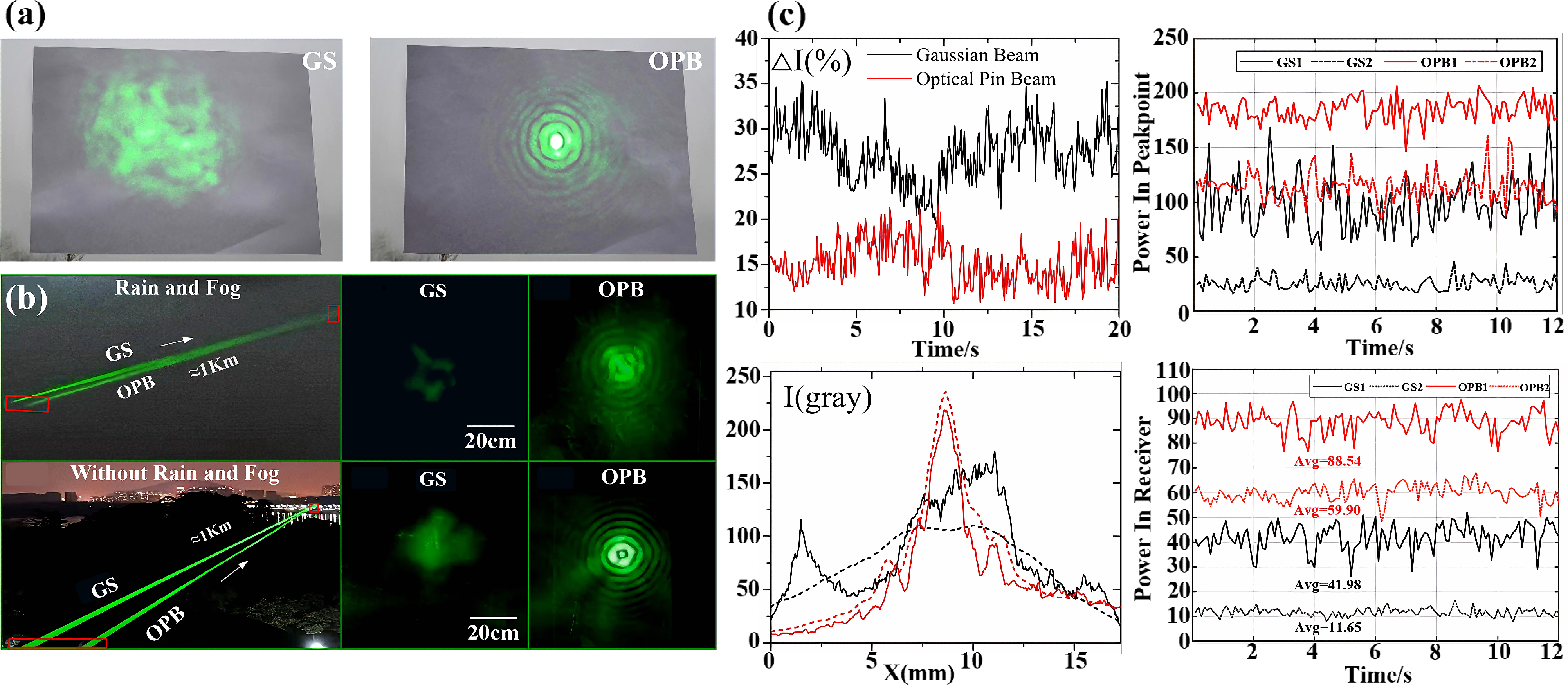}
\end{tabular}
\end{center}
\caption
{ \label{fig:example10}
  (a) Intensity patterns of a GS and an OPB after propagating 1 km in an open outdoor environment\cite{55}. (b) Comparison of beam spot morphology for both beams after 1 km of transmission under clear and rainy-foggy atmospheric conditions\cite{170}. (c) Comparison of spatial and temporal stability between the OPB and GS, demonstrating the superior robustness of the OPB under long-distance and adverse weather conditions\cite{55,170}.}
\end{figure}

\section{Novel Families of OPBs}

In recent years, OPBs have garnered increasing research attention, leading to the emergence of several extended forms. To date, four principal derivatives have been established — the vortex OPB, vector-vortex OPB, inverted OPB, and vortex-inverted OPB. The introduction of these variants enriches the propagation dynamics and functional properties of OPBs, enabling enhanced control over phase, polarization, and spatial structure. Consequently, the OPB family has evolved from its original form into a versatile class of structured beams, offering broad potential for optical communication, imaging, and particle manipulation applications.

\subsection{Vortex OPB (VOPB)}

The analysis of vortex type OPB (VOPB) starts by introducing the vector potential of an arbitrary optical beam, $\vec{U}(r,\theta ,z,t)=\psi (r,\theta ,z)\exp \left( ikz-\omega t \right)\hat{x}$, assuming linear polarization along the $x$-axis directed by the unity vector $\hat{x}$ Under a slowly varying envelope approximation, the linear wave equation describing the beam propagation dynamics can be expressed in cylindrical coordinates as\cite{56}:
\begin{equation}
\label{eq:fov11}
i\frac{\partial \psi }{\partial z}+\frac{1}{2k}\left( \frac{{{\partial }^{2}}\psi }{\partial {{r}^{2}}}+\frac{1}{r}\frac{\partial \psi }{\partial r}+\frac{1}{{{r}^{2}}}\frac{{{\partial }^{2}}\psi }{\partial {{\theta }^{2}}} \right)=0 ,
\end{equation}

In above equation, $\psi (r,\theta ,z)$ denotes the electric field envelope, where $r$ and $\theta $ represent the radial and azimuthal coordinates, and $z$ is the longitudinal distance. In addition, $k=\frac{2\pi }{\lambda }$is the wavenumber of the optical wave, $\lambda $ is the wavelength, $\omega $ is the angular frequency, and   is the time. The integral representation of Eq. (11) is given by the Fresnel integral\cite{56}:
\begin{equation}
\label{eq:fov12}
\psi (r,\theta ,z)=-\frac{ik}{2\pi z}\int_{0}^{r}{\int_{0}^{2\pi }{A(\rho )\exp [i\phi (\rho ,\varphi )]{{e}^{i{{k}^{\frac{{{r}^{2}}+{{\rho }^{2}}-2\rho rcos(\varphi -\vartheta )}{2z}}}}}}}\rho d\rho d\varphi ,
\end{equation}
where $A(\rho )$and $\phi (\rho ,\varphi )=-k{{C}_{\rho }}{{\left( \frac{\rho }{{{w}_{\rho }}} \right)}^{\gamma }}-l\varphi $are the amplitude and the phase of the optical wave on the input plane, with $\rho$ and $\phi$ indicating the radial and azimuthal coordinates at the onset of propagation, ${{C}_{\rho }}$is an arbitrary scaling parameter for the phase, $\gamma $ the power exponent of phase modulation, $l$ the topological charge, and ${{w}_{\rho }}$ the phase normalization factor. To find an asymptotic solution to Eq. (12), we apply the stationary phase method. The beam envelope close to the optical axis can be well approximated by the following expression\cite{56}:

\begin{equation}
\begin{aligned}
\label{eq:fov13}
\psi(r,z) & = i^{-|l|}\sqrt{\frac{2\pi k}{2-\gamma}}A[\rho(z)]J_l\left( kr\left( C\gamma z^{\gamma-1} \right)^{\frac{1}{2-\gamma}} \right)\left( C\gamma z^{\frac{\gamma}{2}} \right)^{\frac{1}{2-\gamma}}\times \\
          & \quad \exp\left\{ i\left[ \frac{kr^{2}}{2z} + \left( C^{2}\gamma^{2}z^{\gamma} \right)^{\frac{1}{2-\gamma}}\frac{k}{2}\left( 1-\frac{2}{\gamma} \right) - \frac{\pi}{4}(1+2|l|) - l\theta \right] \right\},
\end{aligned}
\end{equation}

where ${{J}_{l}}$ is the $l$th-order Bessel function, and $\rho (z)={{\left( {{C}_{\gamma }}z \right)}^{\frac{1}{2-\gamma }}}$, with $C$ equal to $C_{\rho}/w_{\rho}^{\gamma}$.The overall propagation range depends on the maximum radius $\rho_m$ at the onset distance through the relation${{z}_{m}}=\rho _{m}^{2-\gamma }/{{C}_{\gamma }}$. In that work, the value of $\gamma $ lies in the interval between 0 and 2. Additionally, Eq. (13) shows that the peak intensity evolution of a VOPB can be controlled by properly engineering the initial amplitude structure $A[\rho (z)]$ \cite{56}.
\begin{figure}
\begin{center}
\begin{tabular}{c}
\includegraphics[scale=0.9]{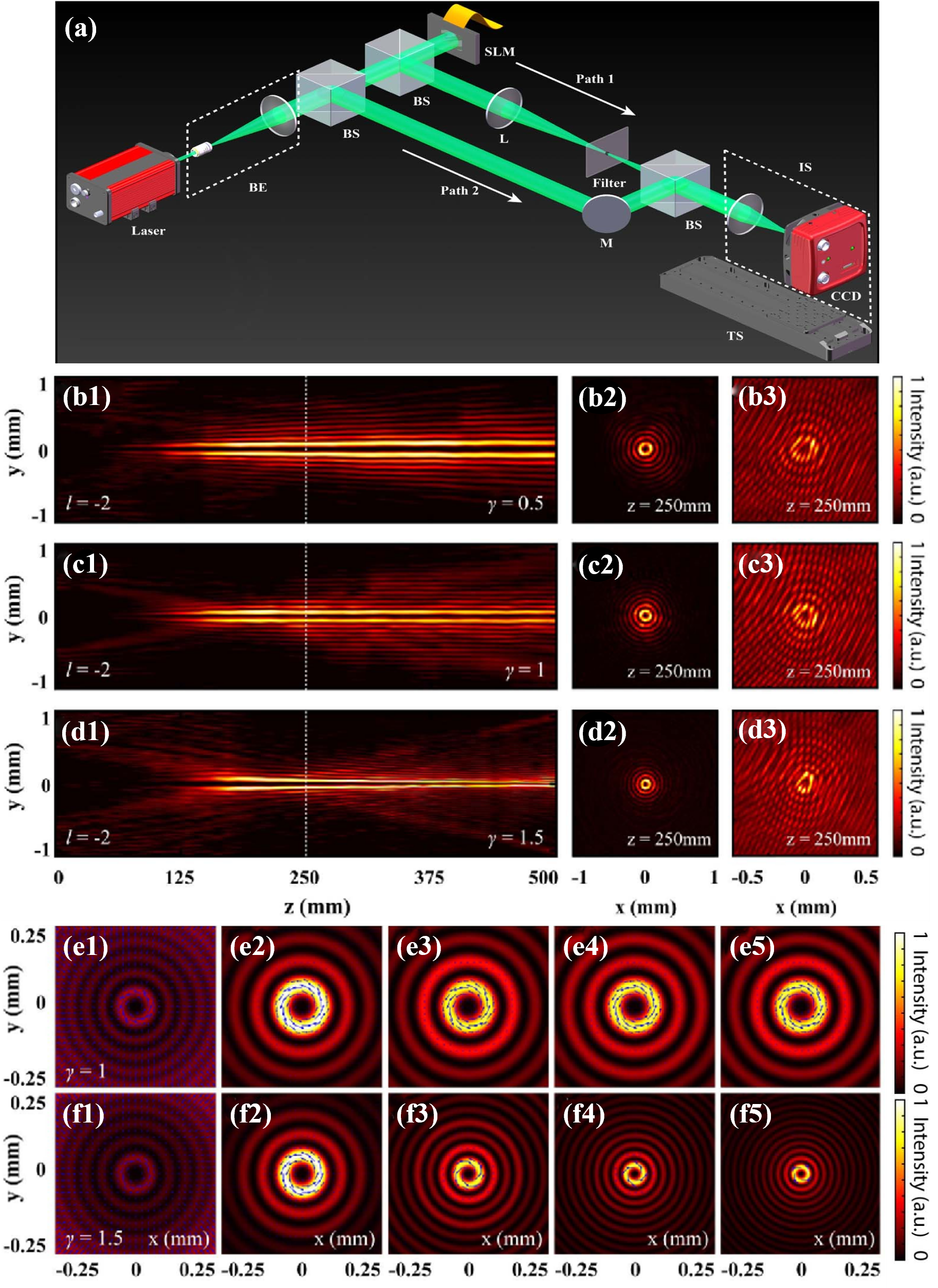}
\end{tabular}
\end{center}
\caption
{ \label{fig:example11}
  Experimental observations of VOPBs in free space for different values of the phase modulation exponent $\gamma$. (a) Schematic of the experimental setup used for the generation and detection of the VOPBs. Path 1 highlights the beamline used for the generation, while Path 2 the beamline used for carrying out the interferograms. (bl)-(d1) Normalized side-view of the beams in the $y-z$ plane for (b1) $\gamma$ = 0.5, (c1) $\gamma$ = 1, and (d1) $\gamma$ = 1.5. (b2)-(d2) Intensity distributions and (b3)-(d3) interferograms recorded at the distance $z$ = 250 mm, marked by dashed white lines in (b1)-(d1). $l$ = -2 is the topological charge. L, lens; M, mirror; BE, beam expander; Bs, beam splitter; IS, imaging system; Ts, translation stage; SLM, spatial light modulator. (e1)-(f5) Numerical calculations of Poynting vectors associated to the VOPBs with topological charge $l$= -2 for two different values of the exponent coefficient $\gamma$: (e1)-(e5) $\gamma$ = 1 and (f1) - (f5) $\gamma$ = 1.5. Background distributions show normalized intensity patterns of the VOPBs at (varying) distances $z$ = 100, 200, 300, 400, and 500 mm. Blue arrows highlight the magnitude and direction of the transverse components of the Poynting vectors in the $x-y$ plane\cite{56}.}
\end{figure}

Developing a quartz-based vortex phase modulator remains technically challenging. In practice, liquid-crystal SLMs) are commonly used to generate VOPBs. However, because VOPBs require simultaneous modulation of both amplitude and phase—while most SLMs provide only phase modulation—the experiment relies on controlling diffraction efficiency to indirectly achieve amplitude shaping\cite{177,178}. The experimental setup is shown in Fig. 11(a): Path 1 generates the beam, while Path 2 forms an interferometric arm for measuring the topological charge. The generation scheme follows approaches similar to those used for Airy beams, bottle beams, and self-focusing beams\cite{179,180,181}. The SLM and spatial Filter are placed at the front and back focal planes of lens L, enabling the modulated amplitude–phase distribution to be transformed into the spatial domain, while the Filter removes unwanted background light, including unmodulated zero-order components.

Experimentally, the VOPB exhibits a hollow intensity profile throughout propagation due to the singular phase at its center. The beam first undergoes an autofocusing process, reaching peak intensity at approximately $z\approx 200$ mm, after which it evolves into a high-order Bessel-like beam. For $\gamma $= 0.5, both the hollow radius and main-lobe width increase during propagation; in contrast, for $\gamma > $ 1 (e.g., $\gamma $ = 1.5), the opposite trend occurs. When $\gamma $ = 1, the beam size remains nearly invariant beyond the focal region. Notably, amplitude modulation at the input plane plays a crucial role: with amplitude modulation, the beam maintains significant intensity even after 500 mm of free-space propagation, whereas without it, the VOPB rapidly decays after reaching its peak intensity. The interferometric arm records fork-shaped interference fringes characteristic of phase singularities, confirming a topological charge of $l$= –2 for all tested cases. These results clearly demonstrate the robust OAM carried by the VOPB.

It is important to note that the specific propagation distances cited above depend strongly on the initial beam size. In practical applications, increasing the beam diameter directly extends the propagation range, which scales proportionally to the square of the beam size\cite{182}. This can be accomplished either by beam expansion—which, however, reduces modulation resolution—or by developing solid-state modulators such as quartz-based VOPB devices. Although challenging, the latter offers the most effective long-term solution for real-world deployments.

Numerical simulations of the Poynting vector further reveal the internal energy flow of the VOPB (Figs. 11(e1)–11(f5)). At early propagation distances, energy flows radially inward from the side lobes toward the center, forming the characteristic focusing vortex structure. At longer distances, the energy circulates around the high-intensity main lobe. For $\gamma $ = 1.5, both the hollow radius and lobe width shrink during propagation, causing the transverse energy flux to become increasingly concentrated (Figs. 11(f1)–11(f5)). In contrast, for $\gamma $ = 1, the energy distribution remains nearly unchanged (Figs. 11(e1)–11(e5)).

The introduction of the VOPB is of significant importance for both fundamental and applied research. It expands the degrees of freedom for particle trapping and manipulation, and in free-space optical communication it offers enhanced data capacity through stable OAM transmission.
\begin{figure}[H]
\begin{center}
\begin{tabular}{c}
\includegraphics[scale=1.2]{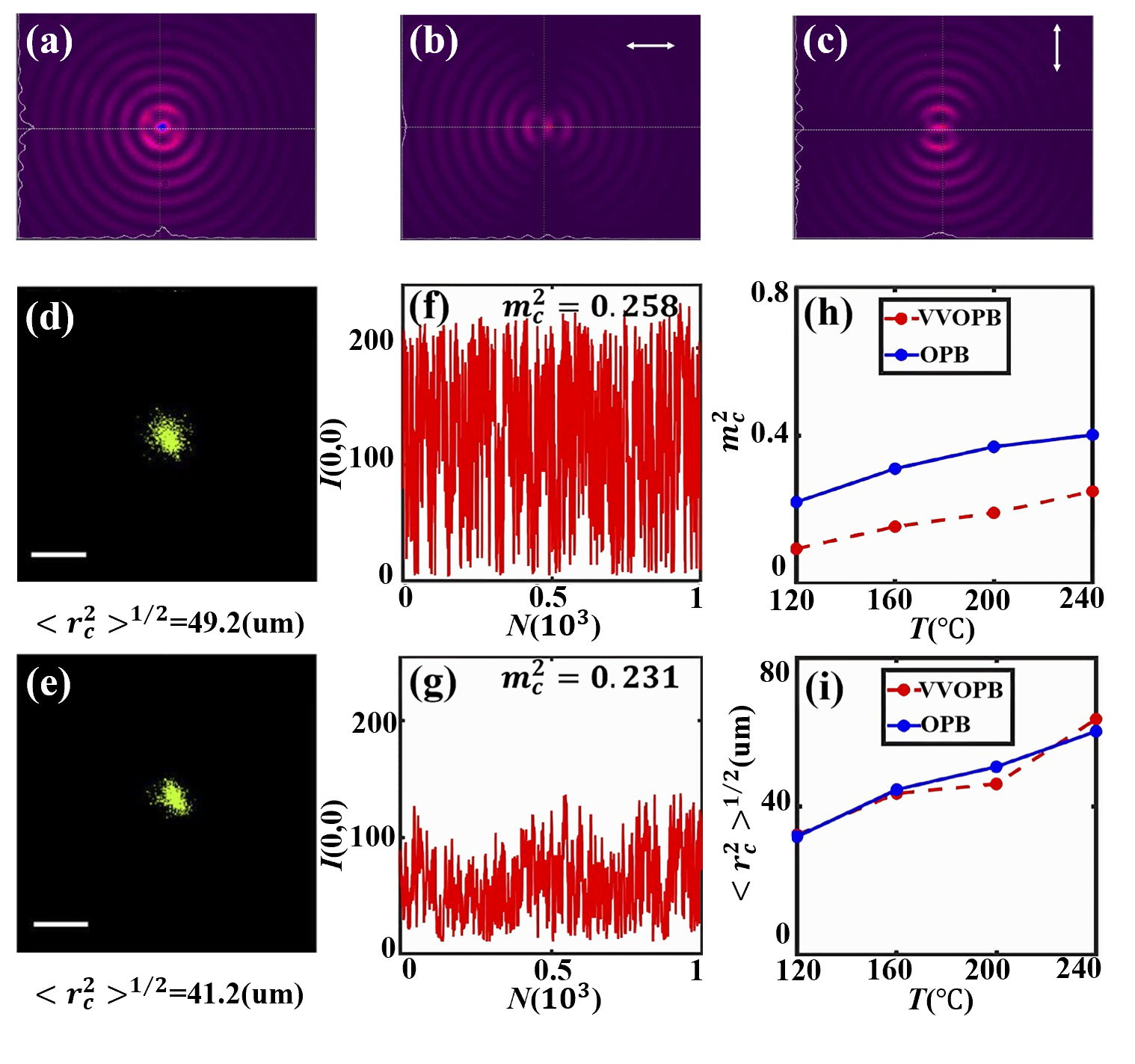}
\end{tabular}
\end{center}
\caption
{ \label{fig:example12}
  (a-c) The total intensity distribution(a), the $x$-component(b), and the $y$-component (c) of the VVOPB spot recorded by a beam profile analyzer. (d-g) distribution of the beam centroid with 1000 frames, (d, e) and Centroid intensity fluctuations with the frame number (f, g), (d) and (f) for the OPB (the first row) and (e) and (g) for the VVOPB (the second row). (h-i) Experimental results of the scintillation index (h) and the beam wander (i) of the OPB and the VVOPB as a function of the temperature of the hot plate\cite{60}.}
\end{figure}
Recent studies, such as that by Liu et al. on VOPB propagation through plasma-sheath turbulence, further highlight the potential of VOPBs as robust information carriers in harsh propagation environments\cite{183}. After the introduction of the VOPB, its propagation behavior in turbulence has drawn considerable attention. In 2022, Cao et al. investigated the OAM spectrum of VOPBs in non-Kolmogorov weak turbulence\cite{184}. They found that the receiving aperture size strongly affects the OAM spectrum. Although VOPBs maintain a stable intensity profile over long distances, their OAM spectrum can broaden under weak turbulence and short propagation, leading to mode crosstalk. However, under longer propagation distances, stronger turbulence, and smaller receiving apertures, the effective interaction cross-section with turbulence is reduced, resulting in significantly improved OAM purity and turbulence suppression. In 2023, Xu et al. performed a 500 m outdoor transmission experiment comparing VOPBs with vortex Gaussian beam\cite{185}. The VOPB exhibited noticeably lower beam distortion and over 50\% reduction in centroid drift, confirming its superior stability in atmospheric turbulence—critical for enhancing coupling efficiency, communication rates, and tracking performance.

More recently, Cai et al. proposed the Vector Vortex OPB (VVOPB) based on polarization-phase coupling\cite{60}. In their experiments, which employed a SLM and a controlled turbulence environment, they demonstrated that the VVOPB exhibits smaller intensity fluctuations than the OPB under identical turbulence conditions. Quantitative measurements of 1000-frame centroid trajectories, scintillation index, and drift (Figs. 12(d-i)) further confirmed that the VVOPB provides stronger turbulence resistance. These results indicate that introducing polarization singularities and topological charge jointly can enhance propagation robustness. VVOPB thus offers a promising strategy for multi-degree-of-freedom beam control in free-space optical communication, lidar, and remote sensing applications.

\subsection{Inverted OPB}

During the transmission of the OPB, its spot size gradually decreases, resembling a pin, hence it is called the optical pin beam, or pin-like beam. Then, the natural question arises: is it possible that there exists a kind of light beam whose spot size is relatively small at the initial stage but gradually increases during the transmission, while maintaining a good suppression effect against disturbances such as atmospheric turbulence? The problem was investigated by Efremidis’ group, who theoretically proposed the inverted OPB (IOPB)\cite{58}. The theoretical analysis is as follows:

We consider a beam propagating along the z-direction and denote the radial coordinate in the transverse $(x,y)$ plane as $r=\sqrt{{{x}^{2}}+{{y}^{2}}}$. We define the radial coordinate   on the initial plane   and assume an input beam profile
\begin{equation}
\label{eq:fov14}
\psi_0(\rho) = A(\rho) e^{i\phi(\rho)},
\end{equation}

where $A(\rho )$ is the amplitude, and the phase has a radial power-law dependence $\phi (\rho )=-kC_{\rho }^{\gamma }$.Also, $k=2\pi /\lambda $ is the wavenumber, $\gamma $ is the power-law exponent, and $C$ determines the speed of the phase variations. Under the assumption of a slowly varying amplitude $A(\rho )$, and following similar calculations as in Refs.\cite{67,76}, we obtain the following relation for the beam profile
\begin{equation}
\label{eq:fov15}
\psi(r,z) = \sqrt{\frac{2\pi k}{2-\gamma}} \left( C\gamma z^{\frac{\gamma}{2}} \right)^{\frac{1}{2-\gamma}} A(\rho(z)) J_0\!\left( \frac{kr\rho(z)}{z} \right) e^{i\Phi},
\end{equation}

where
\begin{equation}
\label{eq:fov16}
\rho(z) = (C\gamma z)^{\frac{1}{2-\gamma}},
\end{equation}

and $\Phi =k{{r}^{2}}/2z+{{\left( {{C}^{2}}{{\gamma }^{2}}{{z}^{\gamma }} \right)}^{1/2-\gamma }}(k/2)\left( 1-2/\gamma  \right)-\pi /4$. In terms of ray optics, Eq.(16) relates the radial displacement $\rho $ of a ray on the input plane to its focal distance $z$. As a result, if we want to generate a beam that maintains its Bessel-like shape up to a maximum propagation distance ${{z}_{\max }}$ with a transmitter aperture diameter ${{D}_{t}}$, then from Eq. (18) we find that
\begin{equation}
\label{eq:fov17}
C({{D}_{t}},{{z}_{\max }})=[1/(\gamma {{z}_{\max }})]{{({{D}_{t}}/2)}^{2-\gamma }}.
\end{equation}

We can derive a simplified real profile for the initial waveform $\psi (\rho ,z=0)$ using Bessel functions. In particular, we express the initial beam amplitude as
\begin{equation}
\label{eq:fov18}
A(\rho )=B(\rho ){{\rho }^{-\gamma /2}}\sqrt{(2-\gamma )/(2\pi k\gamma C)}.
\end{equation}

We then add in Eq. (14) its complex conjugate to obtain a real function. The rays of the complex conjugate are diverging (moving away from the axis) and, thus, do not affect the beam dynamics close to the optical axis. Using large argument Bessel asymptotics, the initial condition can be expressed as
\begin{equation}
\label{eq:fov19}
\psi_0(\rho) = B(\rho) J_0(kC\rho^{\gamma}).
\end{equation}

The initial profile of Eq. (19) takes the form of a Bessel function with a generic power-law dependence from the radius. From Eq. (20) the asymptotic formula of Eq. (15) is simplified to
\begin{equation}
\label{eq:fov20}
\psi (r,z)=B(\rho (z))\sqrt{\frac{2-\gamma }{\gamma }}{{J}_{0}}\left( \frac{kr\rho (z)}{z} \right).
\end{equation}
We note that for constant $B(\rho )$, independently of the value of $\gamma $, the on-axis amplitude of the beam remains constant during propagation. From the argument of the Bessel function in Eq. (20) we see that the full width at half maximum of the beam is
\begin{equation}
\label{eq:fov21}
FWHM(z)=2.2527{{k}^{-1}}{{(C\gamma {{z}^{\gamma -1}})}^{\frac{1}{\gamma -2}}}.
\end{equation}

\begin{figure}[H]
\begin{center}
\begin{tabular}{c}
\includegraphics[scale=0.9]{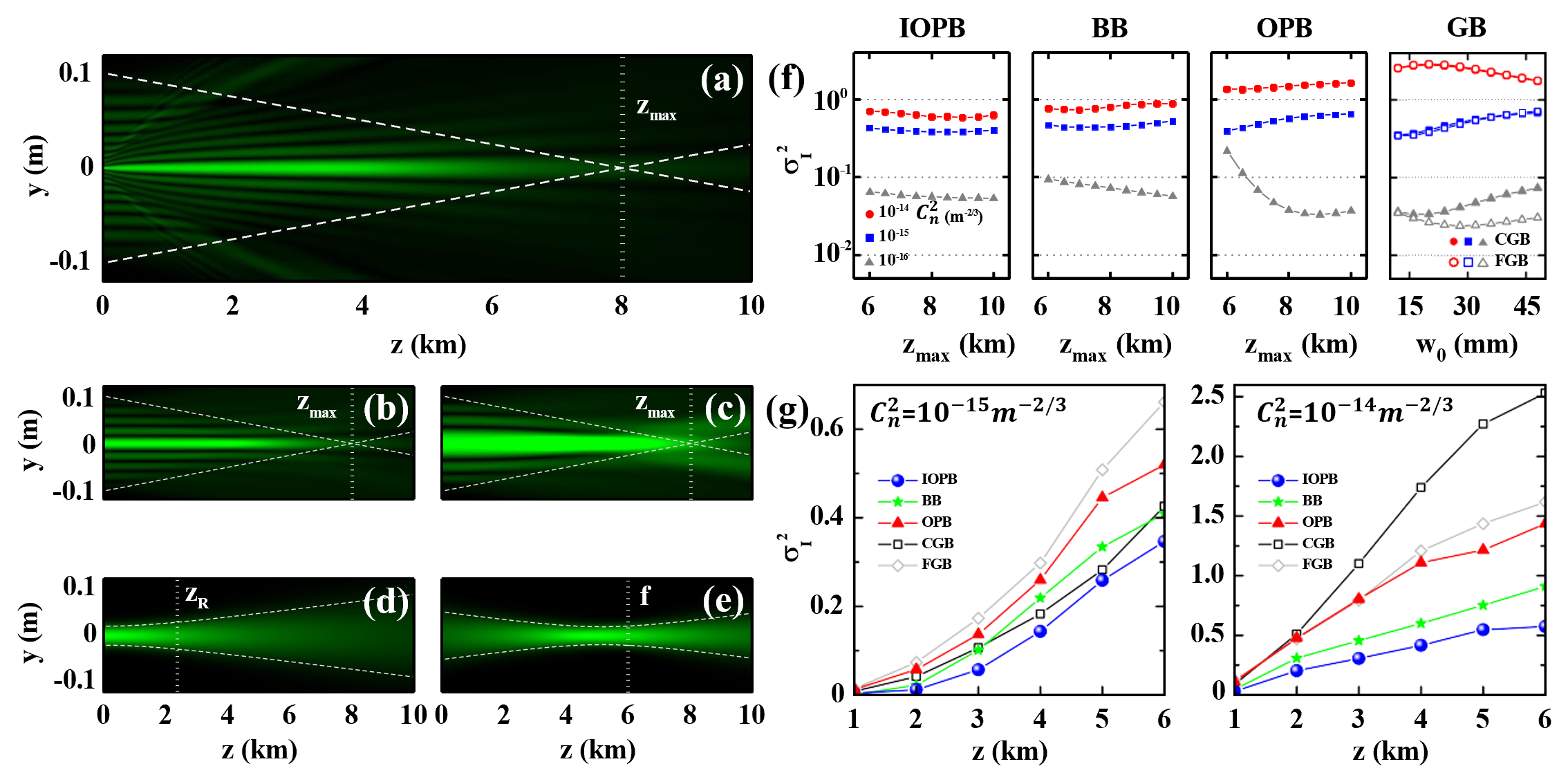}
\end{tabular}
\end{center}
\caption
{ \label{fig:example13}
  (a-e) Field amplitude dynamics of (a) an IOPB with  and comparison with (b) OPB with $\gamma =1$ (BBs), (c) a OPB with $\gamma =1.4$, (d) a CGB, and (e) an FGB. In (a), (b), and (c) ${{z}_{\max }}=8\text{ }\operatorname{km}$is the maximum propagation distance of these beams, and the dashed lines depict the cone generated by the outer rays (starting at $\rho ={{D}_{t}}/2$). In (d), the width of the CGB is ${{w}_{0}}=20\text{ }\operatorname{mm}$corresponding to a Rayleigh length ${{Z}_{R}}\approx 2.4$km (the focal length is $f=\infty $), whereas in (e) the FGB has ${{w}_{0}}=50\text{ }\operatorname{mm}$and $f=6\text{ }\operatorname{km}$ . In (d) and (e) the dashed curves mark the evolution of the beam diameter; (f) Diagrams used for optimal beam selection at a receiver distance $L=6\text{ }\operatorname{km}$, based on the scintillation index for different turbulence strengths ($C_{n}^{2}={{10}^{-16}},{{10}^{-15}},{{10}^{-14}}{{m}^{-2/3}}$). We depict $\sigma _{I}^{2}$versus ${{z}_{\max }}$ for IOPB, BB, and PB, and versus ${{w}_{0}}$ for CGB and FGB; (g) Scintillation index versus propagation distance for beams propagating in moderate turbulence ($C_{n}^{2}={{10}^{-15}}{{m}^{-2/3}}$) and strong turbulence ($C_{n}^{2}={{10}^{-14}}{{m}^{-2/3}}$). All beams are selected to have optimum scintillation at the target distance of $L$= 6 km.\cite{58}.}
\end{figure}

Importantly, for different values of $\gamma $, the generated beams exhibit significantly different behavior. In particular, for$1<\gamma <2$, a regime explored in\cite{55,67}, the width of the beam decreases during propagation leading to a pin resembling beam. On the other hand, for$\gamma =1$the width of the beam remains constant independently of $z-a$ Bessel beam\cite{186}. Finally, for $0<\gamma <1$ the width of the beam increases during propagation resulting to an inverted pin beam profile\cite{58}. 

The behavior of IOPBs is also compared with GS characterized by the input profile
\begin{equation}
\label{eq:fov22}
\psi_0(\rho) = A \exp\left( -\frac{\rho^2}{w_0^2} - \frac{ik\rho^2}{2f} \right),
\end{equation}

where $w_{0}^{2}$ is the beam radius on the input plane and $f$ the focal distance. For $f\to \infty $, the beam is collimated and the only tuning parameter is the input beam radius ${{w}_{0}}$. The closed-form beam dynamics of Eq. (22) can be found for example in [161]\cite{161}.

The transmission behavior of the IOPBs is shown in Fig. 13 and compared with OPBs, Bessel beams (BBs), collimated Gaussian beams (CGBs), and focused Gaussian beams (FGBs). In particular, in Fig. 13(a)-(e), see the dynamics of these classes of beams in the absence of turbulence. As can be see from Fig. 13(a) the IOPB resembles an inverted PIN during transmission, with its spot size gradually increasing as it propagates. Importantly, the IOPB maintains the smallest main lobe as compared to all the other classes of beams. In Fig. 13(f) optimal parameters are selected for these five classes of beams. Specifically, the scintillation index is computed as a function of ${{z}_{\max }}$ for the pin beams and ${{w}_{0}}$ for the GS. We take ensemble average over 1000 realizations and select a 1 inch receiver aperture for different values of the refractive index structure parameter. We note that for different values of the refractive index structure parameter, the scintillation index of GS is minimum for different values of the beam radius. On the other hand, the same values of ${{z}_{\max }}$ are optimal irrespectively of the strength of turbulence. We consider this invariance of the design parameters to be a significant advantage of pin beams over Gaussian beam. Comparing the values of the scintillation index of these five classes of beams, we see that for moderate and especially for strong refractive index fluctuations, all types of pin beams have reduced scintillations as compared to the Gaussian. Furthermore, especially in the moderate and strong fluctuations regime, the IOPBs outperform all the other families of examined beams\cite{58}.

We also examined how well a system, which is designed to be optimal for a specific set of parameters, behaves in different conditions. Specifically, we optimized all five classes of beams according to Fig. 13(f) to propagate over 6 km. We then change the target distance between 1-6 km with 1 km steps. We see that for strong atmospheric turbulence, the pin classes of beams outperform the GS (both focused and collimated). Importantly, for both moderate and strong fluctuations conditions, and for all propagation distances, the IOPB always has the lowest values of the scintillation index. The different between IOPB and the rest of the families is more prominent as the strength of turbulence is increased. 

\begin{figure}[H]
\begin{center}
\begin{tabular}{c}
\includegraphics[scale=1.01]{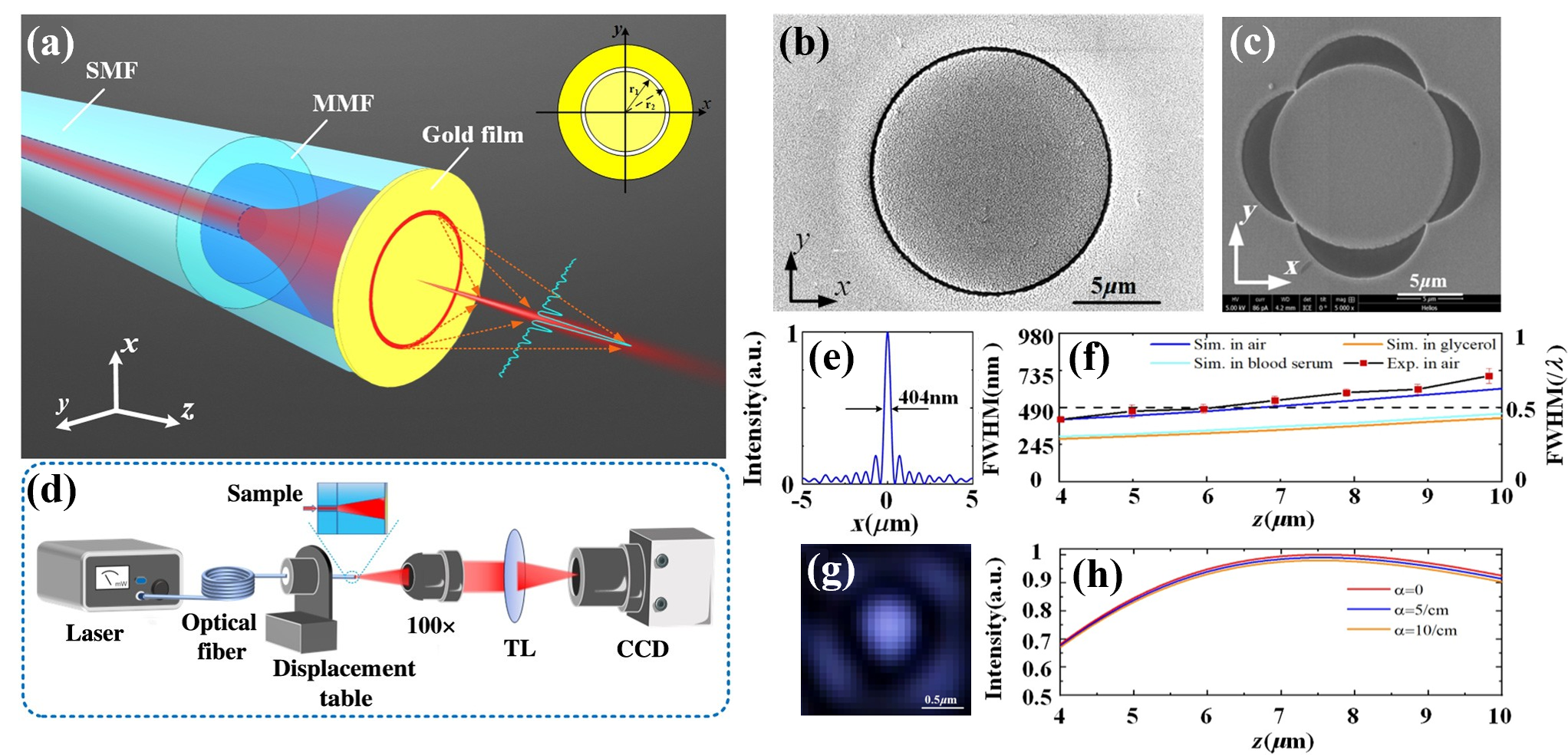}
\end{tabular}
\end{center}
\caption
{ \label{fig:example14}
  (a) Subwavelength IOPB are generated by integrating a simple plasma structure on the end face of the optical fiber. (b-c) SEM images of nano-ring grooves and nano-metal petal structures. (d) Device diagram. (e) The normalized intensity distribution of the output light field on the x-axis. (f) The beam generated by the nano-petal structure has the ability to suppress side petals along the beam direction. (g) The variation of the main lobe bandwidth of IOPB with the propagation distance in different environments. (h) The normalized intensity distribution on the central axis of IOPB in the serum environment under different absorption losses\cite{57}.}
\end{figure}

Optical fiber is an important optical regulation platform, which has the advantages of high integration, flexible regulation, and anti-interference, and is widely used in various systems\cite{187,188,189}. Considering that the optical fiber, as the transmitting end of the light beam, has an extremely small core diameter, it provides an excellent application scenario for the IOPB. Therefore, in 2024, Wang Jian and others, based on theoretical design and experimental preparation, realized a fully fiber-integrated solution\cite{57}. By integrating a plasmonic nanostructure on the end face of the optical fiber, they achieved the efficient generation of subwavelength IOPB. As shown in Fig. 14(a), this solution adopts the structures of a nanoring groove (Fig. 14(b)) and a nanometal (Fig. 14(c)), in which the nanometal structure significantly enhances the ability to suppress the side lobes of the light beam in the x-axis direction. During the experiment, all the subwavelength IOPB generators were fixed on an electronically controlled nano-displacement stage to ensure high-precision regulation. The end face of the optical fiber was defined as the original plane ($z=0$). During the experiment, the end face of the beam generator was gradually moved with an accuracy of 200 nanometers through the nano-displacement stage to systematically observe and record the intensity distribution of the output optical field at different transmission distances. The experimental setup is shown in Fig. 14(d)\cite{57}.

The research further analyzed the variation of the main lobe bandwidth of the subwavelength IOPB with propagation distance in different environments. Results show that even when the beam propagates up to $10\,\mu\mathrm{m}$, its full width at half maximum (FWHM) remains at the subwavelength scale. This characteristic renders the beam particularly suitable for non-invasive single-point illumination or targeted stimulation of biological samples, such as organelles, biomolecules, and nano-pharmaceutical particles, enabling the exploration of light--matter interactions at the single-cell scale \cite{190,191}. Experimental results demonstrate that this fully fiber-integrated device can flexibly adapt to liquid environments including cell culture media, blood, and biological tissue fluids, thereby achieving high-resolution illumination and stimulation of biological cells and molecules. This work provides a fiber-integrated solution with significant application potential for subwavelength-resolution light--matter interaction studies in biological research\cite{234}.

As a special type of pin beam, the IOPB has been shown to have significant advantages in reducing the scintillation index compared with other types of OPBs and GS, and it exhibits better stability especially under moderate to strong turbulence conditions. Based on this characteristic, a Vortex Inverted Optical Pin Beam (VIOPB) carrying OAM was proposed, and demonstrated that this beam can further suppress intensity scintillation under moderate to strong irradiance fluctuation conditions\cite{59}. Through numerical simulations, the optimal scintillation index characteristics of the VIOPB under different topological charges, refractive index structure parameters, and receiver aperture conditions was investigated.

In a turbulence-free environment, the propagation characteristics of the VIOPBs with different topological charges was explored, revealing the unique phase structure and dynamic behavior of this beam, as shown in Figs. 15\cite{59}. As can be seen in Figs. 15(a)-(d) for $0<\gamma <1$ the core radius decreases during propagation leading to a vortex inverted pin beam. Furthermore, by increasing the value of the topological charge $\left| n \right|$, the radius of the hollow core increases. The research further conducted an in-depth analysis of the optimal beam selection as a function of ${{z}_{\max }}$ for different receiver apertures and refractive index structure parameters at a receiving distance of 6 km, and the results are shown in Figs. 15(e1-e4). When the receiver aperture is large and the intensity fluctuations are weak, the VIOPB with a topological charge of $n=0$shows the lowest scintillation index when the parameter ${{z}_{\max }}$ reaches 12 km. As the receiver aperture decreases or the turbulence intensity increases, the beams with larger topological charges (such as $n=1$and $n=3$) are more advantageous. Under strong turbulence conditions, the VIOPB with a topological charge of $n=3$ performs best when ${{z}_{\max }}\approx 10.5$ km. When the receiver aperture is further reduced and the intensity fluctuation is enhanced, the beam with a topological charge of $n=2$ exhibits more prominent performance. The optimal beams, as determined from Fig. 15(e) for propagation distance $L=6$ km, are the compared in Fig. 15(f) for smaller propagation distances. Specifically, it was found that the beams with a smaller topological charge (such as$n=0$, 1) perform better for weak irradiance fluctuations. As the intensity fluctuations increase, the performance of the beams with higher topological charges (such as $n=2$ and $n=3$) is significantly improved, especially showing more prominent advantages under moderate to strong intensity fluctuation conditions.

\begin{figure}
\begin{center}
\begin{tabular}{c}
\includegraphics[scale=0.9]{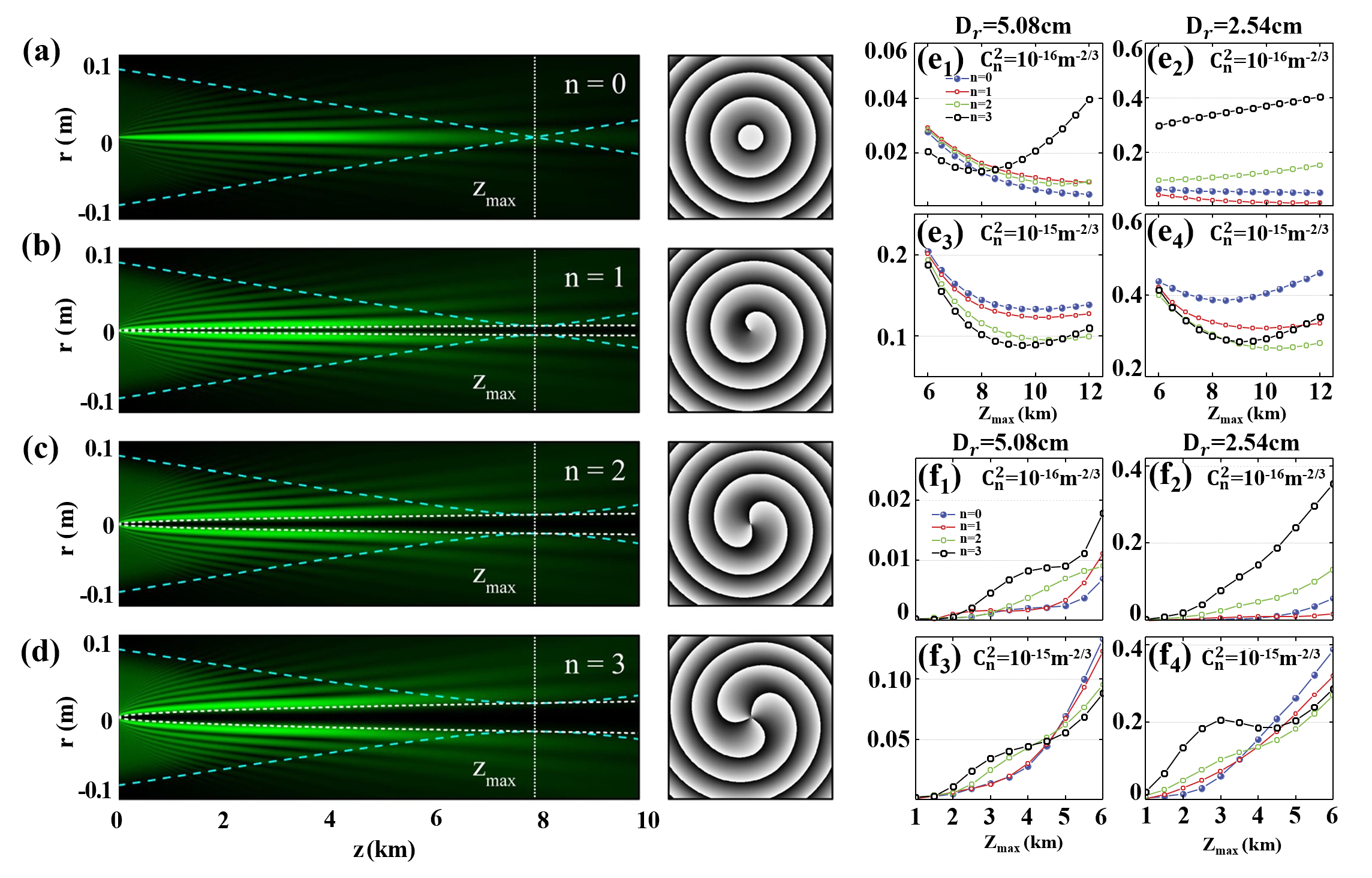}
\end{tabular}
\end{center}
\caption
{ \label{fig:example15}
  The transport patterns of VIOPB when $\gamma =4$and topological charges (a) $n=0$,(b) $n=1$,(c) $n=2$,(d) $n=3$. The phase of each beam after propagation at $L=6$km is shown on the right; (e1)-(e4) At the receiving distance $L=6$km, the selection of the OPB with the optimal $\gamma =0.4$based on the scintillation index;(f1)-(f4) The curves of the scintillation index of VIOPB with different receiving apertures varying with the transmission distance L when $\gamma =0.4$\cite{59}.}
\end{figure}

Overall, the scintillation index value of the VIOPB under moderate to strong fluctuation conditions is lower than that of the ordinary IOPB, and the optimal OAM value depends on the specific environmental parameters and experimental settings. The simplicity of generating the VIOPB and its superior performance have promoted the application of the OPB beams in free-space optical communication.

\section{Emerging Applications of OPBs}

As previously mentioned, OPBs represent a light-field reconstruction technique distinct from BBs or Airy beams, which exhibit relatively well-defined optical field profiles. Due to its unique mechanism where wave vectors mutually cancel during propagation, OPBs demonstrate exceptional stability in transmission and remarkable resistance to disturbances. In principle, this technology holds significant promise for applications involving laser transmission through environments with aberrations, turbulence, or multiple particle scattering. Currently reported applications primarily include laser communication, underwater laser communication, particle trapping, and precision calibration of imaging detectors. This technique leverages its inherent self-stabilizing properties to mitigate signal degradation caused by environmental interference, making it particularly valuable in scenarios requiring robust optical performance under challenging conditions.

\subsection{Application in FSO communication}

Laser communication (LC), as a critical component of modern information technology, offers significantly higher data transmission capacity and lower probability of signal interception compared to traditional radio-frequency systems, demonstrating immense potential for future high-capacity communication networks\cite{88,112,192,193,194,195,196,197,198,199,200}. With the rapid deployment of large-scale satellite constellations like Starlink and China’s "StarNet," the volume of space-based data is growing exponentially, driving an urgent need to advance satellite-to-ground laser communication technologies to enhance downlink rates\cite{201,202}. However, the availability of satellite-ground laser communication links has long been constrained by atmospheric turbulence\cite{202,203,204,205,206,207,208,209}. While adaptive optics (AO) remains the mainstream method for improving link performance, its widespread adoption faces challenges such as high costs and system complexity\cite{6}. In contrast, OPB requires only a single modulation device combined with a beam transformation system to achieve turbulence suppression, significantly reducing both cost and system complexity. This has led to widespread recognition of OPB’s promising applications in laser communication\cite{6,62,66,70,87,185,210,211,212,213,214,215}.

Research on OPB-based laser communication focuses on three key aspects: First, how to apply OAM to enhance carrier channel capacity\cite{6,184,211,216} ; Second, how to leverage self-healing properties to adapt to real-world environments like rain, fog, and water scattering, thereby improving communication performance \cite{60,170,217}; Third, how to implement OPB modulation in practical communication systems for effective transmission and reception. The first two aspects have already been reviewed in preceding sections, covering OAM spectral studies primarily with vortex-type OPB and investigations into self-healing properties of OPB against rain, fog, and water scattering. This section will emphasize current progress in addressing the third challenge.

The basic architecture of an OPB-based laser communication link is illustrated in Fig. 16(a). Its core lies in the transmitter, where a GS is converted into an OPB carrier via a beam transformation system and an OPB modulation mask. Pioneering work on OPB-enabled laser communication was conducted by Hu et al. in 2022, who experimentally validated a 1 Gbit/s On-Off Keying (OOK) laser communication system\cite{62}. Their setup replaced traditional OPB modulators with a liquid-crystal SLM, as shown in Fig. 16(b). Key experimental results include: (1) Power Loss Reduction: Under a limited receiver aperture (Rx = 1 mm), OPB reduced power loss by 8-13 dB compared to conventional GS over propagation distances of 0.45-0.8 m, while maintaining low signal attenuation even in small-aperture conditions (Fig. 16(c)). (2) Bit Error Rate (BER) Improvement: The OPB-based system achieved a BER close to the Forward Error Correction (FEC) threshold, significantly outperforming GS systems (Fig. 16(d)). (3) Enhanced Signal Quality: The eye diagram of the received signal showed marked improvements in eye height and width, particularly under small-aperture reception scenarios (Fig. 16(e)).

These findings underscore the capability of OPBs to mitigate turbulence-induced signal degradation while maintaining high compatibility with compact, low-cost receiver designs—a critical advantage for future satellite-ground and deep-space communication systems.

\begin{figure}[H]
\begin{center}
\begin{tabular}{c}
\includegraphics[scale=0.88]{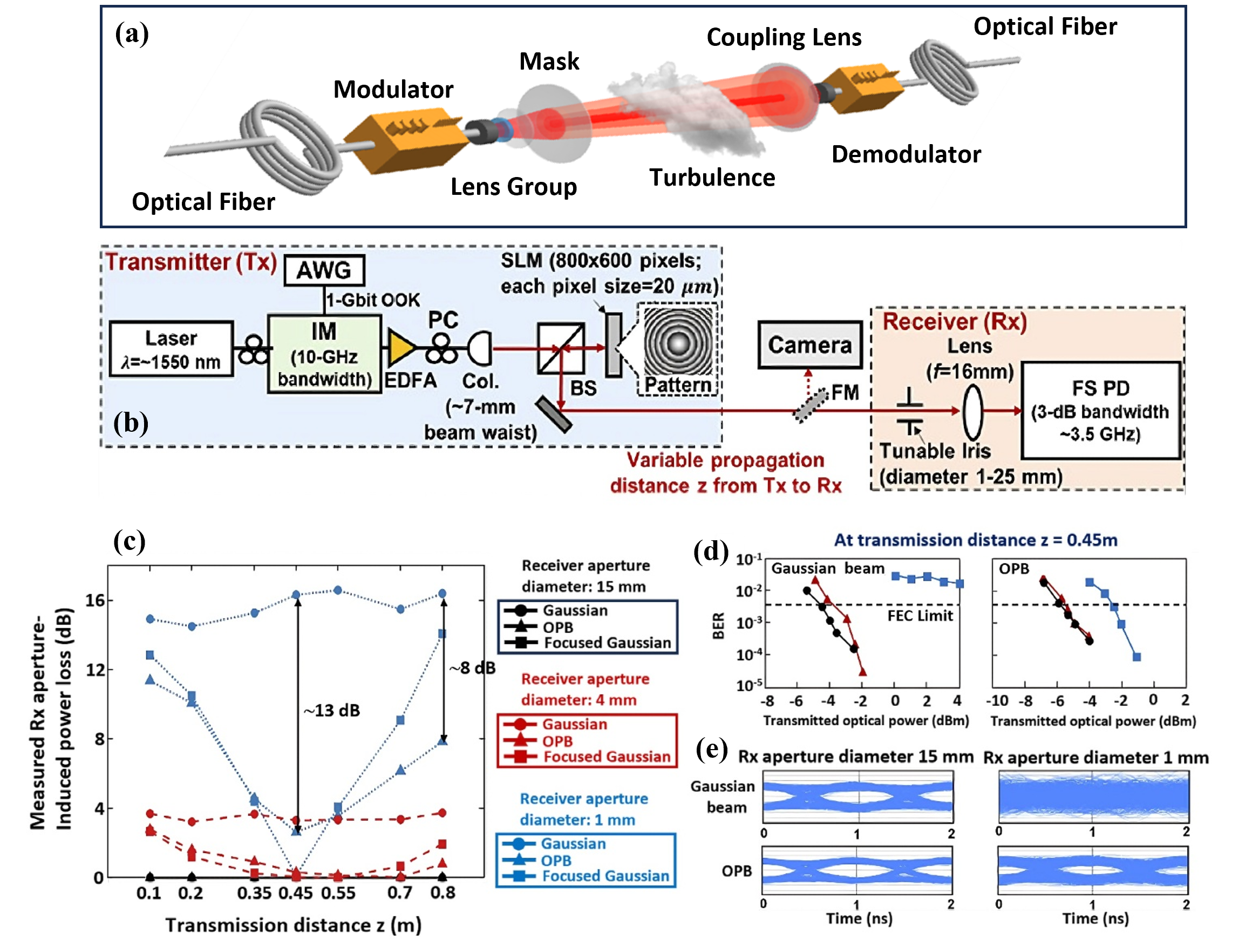}
\end{tabular}
\end{center}
\caption
{ \label{fig:example16}
  (a) The basic architecture of an OPB-based laser communication link. (b) Experimental optical path. (c) Experimentally measured limited-size Rx aperture- induced power loss under different Rx aperture sizes for a Gaussian, focused Gaussian, and pin-like beam at different transmission distances. (d) BER performance for a GS and OPB at $z=0.45$m. (e) Eye diagrams for a GS and OPB at $z=0.45$m\cite{62}.}
\end{figure}

In 2025, Wang and colleagues proposed that OPB holds significant promise for satellite-to-ground laser communication, particularly in uplink laser communication\cite{218}. Due to the complex structures, large size, and weight of adaptive optics components, it is challenging to install them on satellites. In such scenarios, OPB can be employed at the transmitter to mitigate the effects of atmospheric turbulence, effectively addressing communication performance degradation caused by atmospheric scintillation. Their research demonstrated that under predefined conditions, OPB can suppress over 43.56\% of scintillation effects. Specific results, as shown in Fig. 17(a-d), reveal that different beam shape factors and modulation depths yield distinct turbulence suppression efficiencies under varying turbulence conditions. This further validates that OPB parameters should be tailored based on specific turbulence scenarios in practical applications to achieve optimal suppression performance.

\begin{figure}[H]
\begin{center}
\begin{tabular}{c}
\includegraphics[scale=0.88]{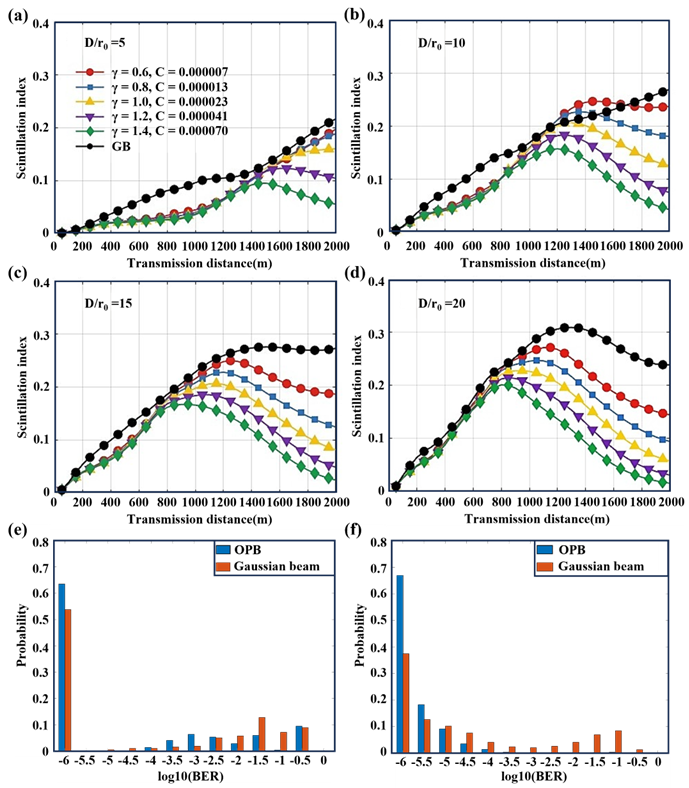}
\end{tabular}
\end{center}
\caption
{ \label{fig:example17}
  (a)-(d) Variation of the scintillation index with propagation distance under different atmospheric turbulence of $D/{{r}_{0}}=5$, $D/{{r}_{0}}=10$, $D/{{r}_{0}}=15$,and $D/{{r}_{0}}=20$\cite{218}. (e)-(f) Probability distribution of BER for OPB and GS when the turbulence intensity is ${{r}_{0}}=1.2$cm, and ${{r}_{0}}=2.1$cm\cite{64}.}
\end{figure}

Also in 2025, Lu et al. conducted a 1 km horizontal link field communication experiment using OPBs\cite{64}. As shown in Fig. 17(e-f), under strong, moderate, and weak turbulence intensities, OPB demonstrated significant improvements in coupling power, communication stability, and BER compared to GS. These results provide clear evidence of OPB’s turbulence suppression capabilities. Furthermore, OPB’s ability to simplify system architecture and reduce complexity suggests its potential for longer-distance laser communication applications. This experiment highlights the practical viability of OPB in real-world scenarios, reinforcing its adaptability and efficiency in mitigating atmospheric turbulence effects.

Researchers from Airbus DS GmbH in Germany, led by Nardo et al., evaluated the performance of OPBs in long-range air-to-air FSO communication links\cite{70}. Their study shows that, under a turbulence strength of $C_{n}^{2}=1.66\times {{10}^{-17}}\text{ }{{m}^{-2/3}}$, replacing GS with OPBs in a 10 km-altitude, 100 km-range air-to-air optical link significantly improves the link performance, as summarized in Table 1. Under identical transmitting aperture, receiving aperture, and transmitted power conditions, the OPB achieves up to 8.6 dB higher received power and a 50\% reduction in scintillation index compared to a collimated GB. In addition, the OPB exhibits a depth-of-focus extending over tens of kilometers, making it particularly advantageous for mobile-target communication, where dynamic alignment and atmospheric fluctuations are critical challenges. The authors further confirmed the quadratic scaling of the effective propagation distance with the transmitting aperture diameter for OPBs, consistent with theoretical predictions.

\begin{table}[ht]
\caption{Turbulent propagation: average beam wander ${{r}_{c}}$, average power received ${{P}_{RX}}$ and scintillation index $\sigma _{I}^{2}$ by a 10cm aperture, w/ and w/o beam wander correction at $z=100$ km.}
\label{tab1:beam_performance}
\begin{center}
\begin{tabular}{|c|c|c|c|c|c|}
\hline
\rule[-1ex]{0pt}{3.5ex} Beam Class & \( r_c \) (m) & \multicolumn{2}{c|}{Receiver on axis} & \multicolumn{2}{c|}{Beam wander corrected} \\
\cline{3-6}
\rule[-1ex]{0pt}{3.5ex} & & \( P_{\text{RX}} \) (dBm) & \( \sigma_r^2 \) & \( P_{\text{RX}} \) (dBm) & \( \sigma_r^2 \) \\
\hline\hline
\rule[-1ex]{0pt}{3.5ex} OPB, \( y = 0.1 \) & 0.23 & 7.00 & 0.3448 & 9.67 & 0.1254 \\
\hline
\rule[-1ex]{0pt}{3.5ex} OPB, \( y = 1 \) & 0.15 & 13.45 & 0.3700 & 17.87 & 0.2038 \\
\hline
\rule[-1ex]{0pt}{3.5ex} OPB, \( y = 1.9 \) & 0.20 & 13.28 & 0.4337 & 18.17 & 0.2175 \\
\hline
\rule[-1ex]{0pt}{3.5ex} Collimated-GS & 0.45 & 4.67 & 0.8274 & 12.32 & 0.1651 \\
\hline
\rule[-1ex]{0pt}{3.5ex} Focused-GS & 0.23 & 15.92 & 1.0894 & 21.87 & 0.1859 \\
\hline
\end{tabular}
\end{center}
\end{table}

To support these conclusions, Nardo et al. developed a unified theoretical propagation model and performed extensive Monte Carlo simulations and link-budget analysis under controlled turbulence conditions. Their results underscore the strong potential of OPBs for airborne and satellite-based optical communication systems. However, they also emphasize a key practical constraint: the quadratic scaling law implies that realizing very long OPB-preservation distances would require transmitting apertures larger than 10 cm, which is often infeasible for airborne or spaceborne platforms due to payload and structural limitations. Nonetheless, even when the OPB eventually transitions away from its ideal structural form beyond the preservation range, its turbulence-suppression advantages persist within that range, meaning OPBs can still offer superior overall link stability compared to conventional beams.
\begin{figure}[H]
\begin{center}
\begin{tabular}{c}
\includegraphics[scale=0.68]{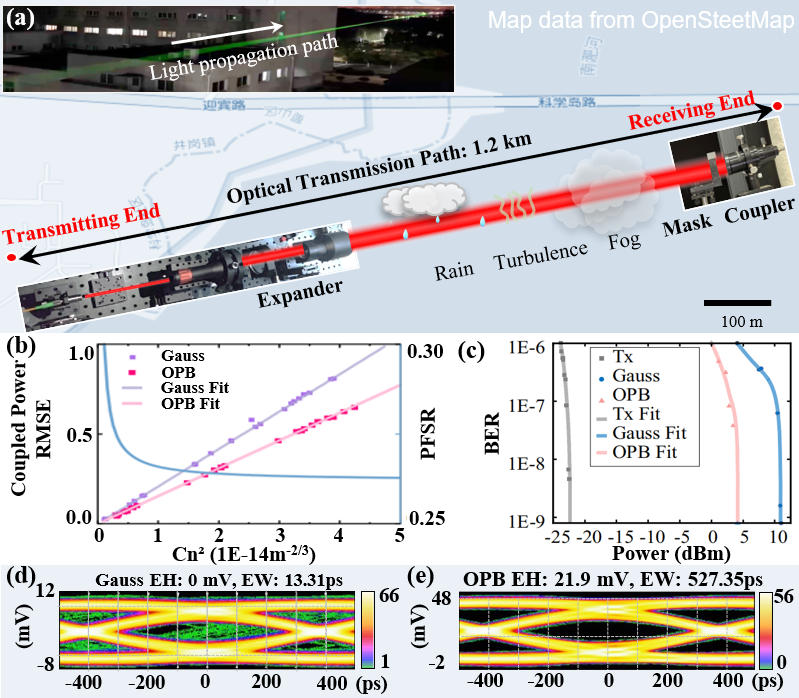}
\end{tabular}
\end{center}
\caption
{ \label{fig:example18}
   High-speed 1-km FSO link performance enabled by the OPB receiver: (a) Layout of the 1-km outdoor link. (b) Root-mean-square error (RMSE) of the coupled optical power at the Gaussian and OPB receivers under various atmospheric conditions (left $y$-axis), where the blue curve denotes the power fluctuation suppression ratio (PFSR) obtained from the fitted trend (right $y$-axis). (c) BER characteristics of the transmitter, Gaussian receiver, and OPB receiver as a function of transmit power. (d-e) Eye patterns of the Gaussian and OPB receivers at 1.25 Gbps with a 15 mm receive aperture\cite{66}.}
\end{figure}
Recently, Guan et al. proposed an OPB receiver-side 100 Gbps laser communication scheme in which a static phase modulation mask is placed before the coupling lens to reshape turbulence-distorted beams into an extended-Rayleigh-length OPB structure\cite{66}. This approach enhances coupling resilience without relying on transmitter-side modulation or adaptive optics, while maintaining simple architecture and high-power compatibility, and remaining applicable to SMF, FMF, and MMF systems. Both simulations and a 1-km outdoor experiment demonstrated that the OPB receiver effectively suppresses wavefront distortions and focal jitter, improving coupled power stability by approximately 26\% and reducing OOK/QPSK BER by up to two orders of magnitude, with correspondingly improved eye diagram quality, as shown in Fig. 18. This work provides a scalable, low-complexity receiver-side solution for high-speed and long-distance free-space optical communication links. Delving into the practical implications, the adoption of OPBs in high-performance FSO system could transform data centers’ interconnectivity, replacing or supplementing fiber links that are susceptible to physical damage or congestion, as commented in the Science News\cite{219}.

\subsection{Application in underwater wireless optical communication}

OPB has been demonstrated to mitigate not only atmospheric turbulence but also aquatic turbulence effects\cite{146,175,220}. Notably, laser transmission in water faces even more severe challenges from turbulence and scattering compared to atmospheric environments. To address this, Xiaotian Han et al. applied OPB to underwater wireless optical communication (UWOC), achieving a 5 Gbps communication rate\cite{65,221}. The experimental optical setup is illustrated in Fig. 19(a), where a SLM loaded with phase masks was used to modulate a GS into zero-order BBs, OPB, and VOPB. Key findings, as shown in Figs. 19(b-d), include:

Auto-Focusing Property: Both OPB and VOPB exhibit self-focusing capabilities, maintaining narrow beam characteristics over a certain range beyond the focal point.

Power Loss Advantage: Under limited receiver aperture conditions, OPB and VOPB show significantly lower optical power loss compared to Gaussian and zero-order BBs. The power loss difference between OPB and Gaussian/Bessel beams reaches its maximum at the designed self-focusing point.

Enhanced Communication Performance: By evaluating signal integrity and BER, OPB and VOPB effectively improve the communication distance and data rate of UWOC systems.

\begin{figure}[H]
\begin{center}
\begin{tabular}{c}
\includegraphics[scale=0.77]{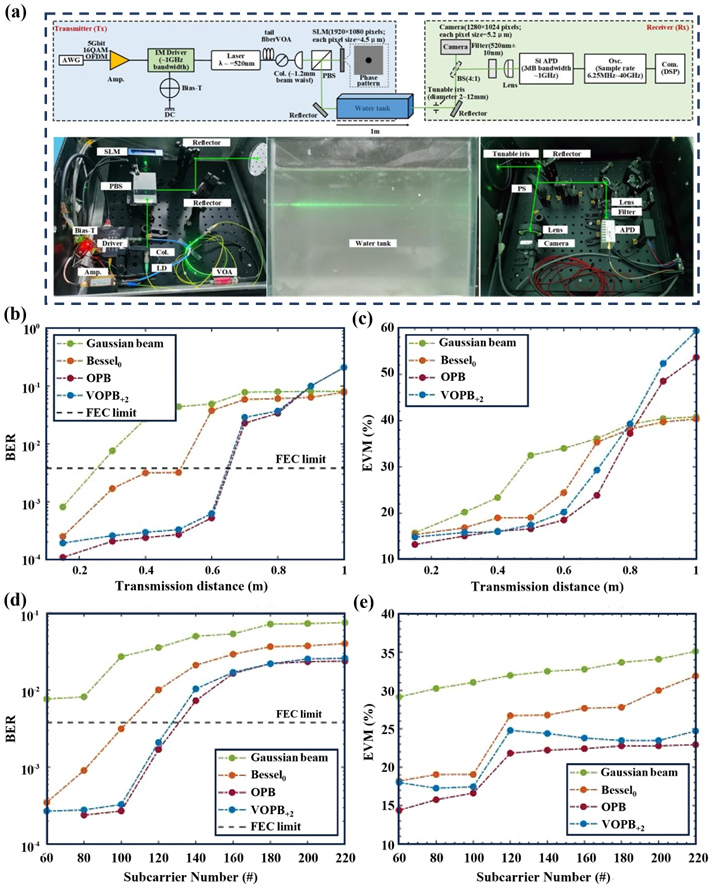}
\end{tabular}
\end{center}
\caption
{ \label{fig:example19}
  Test results of underwater laser communication based on OPB:(a) Schematic diagram and physical image of the underwater laser communication optical path. (b-e) BER (b) and Error Vector Magnitude (EVM) (c) of different beams at various underwater distances, as well as BER (d) and EVM (e) for different subcarrier numbers\cite{65}.}
\end{figure}

The authors also mentioned that the experiment was conducted in a 1-meter water tank, constrained by the SLM’s effective area size. However, the self-focusing distance of OPB and VOPB can be flexibly tailored through design, making them highly valuable for long-range, high-speed UWOC systems with restricted receiver apertures. The authors highlight that OPB opens new possibilities for underwater OAM multiplexing communication and high-resolution imaging in randomly scattering media, such as turbid water\cite{233}. By integrating self-focusing, low power loss, and controllable propagation, OPB and VOPB provide critical technical advantages for next-generation high-performance underwater optical communication systems, particularly in challenging environments with strong scattering and turbulence.

\subsection{Other Emerging Applications}

OPBs possesses a unique light-field structure that evolves dynamically while maintaining an extended depth of focus. Leveraging these properties, Chen et al. investigated the focusing characteristics of OPBs for particle trapping\cite{61}. Their results show that OPBs significantly enhance the stability of particle trapping and rotation, and facilitate multi-particle manipulation along the beam axis. Notably, when a vortex phase is introduced, the trapping mechanism drives particles to circulate around the focal ring, thereby reducing the risk of thermal damage. In parallel, Guo et al. designed non-paraxial pin-like optical beams using a geometric approach. These beams exhibit self-healing, self-reconstruction, and long-distance propagation capabilities, and the authors further analyzed the pulling and pushing optical forces exerted on particles by such beams\cite{68}. Wang et al. introduced a new class of circular Pearcey beams capable of generating tunable OPBs and optical channels; their experiments demonstrated effective manipulation of micron-sized particles using these beams\cite{222}.

The robustness of OPBs, evident in its stable propagation through scattering and disturbances—is not limited to harsh environments such as rain, fog, or turbulence. Similar to how instruments like Zygo interferometers use interference fringes for high-precision surface metrology\cite{223,224,225}, the light field generated by OPBs via the wave-vector elimination principle (the region between ${{z}_{1}}$ and ${{z}_{2}}$ in Fig. 3(a)) features an almost ideal and highly stable intensity distribution. This enables the OPB to serve as a precise optical “ruler” capable of detecting minute structural deviations. In 2025, Zhang Ze et al. employed this high-fidelity light field to measure, for the first time, the pixel-level quantum efficiency of an imaging sensor, thereby realizing a novel imaging technique that surpasses the Nyquist sampling limit—termed hyper-sampling imaging\cite{63}. This method effectively enhances the sampling density so that a single physical pixel can function as sixteen virtual subpixels. As shown in Fig. 20, Fig. 20(a) presents the interference field resulting from transverse wave-vector elimination; Figs. 20(b, c) display the measured quantum efficiency at the pixel and subpixel levels; Figs. 20(d-f) show the original text, QR code, and drone images; and Figs. 20(g-i) illustrate their corresponding hyper-sampled reconstructions.

\begin{figure}[H]
\begin{center}
\begin{tabular}{c}
\includegraphics[scale=1.2]{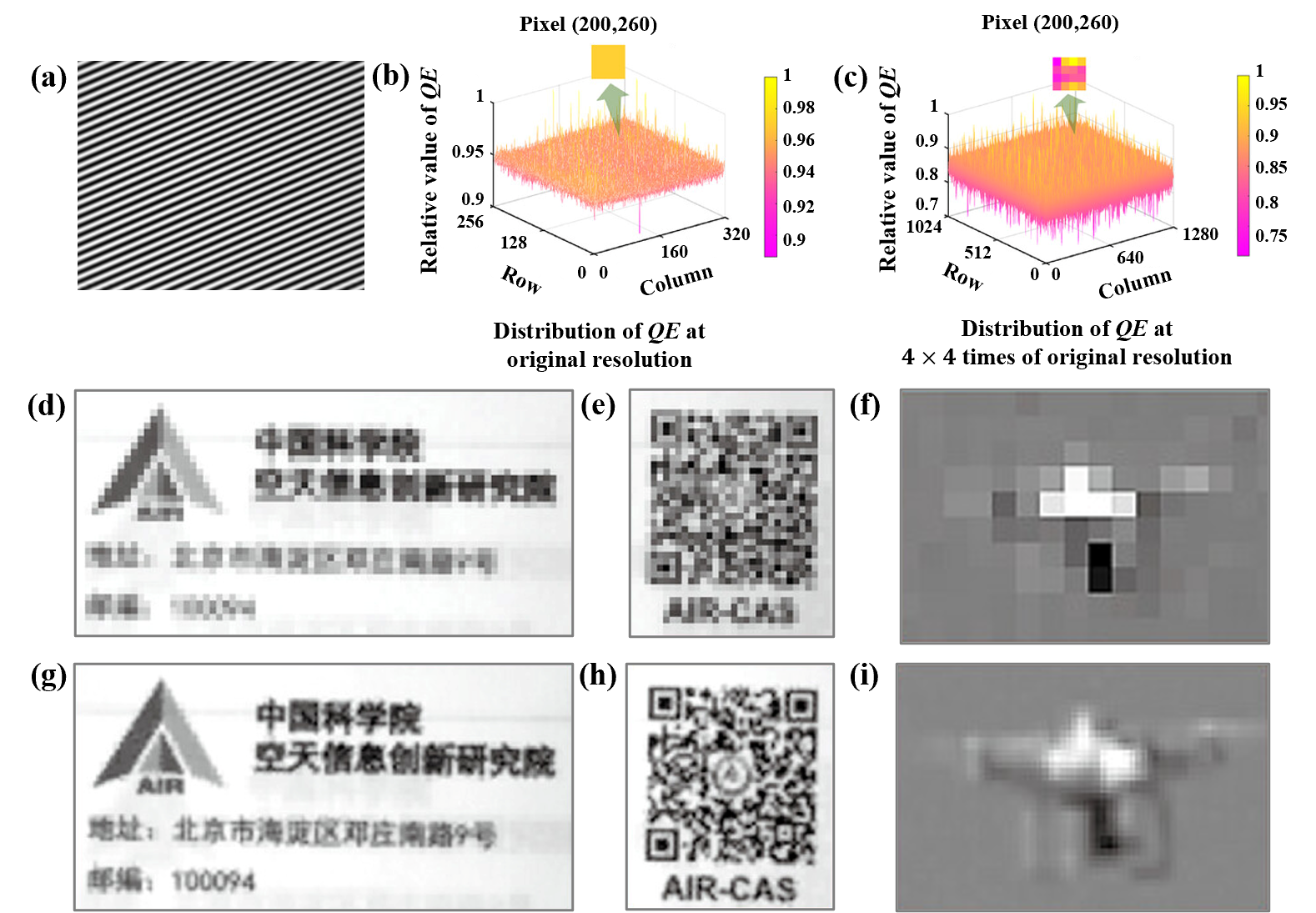}
\end{tabular}
\end{center}
\caption
{ \label{fig:example20}
 Hyper-sampling imaging technology. (a) Interference light field after eliminating the transverse wave vectors. (b) Measured pixel‑level quantum efficiency. (c) Measured sub‑pixel‑level quantum efficiency. (d-f) Original images of Chinese characters (d), QR code (e), and drone (f). (g-i) Corresponding hyper‑sampling reconstructions images of Chinese characters (g), QR code (h), and drone (i)\cite{63}.}
\end{figure}

\section{Summary and Outlook}

In this review, we have systematically examined the fundamental physics, beam construction principles, and experimental implementation methodologies of OPBs, a recently emerging class of structured light fields based on transverse wave-vector elimination. We summarized the propagation dynamics and disturbance-suppression mechanisms that enable OPBs to maintain highly confined, self-stable, and self-healing propagation over long distances—even in the presence of turbulence, scattering, or environmental fluctuations. We further reviewed the development of extended OPB families, including vortex, vector-vortex, inverted, and vortex-inverted OPBs, which greatly expand the degrees of freedom in phase, amplitude, and polarization engineering. Finally, we surveyed representative applications of OPBs in free-space and underwater optical communication, micro-particle manipulation, and hyper-sampling imaging, highlighting their growing role in scenarios where conventional Gaussian, Bessel, or Airy beams face limitations.

Optical pin beams (OPBs) are closely related to, yet fundamentally distinct from, several well-known classes of structured beams, including Bessel, Airy, and frozen beams\cite{226,227,228,229,230,231}. Bessel beams achieve diffraction resistance through conical interference, giving rise to a ring-shaped transverse momentum spectrum and pronounced sidelobes, which inevitably distribute energy away from the central core. Airy beams rely on cubic phase modulation and exhibit self-accelerating trajectories, but their asymmetric structure and extended oscillatory tails limit long-distance stability and robustness in realistic environments. Frozen beams, constructed via superpositions of Bessel modes, enable programmable axial intensity profiles\cite{231,232}; however, their localization still originates from modal interference and therefore inherits the sidelobe structure and sensitivity associated with Bessel constituents.

In contrast, OPBs are not defined by modal superposition or phase-engineered interference, but by a fundamentally different mechanism based on suppressing transverse wave-vector components in momentum space. This wave-vector elimination principle leads to a compact, pin-like intensity profile with minimal sidelobes and an intrinsically extended depth of focus. As a result, OPBs combine strong spatial confinement with exceptional propagation robustness, even in the presence of turbulence, scattering, or environmental perturbations. Unlike frozen beams, whose axial control is externally programmed, OPBs exhibit intrinsic propagation resilience tied directly to their momentum-space structure\cite{14,15,16,17,18,19,20,21,22,23,24,25,26,27,28,29,30,31},as shown in table 2.

\begin{table}[ht]
\caption{Comparison of key features of exemplary  structured beams}
\label{tab:beam_comparison}
\begin{center}
\begin{tabular}{|>{\centering\arraybackslash}m{1.6cm}|>{\centering\arraybackslash}m{2.5cm}|>{\centering\arraybackslash}m{2.2cm}|>{\centering\arraybackslash}m{2.2cm}|>{\centering\arraybackslash}m{2.8cm}|>{\centering\arraybackslash}m{2.2cm}|}
\hline
\rule[-1ex]{0pt}{3.5ex} Beam & Main mechanism & Axial profile & k-space & Key applications & Robustness \\
\hline\hline
\rule[-1ex]{0pt}{3.5ex} OPB & k-vector elimination & Elongated pin focus & Centralized & FSO comm., laser processing & High \\
\hline
\rule[-1ex]{0pt}{3.5ex} Bessel & Conical interference & Central lobe+rings & Annular ring & Optical trapping, imaging & Moderate \\
\hline
\rule[-1ex]{0pt}{3.5ex} Airy & Cubic phase FT & Bending self-healing & Asymmetric & Curved guidance, microscopy & Low \\
\hline
\rule[-1ex]{0pt}{3.5ex} Frozen & Mode superposition & Programmable pattern & Multiple rings & Multi-plane trapping & Limited \\
\hline
\end{tabular}
\end{center}
\end{table}

Looking ahead, OPBs are well positioned to play a transformative role in the next generation of global photonic infrastructure. The rapid deployment of satellite mega-constellations is driving an urgent need for high-capacity, long-distance, and turbulence-resilient free-space optical links\cite{240,241,242,243,244,258,259,260}. In this context, the intrinsic stability and disturbance-suppression capabilities of OPBs make them strong candidates for satellite–satellite, satellite–ground, and air-to-ground optical communication architectures\cite{245,246,247,248,249}. At the same time, the long depth of focus and self-healing characteristics of OPBs offer unique advantages for underwater optical networks, autonomous ocean exploration, and deep-sea communication systems, where conventional beam formats suffer from rapid distortion and scattering\cite{250,251,252,261,262,263}. Beyond communication, OPBs hold promise as versatile optical tools for non-contact optical manipulation in biphotonics, precision energy delivery and micro-processing, and compact structured-light sources integrated directly on chip\cite{253,254,255,256,257}.

Future progress will likely be driven by the convergence of OPB physics with metasurface-based beam engineering, adaptive optics stabilization, and machine-learning-assisted wavefront control, enabling real-time environmental compensation and intelligent self-optimization. Such integration may ultimately translate OPBs from laboratory demonstrations into scalable, miniaturized, and autonomously reconfigurable photonic modules deployable in real-world systems\cite{71,238,239}. We anticipate that continued advances at the intersection of structured-light science, nanophotonic device engineering, and intelligent optical control will unlock a new class of robust, high-performance, and environment-adaptive light fields, positioning OPBs as a cornerstone technology in the evolution of next-decade optical communications, sensing, and advanced photonic instrumentation.

\subsection* {Acknowledgments}
The work was supported by the Deep Earth Probe and Mineral Resources Exploration National Science and Technology Major Project (2024ZD1000700), the Key R\&D Program of Shandong Province (2024CXPT042), the National Funded Postdoctoral Researcher Program (GZC20232760), the National Key R\&D Program of China (2022YFA1404800), and the National Natural Science Foundation of China (Grant Nos. 12134006, W2541003, 62575133, 12074350, 61475161, 62105341).


\bibliography{reference}   

\begin{thebibliography}{100}

\bibitem{1}
H.~Rubinsztein-Dunlop, A.~Forbes, M.~V. Berry, {\em et~al.}, ``Roadmap on
  structured light,'' {\em Journal of Optics} {\bf 19}(1), 013001  (2016).

\bibitem{2}
N.~K. Efremidis, Z.~Chen, M.~Segev, {\em et~al.}, ``Airy beams and accelerating
  waves: an overview of recent advances,'' {\em Optica} {\bf 6}(5), 686--701
  (2019).

\bibitem{3}
Y.~Shen, X.~Wang, Z.~Xie, {\em et~al.}, ``Optical vortices 30 years on: Oam
  manipulation from topological charge to multiple singularities,'' {\em Light:
  Science \& Applications} {\bf 8}(1), 90  (2019).

\bibitem{4}
A.~Forbes, M.~De~Oliveira, and M.~R. Dennis, ``Structured light,'' {\em Nature
  Photonics} {\bf 15}(4), 253--262  (2021).

\bibitem{5}
C.~He, Y.~Shen, and A.~Forbes, ``Towards higher-dimensional structured light,''
  {\em Light: Science \& Applications} {\bf 11}(1), 205  (2022).

\bibitem{6}
A.~E. Willner, K.~Pang, H.~Song, {\em et~al.}, ``Orbital angular momentum of
  light for communications,'' {\em Applied Physics Reviews} {\bf 8}(4)  (2021).

\bibitem{7}
Z.~Wan, J.~Liu, J.~Chen, {\em et~al.}, ``Advances in structured light lasers,''
  {\em Journal of Optics} {\bf 27}(9), 093001  (2025).

\bibitem{8}
X.~Luo, ``Digital optics and optical intelligent agent,'' {\em Applied Physics
  Letters} {\bf 127}(1)  (2025).

\bibitem{9}
W.~Niewiem, S.~Figura, D.~Mergelkuhl, {\em et~al.}, ``Speckle induced by
  reflections in pipe propagation of a structured laser beams,'' {\em Applied
  Optics} {\bf 64}(19), 5433--5443  (2025).

\bibitem{10}
Z.~Lu, D.~Xu, C.~Li, {\em et~al.}, ``Bidirectional high-purity structured light
  beams transformation based on multi-plane light conversion,'' {\em Optics
  Express} {\bf 33}(4), 7155--7170  (2025).

\bibitem{11}
A.~Forbes, ``Structured light from lasers,'' {\em Laser \& Photonics Reviews}
  {\bf 13}(11), 1900140  (2019).

\bibitem{12}
O.~Korotkova, {\em Random light beams: theory and applications}, CRC press
  (2013).

\bibitem{13}
J.~Wang, K.~Li, and Z.~Quan, ``Integrated structured light manipulation,'' {\em
  Photonics Insights} {\bf 3}(3), R05--R05  (2024).

\bibitem{235}
X.~Liu, Q.~Cao, and Q.~Zhan, ``Spatiotemporal optical wavepackets: from
  concepts to applications,'' {\em Photonics Insights} {\bf 3}(4), R08--R08
  (2024).

\bibitem{236}
Q.~Jia, W.~Lyu, W.~Yan, {\em et~al.}, ``Optical manipulation: from fluid to
  solid domains,'' {\em Photonics Insights} {\bf 2}(2), R05--R05  (2023).

\bibitem{14}
C.~J. Zapata-Rodr\'{i}guez and J.~J. Miret, ``Diffraction-free beams in thin
  films,'' {\em Journal of the Optical Society of America A} {\bf 27}(3),
  663--670  (2010).

\bibitem{15}
J.~Durnin, J.~J. Miceli~Jr, and J.~H. Eberly, ``Comparison of bessel and
  gaussian beams,'' {\em Optics letters} {\bf 13}(2), 79--80  (1988).

\bibitem{16}
D.~McGloin and K.~Dholakia, ``Bessel beams: diffraction in a new light,'' {\em
  Contemporary physics} {\bf 46}(1), 15--28  (2005).

\bibitem{17}
S.~M. Saltiel, D.~N. Neshev, R.~Fischer, {\em et~al.}, ``Generation of
  second-harmonic conical waves via nonlinear bragg diffraction,'' {\em
  Physical review letters} {\bf 100}(10), 103902  (2008).

\bibitem{18}
C.~Suchand~Sandeep, A.~Khairyanto, T.~Aung, {\em et~al.}, ``Bessel beams in
  ophthalmology: a review,'' {\em Micromachines} {\bf 14}(9), 1672  (2023).

\bibitem{19}
Y.~Gao, Z.~Huang, J.~Xu, {\em et~al.}, ``Channel modeling for underwater
  scattered light communication based on gaussian and bessel beams,'' {\em
  Chinese Optics Letters} {\bf 23}(6), 060606  (2025).

\bibitem{20}
Z.~Zhai, J.~Huang, X.~Yu, {\em et~al.}, ``High uniformity bessel beams with
  angle-controllable steering,'' {\em Optics Express} {\bf 32}(19),
  33811--33829  (2024).

\bibitem{21}
A.~S. Rao, ``A conceptual review on bessel beams,'' {\em Physica Scripta} {\bf
  99}(6), 062007  (2024).

\bibitem{22}
M.~Balazs and M.~Berry, ``Nonspreading wave packets,'' {\em Am. J. Phys.} {\bf
  47}, 264--267  (1979).

\bibitem{23}
G.~A. Siviloglou, J.~Broky, A.~Dogariu, {\em et~al.}, ``Observation of
  accelerating airy beams,'' {\em Physical Review Letters} {\bf 99}(21), 213901
   (2007).

\bibitem{24}
G.~A. Siviloglou and D.~N. Christodoulides, ``Accelerating finite energy airy
  beams,'' {\em Optics letters} {\bf 32}(8), 979--981  (2007).

\bibitem{25}
H.~Wang, Y.~Liu, J.~Ma, {\em et~al.}, ``Laser coherent combination with
  circular array of airy beams,'' {\em IEEE Photonics Journal} {\bf 16}(1),
  1--5  (2023).

\bibitem{26}
Q.~Jian, J.~Hu, L.~Yan, {\em et~al.}, ``Nonlinear dynamics and self-healing
  properties of elliptical airy beams in kerr media,'' {\em Physical Review A}
  {\bf 112}(2), 023507  (2025).

\bibitem{27}
J.~Chen, H.~Gao, Z.~Liu, {\em et~al.}, ``Multi-modal particle trapping via a
  vortex-loaded dual-ring airy-gaussian beam,'' {\em Photonics Res.} {\bf 14},
  B1  (2026).

\bibitem{28}
Q.~Lin, H.~Zhang, Z.~Hu, {\em et~al.}, ``Airy transform of the new
  power-exponent-phase vortex beam,'' {\em Photonics} {\bf 10}(9), 974  (2023).
\newblock [doi:10.3390/photonics10090974].

\bibitem{29}
S.~Jia, J.~C. Vaughan, and X.~Zhuang, ``Isotropic three-dimensional
  super-resolution imaging with a self-bending point spread function,'' {\em
  Nature photonics} {\bf 8}(4), 302--306  (2014).

\bibitem{30}
J.~Ber{\v{s}}kys and S.~Orlov, ``Accelerating airy beams with particle-like
  polarization topologies and free-space bimeronic lattices,'' {\em Optics
  letters} {\bf 48}(5), 1168--1171  (2023).

\bibitem{31}
N.~Voloch-Bloch, Y.~Lereah, Y.~Lilach, {\em et~al.}, ``Generation of electron
  airy beams,'' {\em Nature} {\bf 494}(7437), 331--335  (2013).

\bibitem{237}
Z.~Zhang, Y.~Hu, J.~Zhao, {\em et~al.}, ``Research progress and application
  prospect of airy beams,'' {\em Chinese Science Bulletin} {\bf 58}(34),
  3513--3520  (2013).

\bibitem{32}
J.~Huang, J.~Mao, X.~Li, {\em et~al.}, ``Integrated optical entangled quantum
  vortex emitters,'' {\em Nature Photonics} , 1--8  (2025).

\bibitem{33}
Z.~Hu, D.~Bongiovanni, Z.~Wang, {\em et~al.}, ``Topological orbital angular
  momentum extraction and twofold protection of vortex transport,'' {\em Nature
  photonics} {\bf 19}(2), 162--169  (2025).

\bibitem{34}
H.~Lin, Y.~Liao, G.~Liu, {\em et~al.}, ``Optical vortex-antivortex
  crystallization in free space,'' {\em Nature Communications} {\bf 15}(1),
  6178  (2024).

\bibitem{35}
S.~Lei, S.~Xia, D.~Song, {\em et~al.}, ``Optical vortex ladder via sisyphus
  pumping of pseudospin,'' {\em Nature communications} {\bf 15}(1), 7693
  (2024).

\bibitem{36}
L.~Allen, M.~W. Beijersbergen, R.~Spreeuw, {\em et~al.}, ``Orbital angular
  momentum of light and the transformation of laguerre-gaussian laser modes,''
  {\em Physical review A} {\bf 45}(11), 8185  (1992).

\bibitem{37}
Y.~Oki, N.~Kidera, and M.~Maeda, ``Enhancement of sensitivity of optogalvanic
  spectroscopy in a flame by laser ionization,'' {\em Optics communications}
  {\bf 110}(1-2), 105--108  (1994).

\bibitem{38}
A.~E. Willner, ``Oam light for communications,'' {\em Optics and Photonics
  News} {\bf 32}(6), 34--41  (2021).

\bibitem{39}
M.~Padgett and R.~Bowman, ``Tweezers with a twist,'' {\em Nature photonics}
  {\bf 5}(6), 343--348  (2011).

\bibitem{40}
G.~Zheng, R.~Horstmeyer, and C.~Yang, ``Wide-field, high-resolution fourier
  ptychographic microscopy,'' {\em Nature photonics} {\bf 7}(9), 739--745
  (2013).

\bibitem{41}
L.~Chen, B.~Gao, X.~Li, {\em et~al.}, ``Optimizing airy needle-like beams for
  long-range axial manipulation and super-resolution imaging,'' {\em ACS
  Photonics} {\bf 11}(9), 3610--3620  (2024).

\bibitem{42}
J.~Ni, L.~Kong, Y.~Yan, {\em et~al.}, ``Autofocusing capabilities of
  double-ring circular airy gaussian beams and their application in particle
  manipulation,'' {\em Optics Express} {\bf 32}(25), 44908--44917  (2024).

\bibitem{43}
A.~Kumari, V.~Dev, T.~M. Hayward, {\em et~al.}, ``Generating optical vortex
  needle beams with a flat diffractive lens,'' {\em Journal of Applied Physics}
  {\bf 136}(11)  (2024).

\bibitem{44}
J.~Wang, M.~Li, X.~Han, {\em et~al.}, ``Hollow gaussian beam-based experimental
  investigation on echo measurements under atmospheric turbulence and central
  obstruction,'' {\em Optics Express} {\bf 32}(17), 30702--30712  (2024).

\bibitem{45}
D.~Liu, B.~Gao, F.~Wang, {\em et~al.}, ``Experimental realization of tunable
  finite square optical arrays,'' {\em Optics \& Laser Technology} {\bf 153},
  108220  (2022).

\bibitem{46}
M.~Wu, S.~Lin, and Y.~Chen, ``Generation of multi-focus abruptly autofocusing
  beams with adjustable focus characteristics,'' {\em Optics Express} {\bf
  30}(2), 1003--1012  (2022).

\bibitem{47}
X.~Wang, Y.~Sun, and L.~Liu, ``Characterization of isotropic laser cooling for
  application in quantum sensing,'' {\em Optics Express} {\bf 29}(26),
  43435--43444  (2021).

\bibitem{48}
Q.~Qin, H.~Jiang, L.~Xu, {\em et~al.}, ``Generating multi-focus circular beams
  via all-dielectric metasurfaces,'' {\em Physica Scripta} {\bf 100}(8), 085525
   (2025).

\bibitem{49}
M.~A. Cox, N.~Mphuthi, I.~Nape, {\em et~al.}, ``Structured light in
  turbulence,'' {\em IEEE Journal of Selected Topics in Quantum Electronics}
  {\bf 27}(2), 1--21  (2020).

\bibitem{50}
C.~Peters, V.~Cocotos, and A.~Forbes, ``Structured light in atmospheric
  turbulence---a guide to its digital implementation: tutorial,'' {\em Advances
  in Optics and Photonics} {\bf 17}(1), 113--184  (2025).

\bibitem{51}
I.~Nape, K.~Singh, A.~Klug, {\em et~al.}, ``Revealing the invariance of
  vectorial structured light in complex media,'' {\em Nature Photonics} {\bf
  16}(7), 538--546  (2022).

\bibitem{52}
A.~V. Degtyarev, M.~M. Dubinin, V.~A. Maslov, {\em et~al.}, ``Propagation of an
  azimuthally polarized terahertz laser beams with a phase singularity,'' {\em
  East European Journal of Physics} (4), 267--273  (2025).

\bibitem{53}
A.~V. Degtyarev, M.~M. Dubinin, V.~Maslov, {\em et~al.}, ``Spatial dynamics of
  a radially polarized terahertz laser beam with a phase singularity,'' {\em
  East European Journal of Physics} (3), 93--102  (2025).

\bibitem{54}
P.~Liu, Y.~Zhao, N.~Li, {\em et~al.}, ``Deep neural networks with adaptive
  solution space for inverse design of multilayer deep-etched grating,'' {\em
  Optics and Lasers in Engineering} {\bf 174}, 107933  (2024).

\bibitem{55}
Z.~Zhang, X.~Liang, M.~Goutsoulas, {\em et~al.}, ``Robust propagation of
  pin-like optical beam through atmospheric turbulence,'' {\em Apl Photonics}
  {\bf 4}(7)  (2019).

\bibitem{56}
D.~Bongiovanni, D.~Li, M.~Goutsoulas, {\em et~al.}, ``Free-space realization of
  tunable pin-like optical vortex beams,'' {\em Photonics Research} {\bf 9}(7),
  1204--1212  (2021).

\bibitem{67}
D.~Li, D.~Bongiovanni, M.~Goutsoulas, {\em et~al.}, ``Direct comparison of
  anti-diffracting optical pin beams and abruptly autofocusing beams,'' {\em
  OSA Continuum} {\bf 3}(6), 1525--1535  (2020).

\bibitem{68}
Q.~Guo, Q.~Kang, X.~Zhang, {\em et~al.}, ``Non-paraxial pin-like optical beams:
  geometric design and pulling forces,'' {\em Optics Express} {\bf 33}(19),
  39522--39532  (2025).

\bibitem{69}
K.~Y. Bliokh, E.~Karimi, M.~J. Padgett, {\em et~al.}, ``Roadmap on structured
  waves,'' {\em Journal of Optics} {\bf 25}(10), 103001  (2023).

\bibitem{57}
Z.~Cai, Z.~Quan, L.~Yuan, {\em et~al.}, ``Generation of subwavelength inverted
  pin beam via fiber end integrated plasma structure,'' {\em Advanced Photonics
  Nexus} {\bf 3}(2), 026003--026003  (2024).

\bibitem{58}
S.~Droulias, M.~Loulakis, D.~G. Papazoglou, {\em et~al.}, ``Inverted pin beams
  for robust long-range propagation through atmospheric turbulence,'' {\em
  Optics Letters} {\bf 48}(21), 5467--5470  (2023).

\bibitem{59}
S.~Droulias, M.~Loulakis, D.~G. Papazoglou, {\em et~al.}, ``Vortex inverted pin
  beams: mitigation of scintillations in strong atmospheric turbulence,'' {\em
  Optics Letters} {\bf 49}(17), 4811--4814  (2024).

\bibitem{60}
J.~Li, H.~Liang, G.~Wu, {\em et~al.}, ``Generation of vector vortex pin-like
  beams and their propagation in turbulent atmosphere,'' {\em APL Photonics}
  {\bf 10}(1)  (2025).

\bibitem{61}
M.~Chen, P.~Wu, Y.~Zeng, {\em et~al.}, ``Trapping dielectric rayleigh particles
  with tightly focused pin-like vortex beam,'' {\em The European Physical
  Journal D} {\bf 76}(2), 20  (2022).

\bibitem{62}
N.~Hu, H.~Zhou, R.~Zhang, {\em et~al.}, ``Experimental demonstration of a
  "pin-like" low-divergence beam in a 1-gbit/s ook fso link using a
  limited-size receiver aperture at various propagation distances,'' {\em
  Optics Letters} {\bf 47}(16), 4215--4218  (2022).

\bibitem{63}
H.~Xue, M.~Shang, Z.~Zhang, {\em et~al.}, ``Hyper-sampling imaging by
  measurement of intra-pixel quantum efficiency using steady wave field,'' {\em
  Laser \& Photonics Reviews} {\bf 19}(4), 2401306  (2025).

\bibitem{64}
H.~Lu, H.~Wang, Y.~Li, {\em et~al.}, ``Research on the horizontal 1 km
  experimental performance based on the optical pin beams,'' in {\em Advanced
  Fiber Laser Conference (AFL 2024)},  G.~Chang and Y.~Feng, Eds., {\em Proc.
  SPIE} {\bf 13544}, 135440D  (2025).
\newblock [doi:10.1117/12.3057636].

\bibitem{65}
X.~Han, W.~Wang, P.~Li, {\em et~al.}, ``Demonstration of anti-diffracting
  optical pin-like beam enabled 5gbit/s ofdm underwater wireless optical
  communication system,'' {\em Optics Communications} {\bf 579}, 131582
  (2025).

\bibitem{66}
M.~Guan, Y.~Liu, H.~Wang, {\em et~al.}, ``High-performance 100 gbps free-space
  optical communication via optical pin beam receiver,'' {\em Communications
  Engineering} {\bf 4}(1), 203  (2025).

\bibitem{70}
C.~C. Foy, J.~Minch, V.~Scalesse, {\em et~al.}, ``Algorithms and principles for
  laser control and tuning for laser communications,'' in {\em Free-Space Laser
  Communications XXXVII},  H.~Hemmati and B.~S. Robinson, Eds., {\em Proc.
  SPIE} {\bf 13355}, 133550H  (2025).
\newblock [doi:10.1117/12.3047903].

\bibitem{71}
Z.~Chen and M.~Segev, ``Highlighting photonics: looking into the next decade,''
  {\em ELight} {\bf 1}(1), 2  (2021).

\bibitem{72}
L.~Zhang, Z.~Xu, T.~Pu, {\em et~al.}, ``Change in the state of polarization of
  gaussian schell-model beam propagating through non-kolmogorov turbulence,''
  {\em Results in physics} {\bf 7}, 4332--4336  (2017).

\bibitem{73}
L.~Cui, ``Effects of anisotropic turbulence on the long term beam spread and
  beam wander of gaussian beam,'' {\em Optik} {\bf 163}, 152--158  (2018).

\bibitem{74}
V.~Dev, A.~N.~K. Reddy, A.~V. Ustinov, {\em et~al.}, ``Autofocusing and
  self-healing properties of aberration laser beams in a turbulent media,''
  {\em Physical Review Applied} {\bf 16}(1), 014061  (2021).

\bibitem{75}
X.~Chen, Y.~Yuan, B.~Yan, {\em et~al.}, ``Propagation and self-healing
  properties of lommel-gaussian beam through atmospheric turbulence,'' {\em
  Optoelectronics Letters} {\bf 17}(9), 572--576  (2021).

\bibitem{76}
M.~Goutsoulas, D.~Bongiovanni, D.~Li, {\em et~al.}, ``Tunable self-similar
  bessel-like beams of arbitrary order,'' {\em Optics Letters} {\bf 45}(7),
  1830--1833  (2020).

\bibitem{77}
J.~Wang, Y.~Zhao, and B.~Geng, ``Impact of atmospheric turbulence on the signal
  intensity of the laser ranging echo of space debris,'' {\em Applied Optics}
  {\bf 63}(36), 9268--9276  (2024).

\bibitem{78}
L.~A. Hall, M.~A. Romer, B.~L. Turo, {\em et~al.}, ``Observation of kilometer
  propagation of space-time wavepackets,'' in {\em 2023 Conference on Lasers
  and Electro-Optics (CLEO)},  1--2  (2023).

\bibitem{79}
L.~A. Hall, M.~Yessenov, M.~A. Romer, {\em et~al.}, ``Long-distance axial
  spectral encoding using space-time wave packets,'' {\em Optics Letters} {\bf
  50}(15), 4698--4701  (2025).

\bibitem{80}
L.~Lu, Z.~Wang, Y.~Huang, {\em et~al.}, ``Utilizing ellipticity to control the
  focus and propagation of an agsm beam,'' {\em Optics Express} {\bf 33}(18),
  37429--37439  (2025).

\bibitem{81}
Z.~Mi, Z.~Zhao, R.~Wei, {\em et~al.}, ``Propagation dynamics of a controllable
  auto-focusing annular tricomi-gaussian beam array,'' {\em Optics
  Communications} , 132104  (2025).

\bibitem{82}
S.~Wang, J.~Wang, M.~Cheng, {\em et~al.}, ``Three transmission properties of
  the perfect vortex beam,'' {\em Optics express} {\bf 32}(19), 34141--34152
  (2024).

\bibitem{83}
R.~Hettel, ``Beam stability at light sources,'' {\em Review of Scientific
  Instruments} {\bf 73}(3), 1396--1401  (2002).

\bibitem{84}
L.~Zhu, A.~Wang, M.~Deng, {\em et~al.}, ``Free-space optical communication with
  quasi-ring airy vortex beam under limited-size receiving aperture and
  atmospheric turbulence,'' {\em Optics Express} {\bf 29}(20), 32580--32590
  (2021).

\bibitem{85}
X.~Ding, X.~Zhao, T.~Wang, {\em et~al.}, ``Research on adaptive receiving
  technology for atmospheric laser communication,'' {\em Optical Engineering}
  {\bf 64}(6), 068102--068102  (2025).

\bibitem{86}
P.~Zhang, J.~Jie, Z.~Liu, {\em et~al.}, ``Chaotic image encryption system as a
  proactive scheme for image transmission in fso high-altitude platform,'' in
  {\em Photonics},   {\bf 12}(7), 635, MDPI  (2025).

\bibitem{87}
A.~E. Willner, H.~Zhou, X.~Su, {\em et~al.}, ``Utilizing structured modal beams
  in free-space optical communications for performance enhancement,'' {\em IEEE
  Journal of Selected Topics in Quantum Electronics} {\bf 29}(6: Photonic
  Signal Processing), 1--13  (2023).

\bibitem{88}
R.~Zhang, X.~Su, H.~Song, {\em et~al.}, ``Automatic turbulence resilience in
  pilot-assisted self-coherent free-space optical communications,'' {\em
  Journal of Lightwave Technology} {\bf 42}(10), 3760--3769  (2024).

\bibitem{89}
J.~Guo, L.~Sun, J.~Liu, {\em et~al.}, ``Beam wander restrained by nonlinearity
  of femtosecond laser filament in air,'' {\em Sensors} {\bf 22}(13), 4995
  (2022).

\bibitem{90}
J.~Wang, Y.~Tan, H.~Li, {\em et~al.}, ``Hollow optical pin-like beam based
  cassegrain system performance enhancement under atmospheric turbulence,''
  {\em Optics Letters} {\bf 50}(6), 2089--2092  (2025).

\bibitem{91}
Z.~Zhong, X.~Zhang, B.~Zhang, {\em et~al.}, ``Mitigation of atmospheric
  turbulence effect by light beams carrying self-rotating wavefront,'' {\em
  Optics Express} {\bf 30}(14), 24421--24430  (2022).

\bibitem{92}
S.~Wang, M.~Cheng, X.~Yang, {\em et~al.}, ``Self-focusing effect analysis of a
  perfect optical vortex beam in atmospheric turbulence,'' {\em Optics Express}
  {\bf 31}(13), 20861--20871  (2023).

\bibitem{93}
B.~Yan, D.~Li, L.~Zhang, {\em et~al.}, ``Filamentation of femtosecond vortex
  laser pulses in turbulent air,'' {\em Optics \& Laser Technology} {\bf 164},
  109515  (2023).

\bibitem{94}
P.~Ju, W.~Fan, W.~Gao, {\em et~al.}, ``Atmospheric turbulence effects on the
  performance of orbital angular momentum multiplexed free-space optical links
  using coherent beam combining,'' {\em Photonics} {\bf 10}(6), 634  (2023).
\newblock [doi:10.3390/photonics10060634].

\bibitem{95}
A.~D. Bulygin, Y.~E. Geints, and I.~Y. Geints, ``Vortex beam in a turbulent
  kerr medium for atmospheric communication,'' {\em Photonics} {\bf 10}(7), 856
   (2023).
\newblock [doi:10.3390/photonics10070856].

\bibitem{96}
Z.~Zhu, M.~Janasik, A.~Fyffe, {\em et~al.}, ``Compensation-free
  high-dimensional free-space optical communication using turbulence-resilient
  vector beams,'' {\em Nature communications} {\bf 12}(1), 1666  (2021).

\bibitem{97}
J.~Zhang, M.~Wang, L.~Diao, {\em et~al.}, ``Experimental study on the
  scintillation index of vortex beam superposition states perturbed by linear
  array acoustic sources in atmospheric environments,'' {\em Photonics} {\bf
  12}(11)  (2025).

\bibitem{98}
M.~Hart, N.~Milton, C.~Baranec, {\em et~al.}, ``A ground-layer adaptive optics
  system with multiple laser guide stars,'' {\em Nature} {\bf 466}(7307),
  727--729  (2010).

\bibitem{99}
F.~Yang, A.~B. Kostinski, Z.~Zhu, {\em et~al.}, ``A single-photon lidar
  observes atmospheric clouds at decimeter scales: resolving droplet activation
  within cloud base,'' {\em npj Climate and Atmospheric Science} {\bf 7}(1), 92
   (2024).

\bibitem{100}
T.~de~Conto, J.~Armston, and R.~Dubayah, ``Characterizing the structural
  complexity of the earth's forests with spaceborne lidar,'' {\em Nature
  Communications} {\bf 15}(1), 8116  (2024).

\bibitem{101}
X.~Zhang, K.~Kwon, J.~Henriksson, {\em et~al.}, ``A large-scale
  microelectromechanical-systems-based silicon photonics lidar,'' {\em Nature}
  {\bf 603}(7900), 253--258  (2022).

\bibitem{102}
R.~Tobin, A.~Halimi, A.~McCarthy, {\em et~al.}, ``Robust real-time 3d imaging
  of moving scenes through atmospheric obscurant using single-photon lidar,''
  {\em Scientific reports} {\bf 11}(1), 11236  (2021).

\bibitem{103}
Z.~Li, Y.~Han, L.~Wu, {\em et~al.}, ``Towards an ultrafast 3d imaging scanning
  lidar system: a review,'' {\em Photonics Research} {\bf 12}(8), 1709--1729
  (2024).

\bibitem{104}
Z.~Tian, M.~Zhao, D.~Yang, {\em et~al.}, ``Optical remote imaging via fourier
  ptychography,'' {\em Photonics Research} {\bf 11}(12), 2072--2083  (2023).

\bibitem{105}
T.~Wen, J.~Hu, Y.~Zhu, {\em et~al.}, ``Multi-axis laser interferometer not
  affected by installation errors based on nonlinear computation,'' {\em
  Applied Sciences} {\bf 13}(19), 10887  (2023).

\bibitem{106}
K.~Wendt, M.~Franke, and F.~H{\"a}rtig, ``Measuring large 3d structures using
  four portable tracking laser interferometers,'' {\em Measurement} {\bf
  45}(10), 2339--2345  (2022).

\bibitem{107}
J.~He, H.~Xie, H.~Liu, {\em et~al.}, ``Multi-baseline bistatic sar
  three-dimensional imaging method based on phase error calibration combining
  pga and eb-isoa,'' {\em Remote Sensing} {\bf 17}(3), 363  (2025).

\bibitem{108}
H.~Meng, G.~Pan, Y.~Pu, {\em et~al.}, ``Holographic particle image velocimetry:
  from film to digital recording,'' {\em Measurement Science and Technology}
  {\bf 15}(4), 673  (2004).

\bibitem{109}
L.~Jiang, S.~Zheng, Q.~Yang, {\em et~al.}, ``A modified omp method for
  multi-orbit three dimensional isar imaging of the space target,'' {\em
  Journal of Systems Engineering and Electronics} {\bf 34}(4), 879--893
  (2023).

\bibitem{110}
J.~Guo, Z.~Zhang, N.~Zhang, {\em et~al.}, ``Sub-ppb nacl aerosol detection at a
  distance of 30 meters by femtosecond laser induced plasma spectroscopy,''
  {\em Optics Express} {\bf 31}(17), 28586--28595  (2023).

\bibitem{111}
S.~Singh, V.~Dev, and V.~Pal, ``Generating asymmetric aberration laser beams
  with controlled intensity distribution,'' {\em Journal of Optics} {\bf
  24}(12), 125601  (2022).

\bibitem{112}
X.~Li, S.~A. Ponomarenko, Z.~Xu, {\em et~al.}, ``Universal self-similar
  asymptotic behavior of optical bump spreading in random medium atop
  incoherent background,'' {\em Optics Letters} {\bf 45}(3), 698--701  (2020).

\bibitem{113}
S.~Jia, J.~C. Vaughan, and X.~Zhuang, ``Isotropic 3d super resolution imaging
  with self-bending point spread function,'' {\em Biophysical Journal} {\bf
  104}(2), 668a  (2013).

\bibitem{114}
G.~Satat, M.~Tancik, and R.~Raskar, ``Lensless imaging with compressive
  ultrafast sensing,'' {\em IEEE Transactions on Computational Imaging} {\bf
  3}(3), 398--407  (2017).

\bibitem{115}
J.~Shen, S.~Wang, K.~Qi, {\em et~al.}, ``The suppression effect of an imaging
  system on the geometric tilt-to-length coupling in a test mass
  interferometer,'' {\em Photonics} {\bf 11}(7)  (2024).

\bibitem{116}
L.~Lai, P.~Dong, H.~Liu, {\em et~al.}, ``Experimental demonstration of constant
  amplitude modulation heterodyne interferometry,'' {\em Optics Letters} {\bf
  49}(11), 2873--2876  (2024).

\bibitem{117}
G.~Hechenblaikner, ``Common mode noise rejection properties of amplitude and
  phase noise in a heterodyne interferometer,'' {\em Journal of the Optical
  Society of America A} {\bf 30}(5), 941--947  (2013).

\bibitem{118}
F.~J. Galdieri, T.~Sutili, N.~Melnikoff, {\em et~al.}, ``Influence of exterior
  acoustic noise on narrow linewidth laser measurements using self-homodyne
  optical fiber interferometer,'' {\em Optik} {\bf 204}, 164101  (2020).

\bibitem{119}
M.~M. Colavita, ``Adverse effects in dual-feed interferometry,'' {\em New
  Astronomy Reviews} {\bf 53}(11-12), 344--352  (2009).

\bibitem{120}
W.~Yang, S.~Shan, M.~Liu, {\em et~al.}, ``Image information transmission based
  on self-rotating beam arrays encoding/decoding,'' {\em Optics Express} {\bf
  33}(13), 28808--28818  (2025).

\bibitem{121}
Z.~Chen, M.~Segev, and D.~N. Christodoulides, ``Optical spatial solitons:
  historical overview and recent advances,'' {\em Reports on Progress in
  Physics} {\bf 75}(8), 086401  (2012).

\bibitem{122}
Y.~S. Kivshar and G.~P. Agrawal, {\em Optical solitons: from fibers to photonic
  crystals}, Academic press  (2003).

\bibitem{123}
J.~C. Bronski, M.~Segev, and M.~I. Weinstein, ``Mathematical frontiers in
  optical solitons,'' {\em Proceedings of the National Academy of Sciences}
  {\bf 98}(23), 12872--12873  (2001).

\bibitem{124}
R.~Bullough, P.~Jack, P.~Kitchenside, {\em et~al.}, ``Solitons in laser
  physics,'' {\em Physica Scripta} {\bf 20}(3-4), 364  (1979).

\bibitem{125}
Y.~Song, X.~Shi, C.~Wu, {\em et~al.}, ``Recent progress of study on optical
  solitons in fiber lasers,'' {\em Applied physics reviews} {\bf 6}(2)  (2019).

\bibitem{126}
A.~Hasegawa, ``Optical solitons in fibers,'' in {\em Optical solitons in
  fibers},  1--74, Springer  (1989).

\bibitem{127}
A.~V. Buryak, P.~Di~Trapani, D.~V. Skryabin, {\em et~al.}, ``Optical solitons
  due to quadratic nonlinearities: from basic physics to futuristic
  applications,'' {\em Physics Reports} {\bf 370}(2), 63--235  (2002).

\bibitem{128}
R.~Davies and M.~Kasper, ``Adaptive optics for astronomy,'' {\em Annual Review
  of Astronomy and Astrophysics} {\bf 50}(1), 305--351  (2012).

\bibitem{129}
M.~J. Booth, ``Adaptive optical microscopy: the ongoing quest for a perfect
  image,'' {\em Light: Science \& Applications} {\bf 3}(4), e165--e165  (2014).

\bibitem{130}
Y.~Guo, Y.~Hao, S.~Wan, {\em et~al.}, ``Direct observation of atmospheric
  turbulence with a video-rate wide-field wavefront sensor,'' {\em Nature
  Photonics} {\bf 18}(9), 935--943  (2024).

\bibitem{131}
P.~Cameron, B.~Courme, D.~Faccio, {\em et~al.}, ``Quantum-assisted adaptive
  optics for microscopy,'' {\em Science} {\bf 383}(6687), adk7825  (2024).

\bibitem{132}
I.~A. Alimi and P.~P. Monteiro, ``Revolutionizing free-space optics: A survey
  of enabling technologies, challenges, trends, and prospects of beyond 5g
  free-space optical (fso) communication systems,'' {\em Sensors (Basel,
  Switzerland)} {\bf 24}(24), 8036  (2024).

\bibitem{133}
Z.~Zhang, S.~Jia, W.~Shao, {\em et~al.}, ``Turbulence compensation based on
  pix2pixgan for the free-space optical communication of orbital angular
  momentum multiplexing,'' {\em Applied Optics} {\bf 64}(5), A1--A11  (2024).

\bibitem{134}
F.~Tarhouni, R.~Wang, and M.-S. Alouini, ``Free space optical mesh networks: A
  survey,'' {\em IEEE Open Journal of the Communications Society}   (2025).

\bibitem{135}
P.~Parvizi, R.~Zou, C.~Bellinger, {\em et~al.}, ``Reinforcement learning
  environment for wavefront sensorless adaptive optics in single-mode fiber
  coupled optical satellite communications downlinks,'' {\em Photonics} {\bf
  10}(12)  (2023).

\bibitem{136}
W.~Wen, ``Quantitative analysis of the effect of atmospheric turbulence on a
  bessel–gaussian beam,'' {\em Photonics} {\bf 10}(8)  (2023).

\bibitem{137}
A.~E.~G. Madsen and J.~Gl{\"u}ckstad, ``Axial holotile: Extended depth-of-focus
  of dynamic holographic light projections,'' {\em Optics Communications} {\bf
  577}, 131441  (2025).

\bibitem{138}
Z.~Sun, J.~Wang, Z.~Li, {\em et~al.}, ``Stable propagation of ince--gaussian
  vector beams through atmospheric turbulence,'' {\em Optics Communications}
  {\bf 515}, 128193  (2022).

\bibitem{139}
Z.~Wang, Z.~Hu, Y.~Zhang, {\em et~al.}, ``Dressed airy vortex beam in a hot
  atomic medium,'' {\em Annalen der Physik} {\bf 535}(5), 2200652  (2023).

\bibitem{140}
Z.-Y. Hu, Z.-N. Tian, H.~Fan, {\em et~al.}, ``Long focusing range and
  self-healing bessel vortex beam generator,'' {\em Optics letters} {\bf
  45}(9), 2580--2583  (2020).

\bibitem{141}
Y.~Zhang, H.~Chang, X.~Chen, {\em et~al.}, ``Design and generation of
  structured array beams with shape-invariant properties,'' {\em New Journal of
  Physics} {\bf 25}(5), 053029  (2023).

\bibitem{142}
K.~Niu, S.~Zhao, Y.~Liu, {\em et~al.}, ``Self-rotating beam in the free space
  propagation,'' {\em Optics Express} {\bf 30}(4), 5465--5472  (2022).

\bibitem{143}
X.~Weng, Q.~Song, X.~Li, {\em et~al.}, ``Free-space creation of ultralong
  anti-diffracting beam with multiple energy oscillations adjusted using
  optical pen,'' {\em Nature Communications} {\bf 9}(1), 5035  (2018).

\bibitem{144}
V.~P. Lukin and I.~Lukin, ``Overview of modern technologies for measuring,
  predicting and correcting turbulent distortions in optical waves,'' {\em
  Computer Optics} {\bf 48}(1), 68--80  (2024).

\bibitem{145}
Y.~Zhu, L.~Zhu, A.~Wang, {\em et~al.}, ``Generation of oam beam with arbitrary
  trajectory using a single phase-only element,'' in {\em Conference on Lasers
  and Electro-Optics},  {\em Proc. {CLEO}}, STh4N.6  (2022).
\newblock [doi:10.1364/CLEO\_SI.2022.STh4N.6].

\bibitem{146}
X.~Kang, X.~Yang, J.~Ma, {\em et~al.}, ``Steady optical beam propagating
  through turbulent environment,'' {\em Optics Express} {\bf 30}(6),
  10063--10070  (2022).

\bibitem{147}
Y.~Xiuting, R.~Yuhang, Z.~Ze, {\em et~al.}, ``Mechanism of suppressing
  transverse wavevectors for antidiffracting propagation via coherent beam
  combination,''

\bibitem{148}
S.~Kumar and N.~Sharma, ``Emerging military applications of free space optical
  communication technology: A detailed review,'' {\em Journal of Physics:
  Conference Series} {\bf 2161}, 012011  (2022).

\bibitem{149}
W.~Jiang, M.~Cheng, L.~Guo, {\em et~al.}, ``Machine learning assisted speckle
  and oam spectrum analysis for enhanced turbulence characterization,'' {\em
  Photonics Research} {\bf 13}(10), B29--B37  (2025).

\bibitem{150}
Z.~H. Tantawy, M.~B. El~Mashade, A.~A. Emran, {\em et~al.}, ``On the
  performance of fso communication system with wdm and mimo structure under
  different turbulent atmospheric conditions,'' {\em Journal of Optical
  Communications} {\bf 45}(s1), s2133--s2149  (2025).

\bibitem{151}
G.~Grechko, A.~Gurvich, V.~Kan, {\em et~al.}, ``Anisotropy of spatial
  structures in the middle atmosphere,'' {\em Advances in Space Research} {\bf
  12}(10), 169--175  (1992).

\bibitem{152}
F.~D. Eaton and G.~D. Nastrom, ``Preliminary estimates of the vertical profiles
  of inner and outer scales from white sands missile range, new mexico, vhf
  radar observations,'' {\em Radio Science} {\bf 33}(4), 895--903  (1998).

\bibitem{153}
C.~Robert, J.-M. Conan, V.~Michau, {\em et~al.}, ``Retrieving parameters of the
  anisotropic refractive index fluctuations spectrum in the stratosphere from
  balloon-borne observations of stellar scintillation,'' {\em Journal of the
  Optical Society of America A} {\bf 25}(2), 379--393  (2008).

\bibitem{154}
R.~Larsson, M.~Karlsson, and P.~A. Andrekson, ``Sensitive optical free-space
  receiver architecture for coherent combining of deep-space communication
  signals through atmospheric turbulence,'' {\em Optics Express} {\bf 32}(25),
  44799--44815  (2024).

\bibitem{155}
A.~I. Martinez, G.~Cavicchioli, S.~Seyedinnavadeh, {\em et~al.},
  ``Self-adaptive integrated photonic receiver for turbulence compensation in
  free space optical links,'' {\em Scientific Reports} {\bf 14}(1), 20178
  (2024).

\bibitem{156}
W.~Ma, C.~Yu, F.~Yang, {\em et~al.}, ``High-sensitivity few-mode heterodyne
  receiver with a few-mode optical fiber amplifier for turbulence resistance in
  free space optical communication,'' {\em Optics Communications} {\bf 554},
  130126  (2024).

\bibitem{157}
A.~Wang, H.~Cao, L.~Zhu, {\em et~al.}, ``Multimode fiber based multi-oam mode
  group receiver for turbulence-resistant free-space optical communications,''
  {\em Optics Express} {\bf 32}(25), 43880--43889  (2024).

\bibitem{158}
S.-M. Song, H.-C. Lim, M.~Choi, {\em et~al.}, ``Analysis of tip/tilt
  compensation of beam wandering for space laser communication,'' {\em Journal
  of Astronomy and Space Sciences} {\bf 40}(4), 237--245  (2023).

\bibitem{159}
A.~A. Farid and S.~Hranilovic, ``Outage capacity optimization for free-space
  optical links with pointing errors,'' {\em Journal of Lightwave technology}
  {\bf 25}(7), 1702--1710  (2007).

\bibitem{160}
R.~Barrios, ``Fading loss for earth-to-space lasercom affected by scintillation
  and beam wander composite channel,'' {\em Optical Engineering} {\bf 59}(5),
  056103--056103  (2020).

\bibitem{161}
L.~Anddrews and R.~Phillips, ``Laser beam propagation through random media
  (bellingham, washington,''  (2005).

\bibitem{162}
H.~Rubinsztein-Dunlop, A.~Forbes, M.~V. Berry, {\em et~al.}, ``Roadmap on
  structured light,'' {\em Journal of Optics} {\bf 19}(1), 013001  (2016).

\bibitem{163}
A.~Belafhal, F.~Khannous, and T.~Usman, ``Closed-forms of integral transforms
  in terms of generalized lauricella hypergeometric series: A. belafhal et
  al.,'' {\em The Ramanujan Journal} {\bf 68}(2), 34  (2025).

\bibitem{164}
M.~P. Lavery, C.~Peuntinger, K.~G{\"u}nthner, {\em et~al.}, ``Free-space
  propagation of high-dimensional structured optical fields in an urban
  environment,'' {\em Science advances} {\bf 3}(10), e1700552  (2017).

\bibitem{165}
O.~Korotkova, ``Light propagation in a turbulent ocean,'' in {\em Progress in
  Optics},   {\bf 64}, 1--43, Elsevier  (2019).

\bibitem{166}
Z.~Chen, U.~Daly, A.~Boldin, {\em et~al.}, ``Weather sensing with structured
  light,'' {\em Communications Physics} {\bf 8}(1), 105  (2025).

\bibitem{167}
A.~J. Vallance, A.~Boldin, U.~J. Daly, {\em et~al.}, ``Structured light probe
  for turbulent media,''  (2025).

\bibitem{168}
X.~Yang, {\em Research on the Mechanism and Method of Stable Light Field
  Transmission in Disordered Media}.
\newblock PhD thesis, University of Chinese Academy of Sciences  (2021).

\bibitem{169}
J.~Broky, G.~A. Siviloglou, A.~Dogariu, {\em et~al.}, ``Self-healing properties
  of optical airy beams,'' {\em Optics express} {\bf 16}(17), 12880--12891
  (2008).

\bibitem{170}
H.~Xiao, H.~Wang, Y.~Liu, {\em et~al.}, ``Auto-recovery property of opb under
  obstacles and inclement weather,'' {\em APL Photonics} {\bf 10}(4)  (2025).

\bibitem{171}
X.~Zhong, X.~Kang, Y.~Liu, {\em et~al.}, ``Optimization design of steady
  optical pin beam using genetic algorithm,'' {\em Optics and Lasers in
  Engineering} {\bf 168}, 107680  (2023).

\bibitem{172}
D.~Hou, J.~Chen, and G.~Guo, ``Analysis and experimental demonstration of
  underwater frequency transfer with diode green laser,'' {\em Review of
  Scientific Instruments} {\bf 91}(7)  (2020).

\bibitem{173}
B.~Cochenour, K.~Dunn, A.~Laux, {\em et~al.}, ``Experimental measurements of
  the magnitude and phase response of high-frequency modulated light
  underwater,'' {\em Applied optics} {\bf 56}(14), 4019--4014  (2017).

\bibitem{174}
C.~T. Geldard, J.~Thompson, and W.~O. Popoola, ``Empirical study of the
  underwater turbulence effect on non-coherent light,'' {\em IEEE Photonics
  Technology Letters} {\bf 32}(20), 1307--1310  (2020).

\bibitem{175}
Y.~Yang, X.~Kang, and L.~Cao, ``Robust propagation of a steady optical beam
  through turbulence with extended depth of focus based on spatial light
  modulator,'' {\em Journal of Physics: Photonics} {\bf 5}(3), 035002  (2023).

\bibitem{176}
X.~Yang, Z.~Zhang, Y.~Ren, {\em et~al.}, ``Propagation of optical pin beams
  through water turbulence,'' in {\em International Conference on
  Optoelectronic and Microelectronic Technology and Application},  J.~Liu, Ed.,
  {\em Proc. SPIE} {\bf 11617}, 116170N  (2020).
\newblock [doi:10.1117/12.2584685].

\bibitem{177}
J.~A. Davis, D.~M. Cottrell, J.~Campos, {\em et~al.}, ``Encoding amplitude
  information onto phase-only filters,'' {\em Applied optics} {\bf 38}(23),
  5004--5013  (1999).

\bibitem{178}
C.-K. Hsueh and A.~A. Sawchuk, ``Computer-generated double-phase holograms,''
  {\em Applied optics} {\bf 17}(24), 3874--3883  (1978).

\bibitem{179}
D.~Abdollahpour, S.~Suntsov, D.~G. Papazoglou, {\em et~al.}, ``Spatiotemporal
  airy light bullets in the linear and nonlinear regimes,'' {\em Physical
  review letters} {\bf 105}(25), 253901  (2010).

\bibitem{180}
T.~Li, D.~Li, X.~Zhang, {\em et~al.}, ``Partially coherent radially polarized
  circular airy beam,'' {\em Optics Letters} {\bf 45}(16), 4547--4550  (2020).

\bibitem{181}
Z.~Zhang, P.~Zhang, M.~Mills, {\em et~al.}, ``Trapping aerosols with optical
  bottle arrays generated through a superposition of multiple airy beams,''
  {\em Chinese Optics Letters} {\bf 11}(3), 033502  (2013).

\bibitem{182}
S.~Miller, ``Light propagation in generalized lens-like media,'' {\em The Bell
  System Technical Journal} {\bf 44}(9), 2017--2064  (1965).

\bibitem{183}
C.~Liu, X.~Zhou, W.~Wang, {\em et~al.}, ``Robust transmission of pin-like
  vortex beams in plasma sheath turbulence,'' {\em Applied Optics} {\bf
  64}(24), 7076--7082  (2025).

\bibitem{184}
J.~Cao, L.~Han, H.~Liang, {\em et~al.}, ``Orbital angular momentum spectrum of
  pin-like optical vortex beams in turbulent atmosphere,'' {\em Journal of the
  Optical Society of America A} {\bf 39}(8), 1414--1419  (2022).

\bibitem{185}
Y.~Xu, B.~Lan, C.~Liu, {\em et~al.}, ``Self-focusing pin-like optical vortex
  beams resist atmospheric turbulence propagation for the space optical
  communication,'' in {\em 3rd International Conference on Laser, Optics, and
  Optoelectronic Technology (LOPET 2023)},  X.~Li and M.~F. Costa, Eds., {\em
  Proc. SPIE} {\bf 12757}, 127572T  (2023).
\newblock [doi:10.1117/12.2690387].

\bibitem{186}
J.~Durnin, ``Exact solutions for nondiffracting beams. i. the scalar theory,''
  {\em Journal of the Optical Society of America A} {\bf 4}(4), 651--654
  (1987).

\bibitem{187}
E.~A. Chan, G.~Adamo, S.~A. Aljunid, {\em et~al.}, ``Plasmono-atomic
  interactions on a fiber tip,'' {\em Applied Physics Letters} {\bf 116}(18)
  (2020).

\bibitem{188}
A.~H. Yang, S.~D. Moore, B.~S. Schmidt, {\em et~al.}, ``Optical manipulation of
  nanoparticles and biomolecules in sub-wavelength slot waveguides,'' {\em
  Nature} {\bf 457}(7225), 71--75  (2009).

\bibitem{189}
Y.~Sun, P.~Parra-Rivas, G.~P. Agrawal, {\em et~al.}, ``Multimode solitons in
  optical fibers: a review,'' {\em Photonics Research} {\bf 12}(11), 2581--2632
   (2024).

\bibitem{190}
J.~Berthelot, S.~S. A{\'c}imovi{\'c}, M.~L. Juan, {\em et~al.},
  ``Three-dimensional manipulation with scanning near-field optical
  nanotweezers,'' {\em Nature nanotechnology} {\bf 9}(4), 295--299  (2014).

\bibitem{191}
L.~Schermelleh, P.~M. Carlton, S.~Haase, {\em et~al.}, ``Subdiffraction
  multicolor imaging of the nuclear periphery with 3d structured illumination
  microscopy,'' {\em science} {\bf 320}(5881), 1332--1336  (2008).

\bibitem{234}
J.~Zhang, H.~Fang, P.~Wang, {\em et~al.}, ``Optical microfiber or nanofiber: a
  miniature fiber-optic platform for nanophotonics,'' {\em Photonics Insights}
  {\bf 3}(1), R02--R02  (2024).

\bibitem{192}
D.~Kedar and S.~Arnon, ``Urban optical wireless communication networks: the
  main challenges and possible solutions,'' {\em IEEE Communications Magazine}
  {\bf 42}(5), S2--S7  (2004).

\bibitem{193}
M.~A. Khalighi and M.~Uysal, ``Survey on free space optical communication: A
  communication theory perspective,'' {\em IEEE communications surveys \&
  tutorials} {\bf 16}(4), 2231--2258  (2014).

\bibitem{194}
H.~Kaushal and G.~Kaddoum, ``Optical communication in space: Challenges and
  mitigation techniques,'' {\em IEEE communications surveys \& tutorials} {\bf
  19}(1), 57--96  (2016).

\bibitem{195}
H.~Hemmati, ``Near-earth laser communications,'' in {\em Near-Earth Laser
  Communications, Second Edition},  1--40, CRC press  (2020).

\bibitem{196}
A.~Trichili, M.~A. Cox, B.~S. Ooi, {\em et~al.}, ``Roadmap to free space
  optics,'' {\em Journal of the optical society of America B} {\bf 37}(11),
  A184--A201  (2020).

\bibitem{197}
J.~Liu, C.~Cai, S.~Wang, {\em et~al.}, ``Rayleigh length extension in
  long-distance free-space optical communications based on lens group
  optimization,'' {\em Optics Express} {\bf 32}(10), 16891--16900  (2024).

\bibitem{198}
Z.~Lu, B.~Yan, K.~Chang, {\em et~al.}, ``Space division multiplexing technology
  based on transverse wavenumber of lommel--gaussian beam,'' {\em Optics
  Communications} {\bf 488}, 126835  (2021).

\bibitem{199}
Y.~Zhang, M.~Wang, and Z.~Zhou, ``Propagation of nonuniformly correlated bessel
  beams in the air--sea turbulent link,'' {\em Journal of the Optical Society
  of America B} {\bf 38}(6), 1900--1908  (2021).

\bibitem{200}
M.~Z. Chaari and S.~Al-Maadeed, ``Testing the efficiency of laser technology to
  destroy the rogue drones,'' {\em Security and Defence Quarterly} {\bf 32}(5),
  31--38  (2020).

\bibitem{201}
Y.~Liu, X.~Dang, and X.~Fu, ``Array detector systems for satellite-to-ground
  atmospheric coherent laser communications: performance evaluation,'' {\em
  Optics Express} {\bf 33}(6), 14288--14303  (2025).

\bibitem{202}
M.-C. Cheng, H.-Y. Hsu, S.~T. Hayle, {\em et~al.}, ``A simulated 1000-km leo
  satellite-to-ground station laser communication using a 1.8-km owc link,''
  {\em Journal of Lightwave Technology}   (2025).

\bibitem{203}
H.~Yu, L.~Li, Y.~Hou, {\em et~al.}, ``Deep learning-based prediction of
  atmospheric turbulence toward satellite-to-ground\break laser
  communication,'' {\em Optics Letters} {\bf 50}(2), 273--276  (2025).

\bibitem{204}
Z.~Yin, X.~Zhou, L.~Lu, {\em et~al.}, ``The influence of atmospheric
  turbulence-induced optical intensity scintillation on the accuracy of
  satellite-to-ground laser one-way timing,'' {\em Advances in Space Research}
  {\bf 75}(3), 2711--2720  (2025).

\bibitem{205}
X.~Hou, Z.~Liu, Y.~Chang, {\em et~al.}, ``Development status and trend analysis
  of satellite laser communication technology,'' {\em Chinese Journal of
  Lasers} {\bf 51}(11), 1101013--1101013  (2024).

\bibitem{206}
R.~Zhang, W.~Zhang, X.~Zhang, {\em et~al.}, ``Research status and development
  trends of high-orbit satellite laser relay links,'' {\em Laser \&
  Optoelectronics Progress} {\bf 58}(5), 0500001--0500001  (2021).

\bibitem{207}
C.~Xu, Y.~Jin, L.~Li, {\em et~al.}, ``Satellite-ground integrated wireless
  transmission technology for 6g,'' {\em Journal of Electronics \& Information
  Technology} {\bf 43}(1), 28--36  (2021).

\bibitem{208}
Z.~Yun, W.~Han, D.~Binbin, {\em et~al.}, ``Research progress and fronts in
  satellite-to-ground laser communication,'' {\em Chinese Journal of Space
  Science} {\bf 45}(2), 612--628  (2025).

\bibitem{209}
C.~Ma, Y.~Zhou, Y.~Jiang, {\em et~al.}, ``Performance of strong turbulence
  fluctuation suppression in coherent laser communication with array
  reception,'' {\em Acta Optica Sinica} {\bf 45}(9), 0906001--0906001  (2025).

\bibitem{210}
C.~Liu, R.~Wang, B.~Lan, {\em et~al.}, ``Research progress and prospects of
  pin-like beams for atmospheric turbulence suppression (cover
  article{\textperiodcentered} invited),'' {\em Infrared and Laser Engineering}
  {\bf 53}(5), 20240098--20240098  (2024).

\bibitem{211}
A.~E. Willner, H.~Song, K.~Zou, {\em et~al.}, ``Orbital angular momentum beams
  for high-capacity communications,'' {\em Journal of Lightwave Technology}
  {\bf 41}(7), 1918--1933  (2022).

\bibitem{212}
S.~Wang, J.~Xu, Y.~Yang, {\em et~al.}, ``Optimization of wireless optical
  communication using perfect vortex beam,'' {\em Optics Communications} {\bf
  556}, 130258  (2024).

\bibitem{213}
R.~Wang, B.~Lan, C.~Liu, {\em et~al.}, ``Performance evaluation of optical
  pin-like beams through atmospheric turbulence,'' {\em Physica Scripta}
  (2025).

\bibitem{214}
J.~Xu, P.~Zhang, H.~Chen, {\em et~al.}, ``Atmospheric turbulence transmission
  characteristics of inverted pin-like beams with nonuniform correlation,''
  {\em Acta Optica Sinica} {\bf 45}(12), 1201003--1201003  (2025).

\bibitem{215}
Z.~Jiang, X.~Su, H.~Zhou, {\em et~al.}, ``Experimental demonstration of a
  narrow and low-divergence ``pin-like'' beam in a 2-gbit/s ook fso link under
  turbulence effects when using a limited-size receiver aperture,'' in {\em
  ECOC 2024: 50th European Conference on Optical Communication},  1896--1899
  (2024).

\bibitem{216}
H.~Wang, M.~Guan, H.~Xiao, {\em et~al.}, ``Quadratic vortex optical pin beam
  with oam mode anti-degradation characteristic,'' {\em IEEE Photonics Journal}
    (2025).

\bibitem{217}
A.~Yu and G.~Wu, ``Self-healing properties of optical pin beams,'' {\em Journal
  of the Optical Society of America A} {\bf 40}(11), 2078--2083  (2023).

\bibitem{218}
R.~Wang, B.~Lan, C.~Liu, {\em et~al.}, ``Method for suppressing scintillation
  in up-link optical communication using optical pin-like beams propagating
  through atmospheric turbulence,'' {\em Photonics} {\bf 12}(7), 739  (2025).
\newblock [doi:10.3390/photonics12070739].

\bibitem{219}
{Bioengineer.org}, ``100 gbps free-space optical communication breakthrough,''
  {\em Bioengineer.org}   (2025).
\newblock [Online news article; accessed 17-December-2025].

\bibitem{220}
J.~Nie, L.~Tian, H.~Wang, {\em et~al.}, ``Adaptive beam shaping for enhanced
  underwater wireless optical communication,'' {\em Optics express} {\bf
  29}(17), 26404--26417  (2021).

\bibitem{221}
X.~Han, W.~Nie, P.~Li, {\em et~al.}, ``Analysis of research status and
  development trends in underwater wireless optical communication (invited),''
  {\em Acta Optica Sinica} {\bf 45}(13), 1306016--1306016  (2025).

\bibitem{233}
H.~Wang, W.~Wang, X.~Xu, {\em et~al.}, ``High-spatiotemporal-resolution
  structured illumination microscopy: principles, instrumentation, and
  applications,'' {\em Photonics Insights} {\bf 4}(1), R01--R01  (2025).

\bibitem{222}
W.~Wang, Y.~Wu, Z.~Liu, {\em et~al.}, ``Tunable optical pins and optical
  channels,'' {\em Optics Communications} , 132406  (2025).

\bibitem{223}
X.~Chen, H.~Yu, H.~Pan, {\em et~al.}, ``A layered split polishing path planning
  method used in awjp based on optical glass surface morphology,'' {\em
  International Journal of Precision Engineering and Manufacturing} {\bf
  26}(2), 269--282  (2025).

\bibitem{224}
H.~Liu, Z.~Lu, and F.~Li, ``Using diffractive optical element and zygo
  interferometer to test large-aperture convex surface,'' {\em Optics \& Laser
  Technology} {\bf 37}(8), 642--646  (2005).

\bibitem{225}
R.~Shukla, D.~Udupa, and M.~Aggarwal, ``Zygo interferometer for measuring
  refractive index of photorefractive bismuth silicon oxide (bi12sio20)
  crystal,'' {\em Optics \& Laser Technology} {\bf 30}(6-7), 425--430  (1998).

\bibitem{226}
M.~Zamboni-Rached, ``Stationary optical wave fields with arbitrary longitudinal
  shape by superposing equal frequency bessel beams: Frozen waves,'' {\em
  Optics Express} {\bf 12}(17), 4001--4006  (2004).

\bibitem{227}
E.~Recami and M.~Zamboni-Rached, ``Localized waves: a review,'' {\em Advances
  in Imaging and Electron Physics} {\bf 156}, 235--353  (2009).

\bibitem{228}
J.~A.~V. Mendon{\c{c}}a, M.~Zamboni-Rached, and E.~Recami, ``Arrays of frozen
  waves: Some theory and experiments,'' {\em Optics Communications} {\bf 482},
  126576  (2021).

\bibitem{229}
L.~A. Ambrosio, ``Analytical description of on-axis zero-order continuous
  frozen waves in the generalized lorenz--mie theory,'' {\em Journal of
  Quantitative Spectroscopy and Radiative Transfer} {\bf 296}, 108442  (2023).

\bibitem{230}
S.~Deng, D.~Yang, Y.~Zheng, {\em et~al.}, ``Transmittance of finite-energy
  frozen beams in oceanic turbulence,'' {\em Results in Physics} {\bf 15},
  102802  (2019).

\bibitem{231}
M.~Zamboni-Rached, ``Stationary optical wave fields with arbitrary longitudinal
  shape by superposing equal frequency bessel beams: Frozen waves,'' {\em
  Optics Express} {\bf 12}(17), 4001--4006  (2004).

\bibitem{232}
T.~A. Vieira, M.~R. Gesualdi, and M.~Zamboni-Rached, ``Frozen waves:
  experimental generation,'' {\em Optics Letters} {\bf 37}(11), 2034--2036
  (2012).

\bibitem{240}
W.~Shao, Y.~Wang, S.~Jia, {\em et~al.}, ``Terabit fso communication based on a
  soliton microcomb,'' {\em Photonics Research} {\bf 10}, 2802  (2022).

\bibitem{241}
G.~zhong Ding, B.~ke~Ma, J.~qi~Sun, {\em et~al.}, ``The impact of rainfall on
  the performance of free space laser communication systems,'' {\em
  Telecommunication Systems} {\bf 88}(3), 108--108  (2025).

\bibitem{242}
S.~M. Venkata and R.~G. K, ``Implementation of 23 gbps optical wireless link
  for 750 km inter-aircraft communications,'' {\em Engineering Research
  Express} {\bf 5}(1)  (2023).

\bibitem{243}
aff>R\&, ``Adaptive probabilistic shaped modulation for high-capacity
  free-space optical links,'' {\em Journal of Lightwave Technology} {\bf
  PP}(99), 1--1  (2020).

\bibitem{244}
M.~Sheng, D.~Zhou, S.~Ji, {\em et~al.}, ``Effects of space environment on
  satellite mega-constellations: From nodes and links to network performance,''
  {\em Engineering} {\bf 54}, 93--102  (2025).

\bibitem{258}
H.~Wang, Z.~Zhang, C.~Cui, {\em et~al.}, ``Improving rotational doppler
  velocimetry accuracy and spectral characteristics of vortex beam in maritime
  atmospheric turbulence,'' {\em IEEE Photonics Journal} {\bf 16}(4), 1--9
  (2024).

\bibitem{259}
Z.~Xiao, J.~Zhou, L.~Wang, {\em et~al.}, ``Propagation properties and
  spatial-mode uwoc performance of quasi-perfect optical vortex beam in an
  improved oceanic channel,'' {\em Optics \& Laser Technology} {\bf 188},
  112957  (2025).

\bibitem{260}
A.~Shikder, N.~Choudhary, and N.~K. Nishchal, ``Free-space propagation
  properties of gaussian vortex and bessel--gaussian vortex beams having
  fractional topological charges in turbulent medium,'' {\em Journal of Optics}
  {\bf 27}(6), 065603  (2025).

\bibitem{245}
H.~Jang, H.~Song, and H.~Jang, ``Evaluation of daylight background noise for
  satellite-to-ground free-space optical communication during daytime
  operation,'' {\em Photonics Research} {\bf 13}, 2630  (2025).

\bibitem{246}
J.~H. Lee, P.~Kim, J.-Y. Lee, {\em et~al.}, ``Beamforming for beam-squint
  effect mitigation in leo satellite communication systems,'' {\em ICT Express}
  {\bf 11}(6), 1103--1109  (2025).

\bibitem{247}
R.~Saiyyed, M.~Sindhwani, S.~Sachdeva, {\em et~al.}, ``Enabling high-speed and
  large-capacity data transmission in optical inter-satellite communication
  links under various conditions,'' {\em Journal of Optics} (prepublish), 1--12
   (2024).

\bibitem{248}
Y.~Li, H.~Zhang, P.~Huang, {\em et~al.}, ``Demonstration of 10 gbps
  satellite-to-ground laser communications in engineering,'' {\em The
  Innovation} {\bf 5}(1), 100557  (2024).

\bibitem{249}
A.~B. Wondmagegn, D.~Won, Q.~T. Do, {\em et~al.}, ``Performance analysis of
  fso-based communications in space–air–ground integrated networks: A
  comprehensive survey,'' {\em Computer Networks} {\bf 271}, 111579  (2025).

\bibitem{250}
A.~S. S. and S.~C. Dhongdi, ``Review of underwater mobile sensor network for
  ocean phenomena monitoring,'' {\em Journal of Network and Computer
  Applications} {\bf 205}, 103418  (2022).

\bibitem{251}
Z.~Lu, Z.~Li, X.~Lin, {\em et~al.}, ``170 gbps pdm underwater visible light
  communication utilizing a compact 5-<inline-formula><math display="inline"
  id="m1" xmlns:mml="http://www.w3.org/1998/math/mathml"><mrow><mi
  mathvariant="bold-italic">lambda</mi></mrow></math></inline-formula> laser
  transmitter and a reciprocal differential receiver,'' {\em Photonics
  Research} {\bf 13}, 1654  (2025).

\bibitem{252}
L.~Yang, ``<italic toggle="yes">photonics research</i> interview with professor
  yidong huang,'' {\em Photonics Research} {\bf 12}, 2521  (2024).

\bibitem{261}
Y.~Yang, A.~Forbes, and L.~Cao, ``A review of liquid crystal spatial light
  modulators: devices and applications,'' {\em Opto-Electronic Science} {\bf
  2}(8), 230026--1  (2023).

\bibitem{262}
Y.~Yang and L.~Cao, ``Quantum hyper-entangled system with multiple qubits based
  on spontaneous parametric down-conversion and birefringence effect,'' {\em
  Optical and Quantum Electronics} {\bf 56}(1), 12  (2024).

\bibitem{263}
J.~Gao, X.~Lu, X.~Zhao, {\em et~al.}, ``Rotational doppler effect using
  ultra-dense vector perfect vortex beams,'' {\em Photonics Research} {\bf
  13}(2), 468--476  (2025).

\bibitem{253}
P.~Gupta, Pravesh, R.~Rawal, {\em et~al.}, ``Biophotonics for early and
  recurrent oral cancers and pre-malignant lesions: Clinical cases and future
  directions for india,'' {\em Photodiagnosis and Photodynamic Therapy} {\bf
  56}(S), 105172--105172  (2025).

\bibitem{254}
A.~Gualerzi, S.~Picciolini, C.~Carlomagno, {\em et~al.}, ``Biophotonics for
  diagnostic detection of extracellular vesicles,'' {\em Advanced Drug Delivery
  Reviews} {\bf 174}(prepublish)  (2021).

\bibitem{255}
R.~C. Cowie and M.~Schubert, ``Light sheet microscope scanning of biointegrated
  microlasers for localized refractive index sensing,'' {\em Photonics
  Research} {\bf 12}, 1673  (2024).

\bibitem{256}
D.~G. Grier, ``A revolution in optical manipulation,'' {\em Nature} {\bf 424},
  810--816  (2003).

\bibitem{257}
T.~Sneh, S.~Corsetti, M.~Notaros, {\em et~al.}, ``Optical tweezing of
  microparticles and cells using silicon-photonics-based optical phased
  arrays,'' {\em Nature Communications} {\bf 15}(1), 8493--8493  (2024).

\bibitem{238}
T.~Pan, X.~Zhang, H.~Xin, {\em et~al.}, ``Progress in the application of
  light-driven micro/nanorobots in precision medicine (invited),'' {\em Acta
  Optica Sinica (Online)} {\bf 2}, 1816001  (2025).

\bibitem{239}
Y.~Chen, C.~Zhai, X.~Gao, {\em et~al.}, ``Optical manipulation of
  ratio-designable janus microspheres,'' {\em Photonics Research} {\bf 12},
  1239  (2024).

\end{thebibliography}
\bibliographystyle{spiejour}   


\listoffigures
\listoftables

\end{spacing}
\end{document}